\documentclass[authoryear]{elsarticle}
\usepackage[english]{babel}
\usepackage{fancyhdr} % um Header und Footer zu designen
\usepackage{url} % zum korrekten Einf\"ugen von URL mit dem \url Befehl, das Package hyperref erzeugt dazu noch Links
\usepackage{caption} % um Bilder neben Bildern oder Tabellen neben Tabellen zu erzeugen
\usepackage{subcaption} % um Bilder neben Bildern oder Tabellen neben Tabellen zu erzeugen
\usepackage{natbib} %erlaubt es, andere Zitatdarstellungen vorzunehmen als numerische
\usepackage{multirow} %erlaubt es, in einer Tabelle mehrerer Zeilen zu einer zusammenzufassen
\usepackage{float} % notwendig, um Abbildungen an der Stelle zu platzieren, an der sie im Quelltext stehen
\usepackage{amssymb}
\usepackage[intlimits]{amsmath} % beides sind Mathematikpakete
\usepackage{amsthm} % Um Theoreme, Beweise etc. zu verwalten
\usepackage[german]{nomencl}
\usepackage{booktabs} % f\"ur bessere Tabellenlinien
\usepackage{geometry}
\usepackage{enumitem}
\usepackage{array}
\makenomenclature
\usepackage[pdfborder={000},colorlinks=true,linkcolor=blue,citecolor=blue]{hyperref}
\usepackage{listings}
\usepackage[usenames,dvipsnames]{color}
\usepackage{tikz}
\usetikzlibrary{mindmap,trees, backgrounds}

\definecolor{b}{rgb}{0,0,.8}	%%omega-blau
\definecolor{g}{rgb}{0,.6,0}	%%Tau-grün
\definecolor{n}{rgb}{0,0,0}	%%normal-schwarz
\definecolor{h}{rgb}{0.4,0.2,0.2}	%%hint
\definecolor{v}{rgb}{0.2,0.6,0}

\newcommand{\T}{{\mathbb T}}

\newcommand{\GG}{{\mathcal{G}}}

\newcommand{\TT}{{\mathcal{T}}}

\newcommand{\XX}{{\mathcal{X}}}

\newcommand{\bsY}{\boldsymbol Y}

\newcommand{\bsone}{\boldsymbol 1}

\newcommand{\bsmu}{\boldsymbol \mu}
\newcommand{\bseps}{\boldsymbol \varepsilon}

\newcommand{\bsPhi}{\boldsymbol \Phi}

\newcommand{\eps}{{\varepsilon}}

\newcommand{\ov}\overline
\newcommand{\what}{\widehat}
\newcommand{\wtilde}{\widetilde}

\newcommand{\rig}\right
\newcommand{\lef}\left
\newcommand{\nf}\normalfont

\newcommand{\MAE}{\text{MAE}}

\newcommand{\RMSE}{\text{RMSE}}

\newcommand{\EXAA}{\text{EXAA}}

\definecolor{rickgreen}{rgb}{0,0.6,0}

\begin{document}
\title{Forecasting day ahead electricity spot prices: \\ The impact of the EXAA to other European electricity markets}

\author{Florian~Ziel}
\ead{ziel@europa-uni.de}

\author{Rick~Steinert}
\ead{steinert@europa-uni.de}

\author{Sven~Husmann}
\ead{husmann@europa-uni.de}
\address{Europa-Universit\"at Viadrina, Gro\ss e Scharrnstra\ss e 59, 15230 Frankfurt (Oder), Germany}

\begin{keyword}
Electricity price \sep EXAA\sep AR-Model \sep Forecasting \sep European electricity markets 
\end{keyword}
\begin{frontmatter}
\lhead{\nouppercase{\leftmark}}
\begin{abstract}
In our paper we analyze the relationship between the day-ahead electricity price of the Energy Exchange Austria (EXAA) and other day-ahead electricity prices in Europe. We focus on markets, which settle their prices after the EXAA, which enables traders to include the EXAA price into their calculations.  For each market we employ econometric models to incorporate the EXAA price and compare them with their counterparts without the price of the Austrian exchange. By employing a forecasting study, we find that electricity price models can be improved when EXAA prices are considered.

%
%In this paper we show that the hardly analysed electricity spot price of the Austrian EXAA has 
%a huge impact to other European markets. As the Austrian spot market trades earlier than
%all the other spot markets, this information can be used for forecasting other electricity market.
%We show evidence for the huge improvements of the several spot prices 
%EEX (German and Austria, Swiss, France), if the EXAA acution results are considered as well. 
%The benifit is largest for the german EEX spot price. For modelling and forecasting we use
%autoregressive time series models to show that even by the use of simple models the improvement is great.

\end{abstract}
\end{frontmatter}

\section{Introduction} \label{Introduction}
Electricity is a standardized cross-border traded commodity. Especially in Europe, where an ongoing market integration between countries proceeds rapidly, national markets cannot be considered as one isolated trading place. \par

Several authors have studied the relation of European electricity markets empirically within the past years. For instance, using price data of forward markets \cite{bunn2010integration} showed by analyzing cross-market interactions of some
of the major European electricity markets that they are integrated. Moreover, they provide evidence for an increase in this integration over time. The German electricity market turned out to be the most integrated market. According to their study the integration is not necessarily reliant on sharing a geographical border: The Spanish and German market, for instance, seemed to transmit shocks as well. The important role of the German electricity market for other European markets was also pointed out by \cite{bollino2013integration}. Using cointegration techniques they find that the German electricity price embodies a price signal for the other investigated European markets, e.g. France and Italy. \par 
Even though the hypothesis of market integration for some of the major markets seems to be satisfied (see also \cite{bosco2010long}, \cite{kalantzis2010market} or \cite{houllier2012fractional}), it is debatable if this holds true for every European market. \cite{zachmann2008electricity} as well as \cite{huisman2013history} argued that especially some of the Scandinavian electricity prices are behaving differently. This issue was also analyzed in detail by \cite{ferkingstad2011causal}. They were able to show that at least for the weekly time series of Nordic and German electricity prices a connection through gas prices is present. \par 
In our paper we exploit those findings by combining them with the different specifications of the European exchanges. As the price results for the day-ahead auction of each of these markets are revealed at different points in time, even though the same trading period is considered, we use the relationship of those markets to improve common modeling approaches. As basic electricity exchange we focus on the Energy Exchange Austria (EXAA) for two reasons. First, the EXAA discloses day-ahead prices prior to most of the other European exchanges which are connected with Germany and Austria. Second, the EXAA contains a special case of price relations, where not only the same time period is traded prior to other markets but also the same market region. This holds true for the European Power Exchange (EPEX) and the EXAA, as both cover Germany and Austria.
The EXAA reveals their prices approximately at 10:20 pm, whereas offers to the EPEX can be submitted until 12:00 pm. These EXAA prices are the prices for all days ahead up to the next working day, which can be e.g. three days at once for weekends.  If there is a systematic relation between both markets present, traders could use the price information of the EXAA to adjust their bidding structure. This approach is applied to many European exchanges. As the EXAA covers Germany and Austria, we focus only on these European markets, which are directly connected with Germany and Austria.  \par 
The existent literature concerning the usage of the early price disclosure of the EXAA is very scarce. It was only discussed in the framework of forward risk premiums, for instance in \cite{ronn2009intra} or \cite{erni2012day}. In these studies the EXAA price is usually regarded as an early price signal for electricity of the EPEX or European Energy Exchange (EEX) respectively. \cite{viehmann2011risk} for instance considers the EXAA prices as a snapshot for the German and Austrian electricity price traded via over-the-counter (OTC) business. However, a direct application of the EXAA price into modeling the electricity price of other European markets has, to our knowledge, not been done so far. \par
Our paper closes this gap by considering the time series of EXAA electricity prices as an external regressor. 
Because, in an econometric modeling framework, autoregressive models turn out to be of superior model performance,
we will estimate the most common basic approaches and compare them with their counterparts when the EXAA electricity price is used as influencing variable. \par
Therefore we organized our paper as follows. In section 2 we will describe the different data sets and exchange specifications. Moreover, we will provide information about the necessary data arrangement for using the EXAA price as a regressor. The subsequent section will then introduce the econometric models which are applied to our data set. In section 4 we will employ a comprehensive forecasting study for every examined market place. Section 5 discusses our findings and analyses the temporal structure of the relationship. We will focus on short term forecasting of one day ahead.
The last section summarizes our results and grants insights for future research. 

 Throughout the whole paper we will use bold symbols for multivariate expressions and normal symbols 
for univariate objects.

\section{The considered electricity markets} 
%data range 15.08.2007-12.08.2014
In order to measure the impacts of the EXAA day-ahead price on other electricity exchanges, it is mandatory to determine a feasible set of these exchanges. In our case we decided to use exchanges, which are directly connected with Germany and Austria, as the EXAA covers both countries. According to the transparency platform of the European Network of Transmission System Operators for Electricity (ENTSOE) there are 10 different countries with a cross border physical flow to either Germany or Austria. These are Switzerland, Czech Republic, Denmark, France, Hungary, Italy, Netherlands, Poland, Slovenia and Sweden. Denmark operates two different interconnectors with such a cross-border physical flow to the investigated region. This specifically means that we ignore countries like Belgium or Slovakia that are geographically connected to Germany and Austria, but have no cross border physical flow. To each country and interconnector there can be a specific spot market price of electricity assigned. As 
the German electricity 
price is traded at two different exchanges, there are 12 different possible electricity price time series in consideration. \par
However, it is arguable that also indirectly connected markets may be affected by the EXAA. This would also coincide with the findings of \cite{bunn2010integration}. But analyzing every indirectly related market would lead in a tremendously big study. One could even think of the Russian electricity market being affected by the EXAA – there would simply be almost no end to markets in focus. Therefore we decided to employ a criterion which, in our opinion, is intuitive and comprehensible and also grants a manageable size of markets to consider. Using the direct connection in terms of an interconnector fulfills these goals very well. Despite that fact we also analyzed the electricity price data of Spain, Norway and Belgium exemplarily, but the effects were not significant. We also avoided specifically the Nordpool spot price as it is only a reference price that is computed from several Scandinavian prices, like Danish, Swedish and Norway etc. and therefore not traded directly.
\par
In order to use the information of the EXAA price disclosure as a snapshot for the market, it is mandatory to filter out those electricity markets, which allow for order submission post to the EXAA price results. \par
The set of investigated countries is shown in Table \ref{tab_exchanges}. It presents an overview about the name of the exchange, the used abbreviation within this paper, the latest submission time point, the time point of price results and the data source for the information gathered on these exchanges, including the time series of prices. Information in brackets corresponds to the name of a specific price zone of the region. Except for Poland they are chosen to represent the special zone which is directly connected with Germany and Austria. The two exchanges beneath the double horizontal line represent those markets, which are connected to Germany and Austria but did not meet our second criterion, as they did not allow for order submission post to the price disclosure of the EXAA. \par
\begin{figure}[h!]
\begin{center}
\resizebox{\textwidth}{0.4\textheight}{
\begin{tikzpicture}[mindmap,
  level 1 concept/.append style={level distance=130,sibling angle=30},
  extra concept/.append style={color=blue!50,text=black}]
   
\definecolor{col_epex}{rgb}{0.4,.8,1}
\definecolor{col_epex_exaa}{rgb}{0.4,0.4,1}
\definecolor{col_hupx}{rgb}{1,0.3,0.6}
\definecolor{col_ote}{rgb}{1,0.2,0.1}
\definecolor{col_polpx}{rgb}{1,.6,0}
\definecolor{col_nordpool}{rgb}{0.8,1,0.2}
\definecolor{col_apx}{rgb}{0,1,0.2}
\definecolor{col_others}{rgb}{.7,.7,.7}

   \begin{scope}[mindmap, concept color=col_epex_exaa, text=black]

    \node [concept, text=black, minimum size=30, scale=1] (ger) at (0,0) {Germany}
      child [concept, color=col_epex_exaa!100, grow=-60, level distance=120]{
        node [concept, minimum size=30, scale=1.3, text=black] (aus) {Austria}
	  child [concept color=col_hupx!100, grow=-15, level distance=100]{
	    node [concept, text=black] (hu) {Hungary}
% 	    child [concept color=col_others!100, grow=-60, level distance=55]{
% 	      node [concept, text=black] (cro) {\textit{Croatia}}
% 	    }
	  }
	  child [concept color=col_others!100, grow=-50, level distance=90]{
	    node [concept, text=black] (slov) {\textit{Slovenia}}
	  }
	  child [concept color=col_others!100, grow=-140, level distance=100]{
	    node [concept, text=black] (it) {\textit{Italy Nord}}
	  }
        }
      child [concept color=col_ote!100, grow=-17, level distance=115]{
        node [concept] (cz) {Czech}
% % 	child [concept color=col_ote!100, grow=-35, level distance=70]{
% % 	  node [concept] (sk) {Slovakia}
% % 	}
      }
      child [concept color=col_polpx!100, grow=20, level distance=120]{
        node [concept] (pol) {Poland}
      }
      child [concept color=col_nordpool!100, grow=55, level distance=155]{
        node [concept] (sw) {Sweden4}
      }
      child [concept color=col_nordpool!100, grow=80, level distance=110]{
        node [concept] (dke) {Denmark East}
      }
      child [concept color=col_nordpool!100, grow=115, level distance=120]{
        node [concept] (dkw) {Denmark West}
      }
      child [concept color=col_apx!100, grow=150, level distance=120]{
        node [concept] (ned) {Netherlands}
%         child [concept color=col_nordpool!100, grow=100, level distance=70]{
% 	  node [concept] (norw) {\textit{Norway2}}
% 	}
%         child [concept color=col_apx!100, grow=230, level distance=80]{
% 	  node [concept] (uk) {\textit{United Kingdom}}
% 	}
%         child [concept color=col_apx!100, grow=280, level distance=70]{
% 	  node [concept] (bel) {\textit{Belgium}}
% 	}
      }
      child [concept color=col_epex, grow=215, level distance=140]{
        node [concept] (fr) {France}
      }
      child [concept color=col_epex, grow=250, level distance=100]{
        node [concept] (switz) {Switzerland}
      }
;

   \end{scope}

    \begin{scope}
\node[rectangle, minimum width=1cm,minimum height=1cm, scale=.8] at (7,5.2) {Spot Markets} ;
\node[rectangle, fill=col_epex_exaa, minimum width=1cm,minimum height=1cm, scale=.8] at (7,4.5) {EPEX+EXAA} ;
\node[rectangle, fill=col_epex, minimum width=1cm,minimum height=1cm, scale=.8] at (7,3.5) {EPEX} ;
\node[rectangle, fill=col_apx, minimum width=1cm,minimum height=1cm, scale=.8] at (7,2.5) {APX} ;
\node[rectangle, fill=col_nordpool, minimum width=1cm,minimum height=1cm, scale=.8] at (7,1.5) {Nordpool} ;
\node[rectangle, fill=col_polpx, minimum width=1cm,minimum height=1cm, scale=.8] at (7,0.5) {POLPX} ;
\node[rectangle, fill=col_ote, minimum width=1cm,minimum height=1cm, scale=.8] at (7,-0.5) {OTE} ;
\node[rectangle, fill=col_hupx, minimum width=1cm,minimum height=1cm, scale=.8] at (7,-1.5) {HUPX} ;
\node[rectangle, fill=col_others, minimum width=1cm,minimum height=1cm, scale=.8] at (7,-2.5) {Others} ;

\end{scope}

  \begin{pgfonlayer}{background}
    \draw [circle connection bar, color=col_epex_exaa!50]
       (aus) to [circle connection bar switch color=from (col_epex_exaa!100!black) to (col_ote!100!black)] (cz);
    \draw [circle connection bar, color=col_epex_exaa!50]
       (aus) to [circle connection bar switch color=from (col_epex_exaa!100!black) to (col_epex!100!black)] (switz);
  %additional ones
   \draw [circle connection bar, color=col_epex_exaa!50]
        (cz) to [circle connection bar switch color=from (col_ote!100!black) to (col_polpx!100!black)] (pol);
   \draw [circle connection bar, color=col_epex_exaa!50]
        (pol) to [circle connection bar switch color=from (col_polpx!100!black) to (col_nordpool!100!black)] (sw);
   \draw [circle connection bar, color=col_epex_exaa!50]
        (sw) to [circle connection bar switch color=from (col_nordpool!100!black) to (col_nordpool!100!black)] (dke);
   \draw [circle connection bar, color=col_epex_exaa!50]
        (dke) to [circle connection bar switch color=from (col_nordpool!100!black) to (col_nordpool!100!black)] (dkw);
%    \draw [circle connection bar, color=col_epex_exaa!50]
%         (dkw) to [circle connection bar switch color=from (col_nordpool!100!black) to (col_nordpool!100!black)] (norw);
%    \draw [circle connection bar, color=col_epex_exaa!50]
%         (fr) to [circle connection bar switch color=from (col_epex!100!black) to (col_apx!100!black)] (bel);
%    \draw [circle connection bar, color=col_epex_exaa!50]
%         (fr) to [circle connection bar switch color=from (col_epex!100!black) to (col_apx!100!black)] (uk);
   \draw [circle connection bar, color=col_epex_exaa!50]
        (fr) to [circle connection bar switch color=from (col_epex!100!black) to (col_epex!100!black)] (switz);
   \draw [circle connection bar, color=col_epex_exaa!50]
        (fr) to [circle connection bar switch color=from (col_epex!100!black) to (col_others!100!black)] (it);
   \draw [circle connection bar, color=col_epex_exaa!50]
        (switz) to [circle connection bar switch color=from (col_epex!100!black) to (col_others!100!black)] (it);
   \draw [circle connection bar, color=col_epex_exaa!50]
        (it) to [circle connection bar switch color=from (col_others!100!black) to (col_others!100!black)] (slov);
%    \draw [circle connection bar, color=col_epex_exaa!50]
%         (cro) to [circle connection bar switch color=from (col_others!100!black) to (col_others!100!black)] (slov);
%   \draw [circle connection bar, color=col_epex_exaa!50]
%         (hu) to [circle connection bar switch color=from (col_hupx!100!black) to (col_ote!100!black)] (sk);
%   \draw [circle connection bar, color=col_epex_exaa!50]
%         (pol) to [circle connection bar switch color=from (col_others!100!black) to (col_ote!100!black)] (sk);

       ;
\end{pgfonlayer}

\end{tikzpicture}
} % end resizebox
\end{center}
\caption{Map of the investigated electricity markets with their connections}
\label{fig_map}
\end{figure}
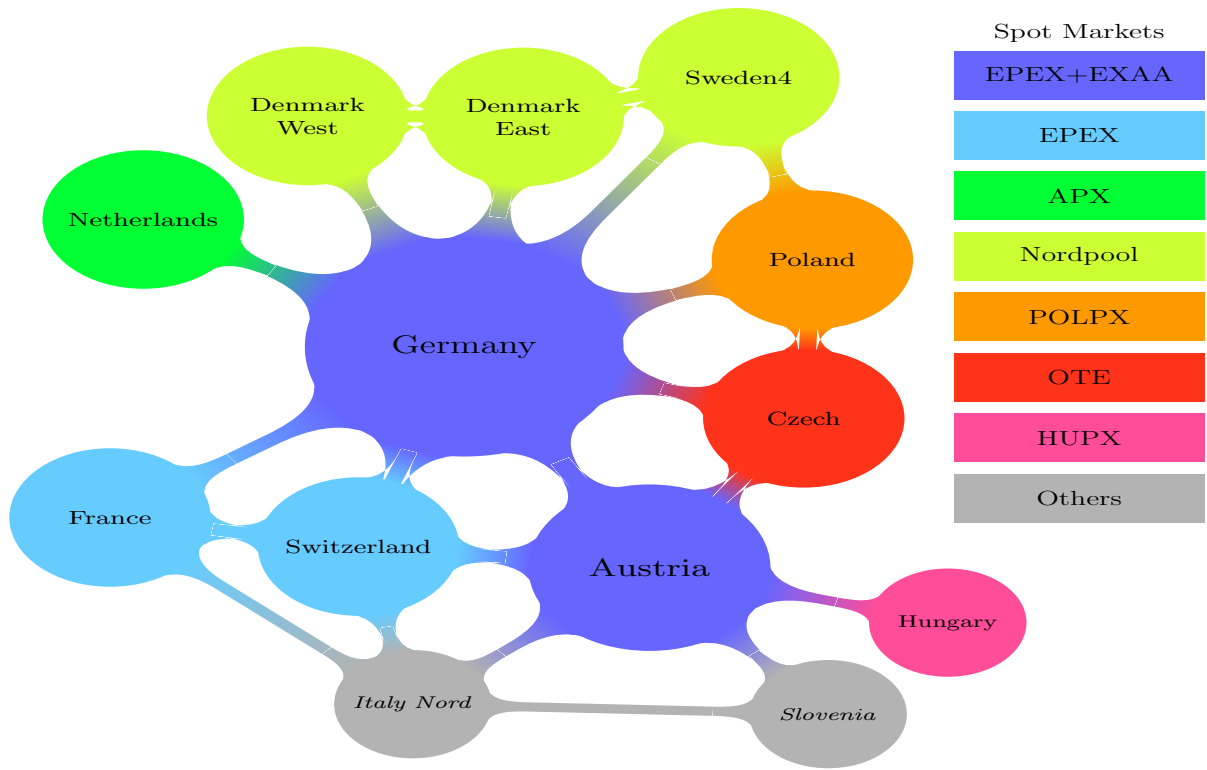

Figure  \ref{fig_map} illustrates those markets and provides additional information on the connection between the regions. Such a connection is depicted as a link between two bubbles. Any colored bubble represents a region which was finally included after the filtering described above. Hence, our dataset contains 10 different time series, for which we will utilize a possible relationship with the EXAA. \par 

%  [1] "EXAA.DE&AT" "EPEX.DE&AT" "EPEX.CHE"   "EPEX.FR"    "BELPEX.BE" 
%  [6] "NP.DK.West" "NP.DK.East" "NP.SW4"     "POLPX.PL"   "OTE.CZ"    
% [11] "HUPX.HU"
\begin{table}[tbh]
\centering
%\resizebox{\textwidth}{!}{
\begin{tabular}{rllccl}
  %\hline
 \textbf{Exchange} & \textbf{Region}& \textbf{Abbreviation}& \textbf{Sub.}&  \textbf{Res.} & \textbf{Data source}\\ \hline
 EXAA &Germany\&Austria & EXAA.DE\&AT &10:12 & 10:20 & exaa.at\\
 EPEX &Germany\&Austria& EPEX.DE\&AT & 12:00 & 12:42 &epexspot.com\\
 EPEX &Switzerland  & EPEX.CHE & 11:00 & 11:10 &epexspot.com\\
 EPEX &France & EPEX.FR & 12:00 & 12:42 &epexspot.com\\ 
%  BELPEX & Belgium &  12:00 & 13:05 & every day (???)\\ 
 APX & Netherlands & APX.NL & 12:00 & 12:55 & apxgroup.com\\ 
% GME Italy (several prices for different regions, e.g. Sicily, Sardina, Northern Italy,...) & till 9:15 & 10:45 & ? \\

%  BSP Serbia & till 10:15 &  10:35 & all days \\
 Nordpool & Denmark (West) & NP.DK.West & 12:00  & 12:42 & nordpoolspot.com\\
 Nordpool & Denmark (East) & NP.DK.East & 12:00  & 12:42 & nordpoolspot.com\\
 Nordpool & Sweden (4) & NP.SW4 & 12:00  & 12:42 & nordpoolspot.com \\
  POLPX & Poland (Auction I) & POLPX.PL & 10:30  &  10:35  & tge.pl \\
 OTE &  Czech Republic& OTE.CZ &11:00 & 11:30 & ote-cr.cz \\ % 14:30 (from document), but 11:30 on website???
 HUPX &  Hungary & HUPX.HU &11:00 & 11:30 & hupx.hu \\ % 14:30 (from document), but 11:30 on website???
%  OMIT & Spain & 12:00 (10:00 till 14th Oct 2013) & 14:00 & every day\\  % (+PT, as PT has WET)
 \midrule % In-table horizontal line
\midrule % In-table horizontal line 
   GME & Italy (Nord) & GME.ITN   & 9:15 &  10:45 & mercatoelettrico.org \\
  BSP & Slovenia & BSP.SL & 9:40 &  10:30 & bsp-southpool.com \\
%  Nordpool & Norway & 12:00  & 12:30-45 &every day\\
%  Nordpool & \textit{System price}& 12:00  & 12:30-45 &every day\\
 \end{tabular}
% }
\caption{Summary of the considered electricity markets, times in CET (UTC+1/+2). (Sub. = submission, Res. = price results)}
\label{tab_exchanges}
\end{table}

\begin{figure}[hbt!]
\centering
 \includegraphics[width=0.49\textwidth]{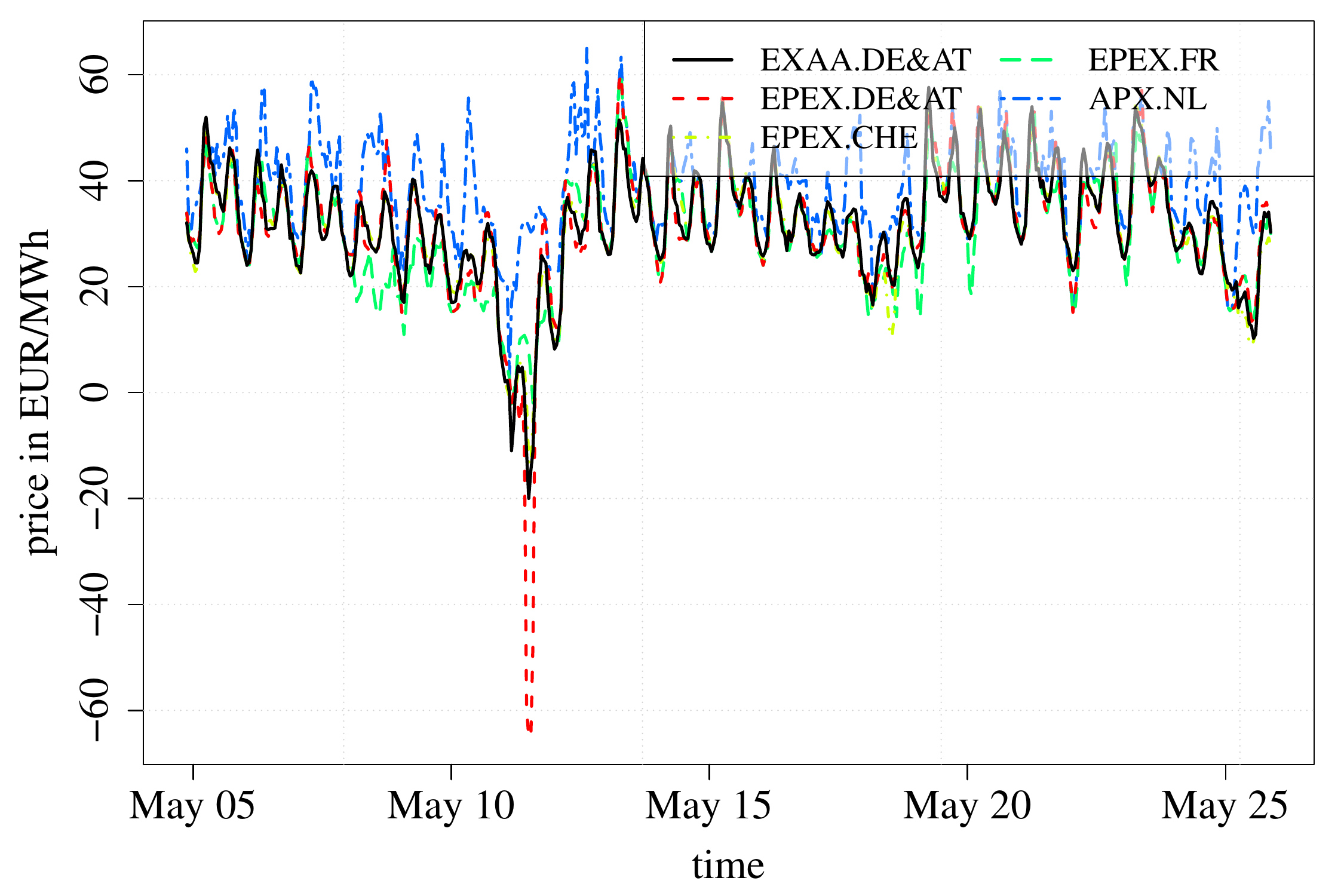}
 \includegraphics[width=0.49\textwidth]{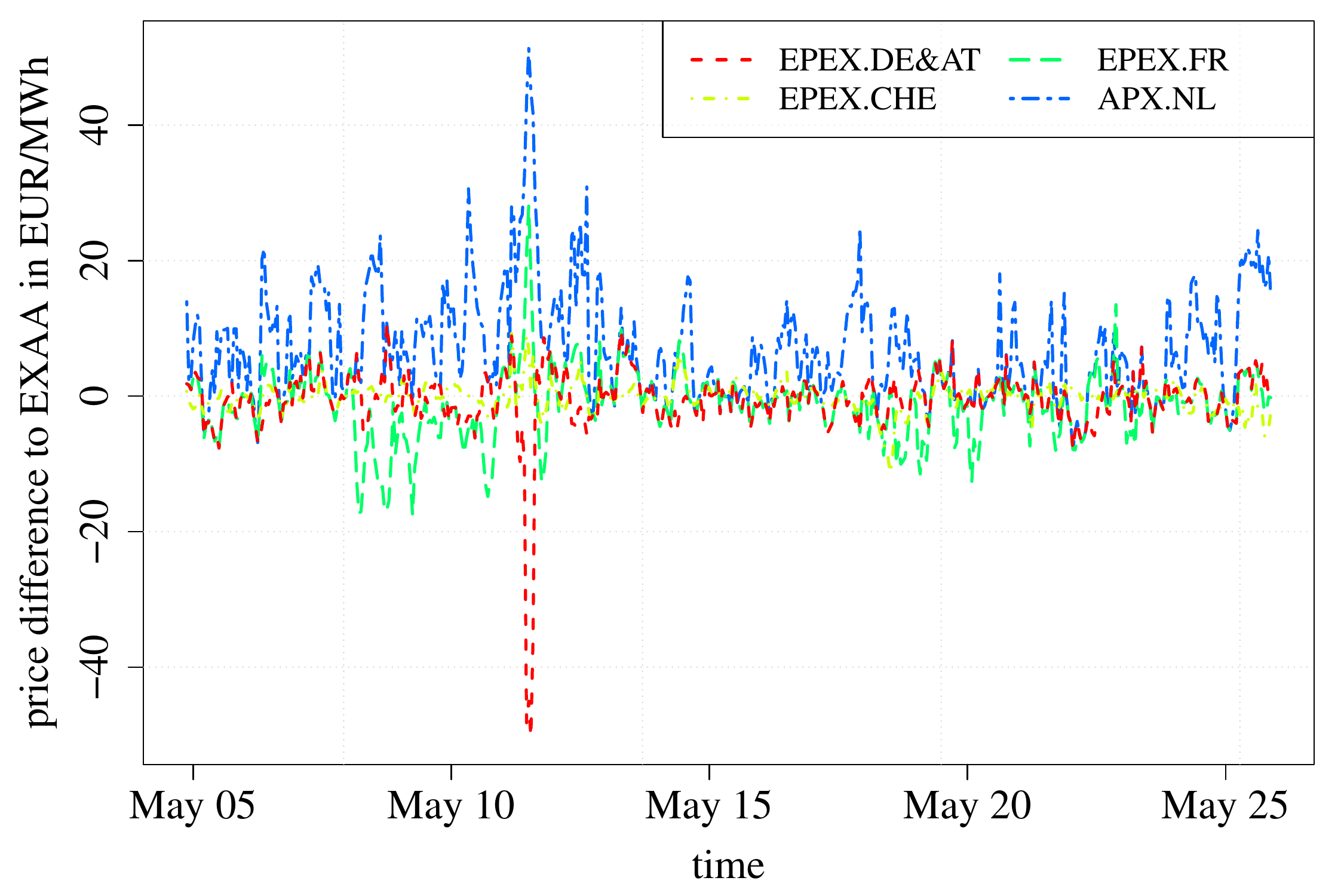}
 \includegraphics[width=0.49\textwidth]{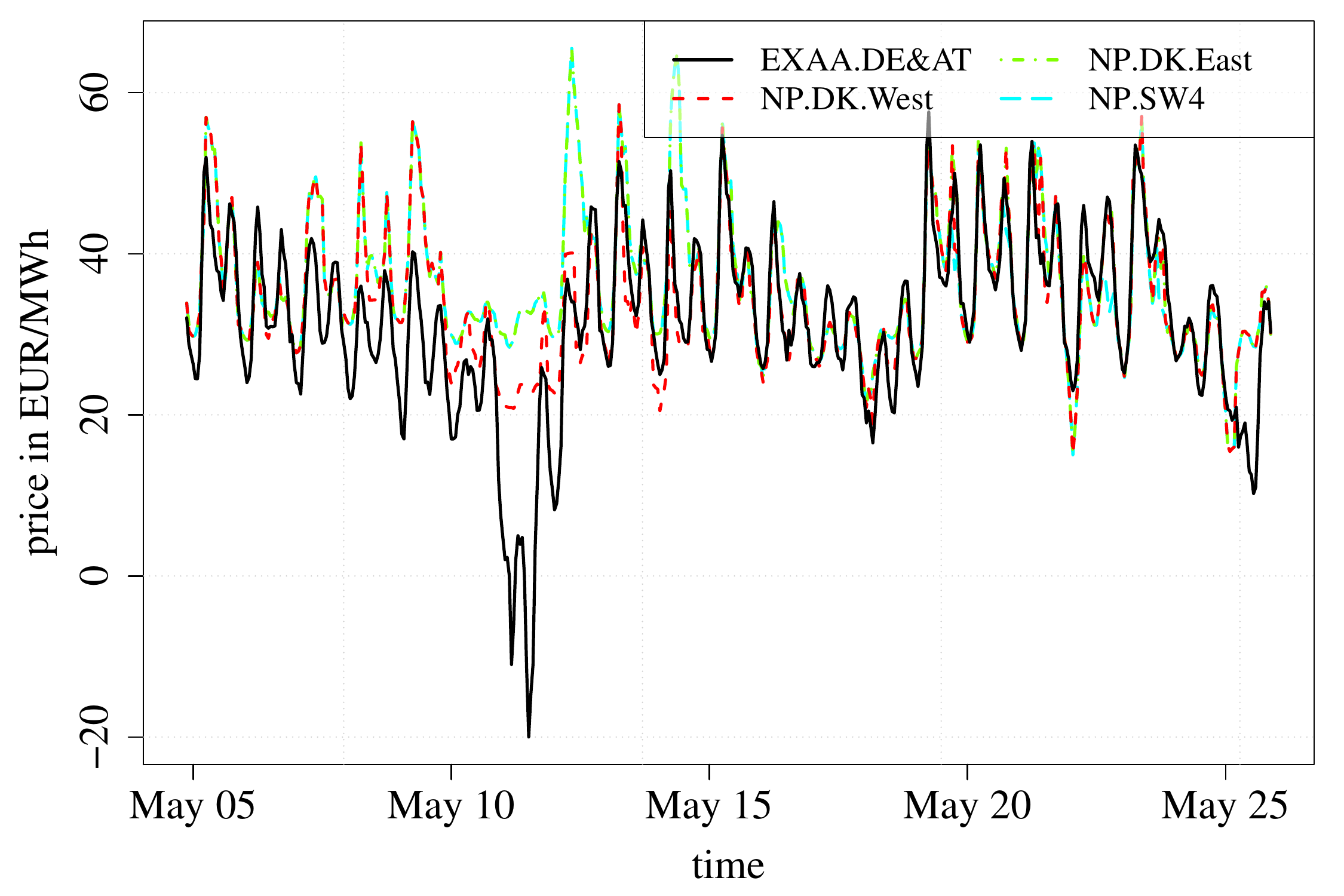}
 \includegraphics[width=0.49\textwidth]{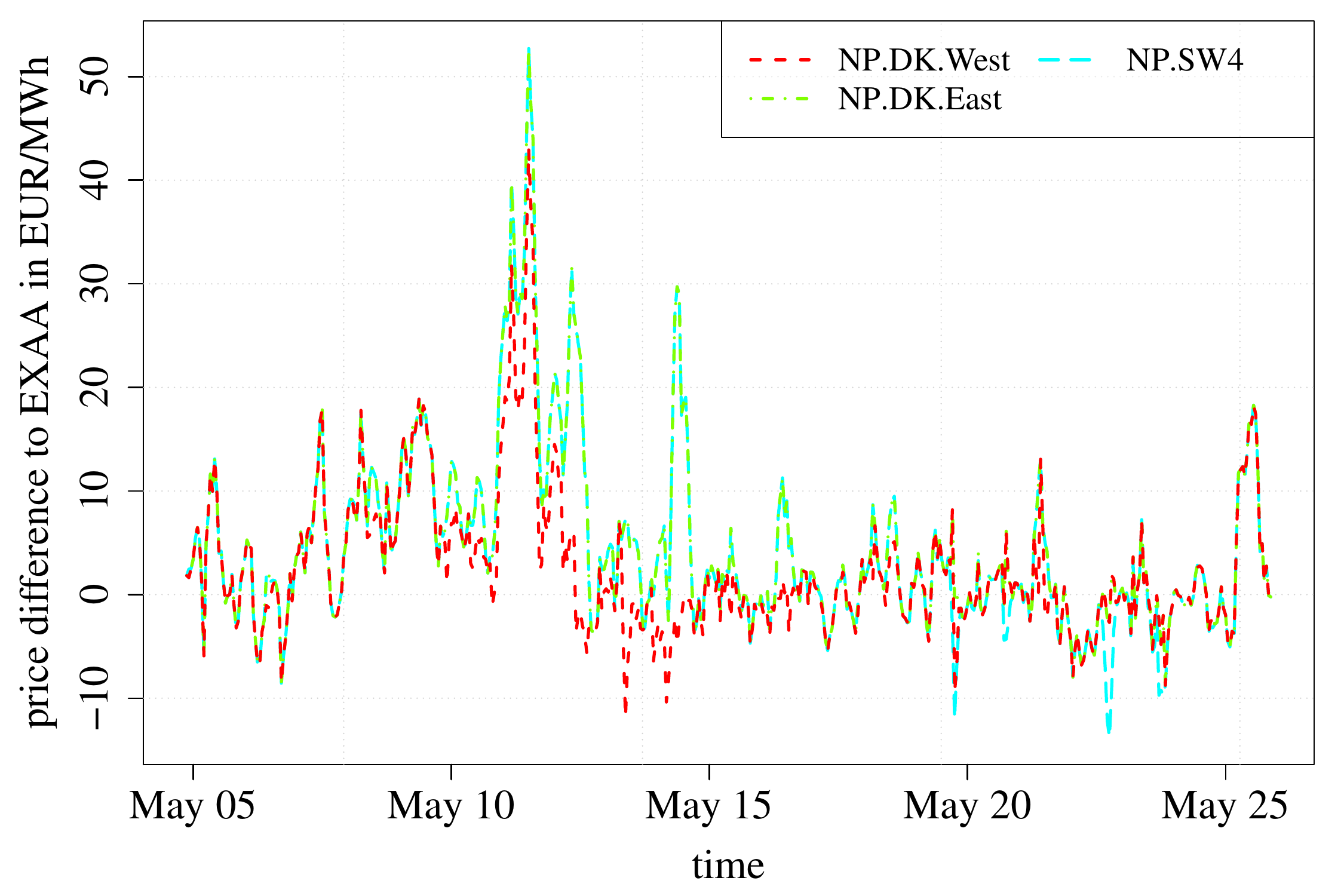}
 \includegraphics[width=0.49\textwidth]{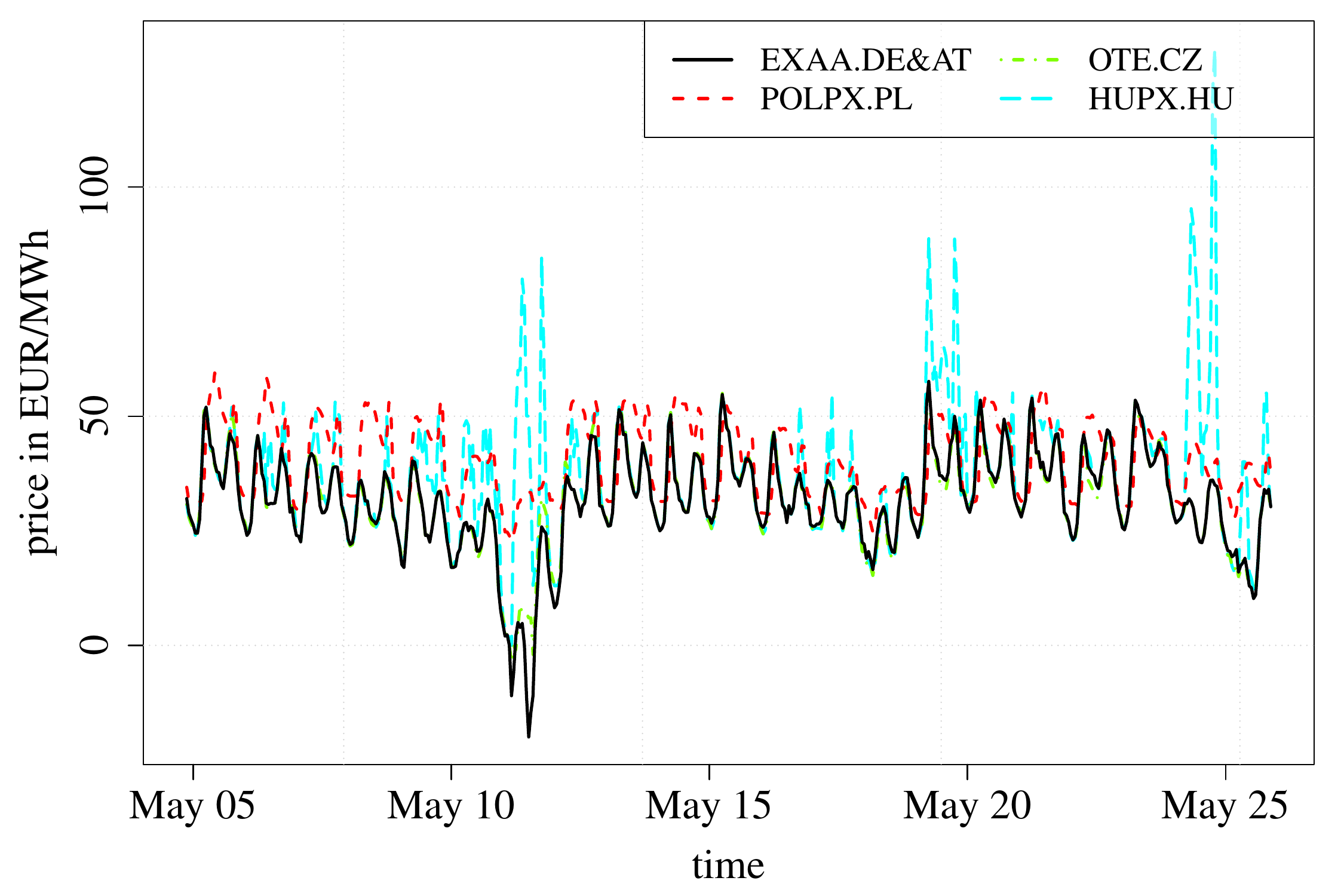}
 \includegraphics[width=0.49\textwidth]{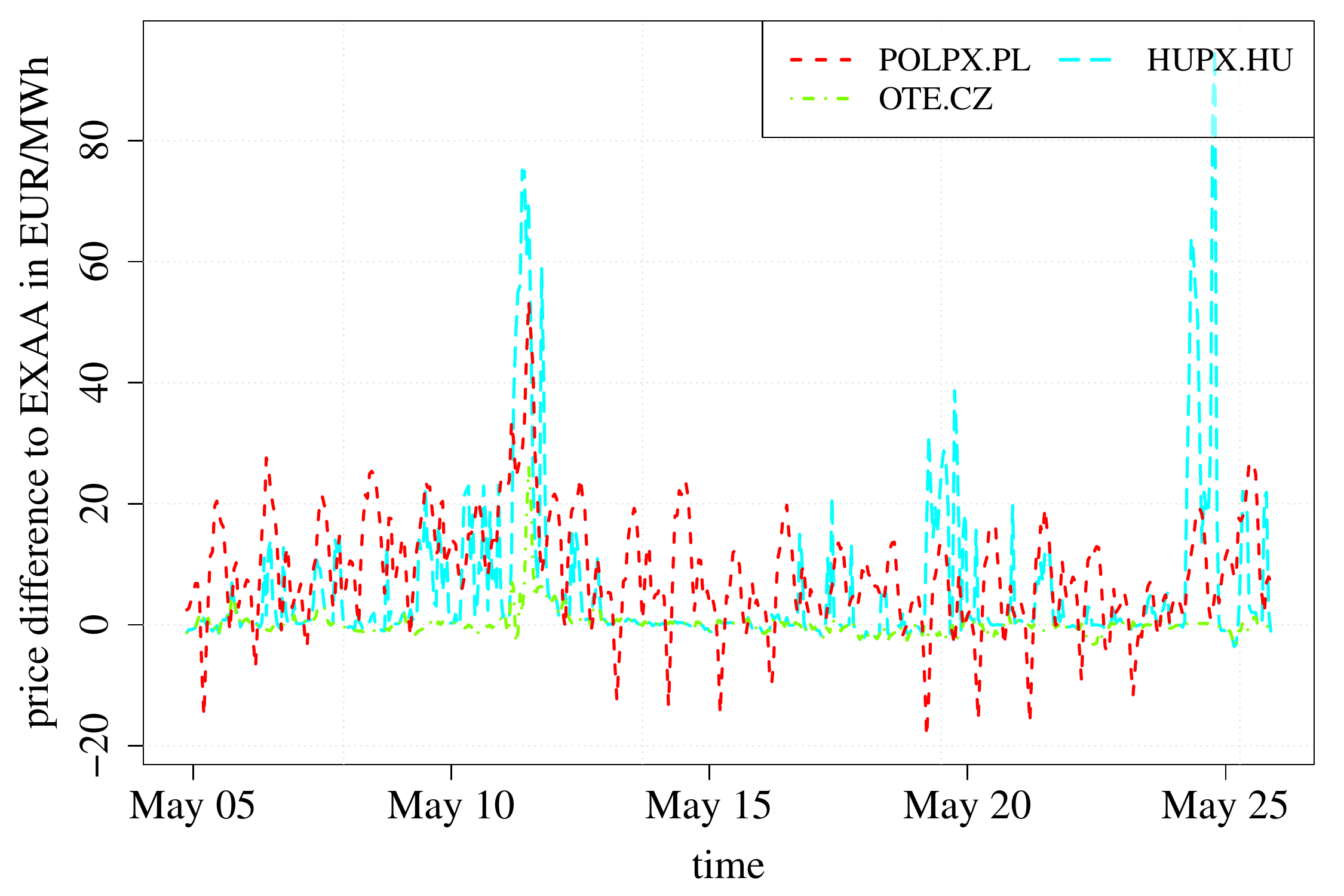}
 \caption{Prices of considered electricity markets and price differences to the EXAA price within three weeks of May 2014.}
 \label{fig_price}
\end{figure}

The time series of the investigated electricity prices for an exemplary time horizon in May 2014 can be obtained from Figure \ref{fig_price}. As a first impression of the behavior of the time series, it shows that some of the prices exhibit similar patterns and price levels, e.g. EPEX (France) and EXAA, whereas others seem not to have a visible relationship, e.g. POLPX and EXAA. The right-hand side of the figure displays explicitly the differences between the EXAA and the other time series. It can be seen that they often follow a recurring structure, which seems to be most dominant for the EXAA and POLPX difference. Our whole dataset contains the hourly day-ahead electricity spot prices, which were downloaded from the websites of the different exchanges for the time period of the 15.08.2007-12.08.2014. In order to guarantee comparability some minor adjustments had to be done. \par 

In case of the EXAA, data for the 13.11.2012 is not available, as one day before the data processing center suffered from a blackout and therefore no trading took place. To prevent excluding this day from our other time series, we decided to replace the prices of the 13.11.2012 with the prices of the previous week. \par 
Moreover, for the HUPX and the Nordpool region Sweden (4) prices were not available for every day of the whole time period. Hence, we were only able to apply our methodology to their individual available time horizons, which is the 01.01.2012-12.08.2014 for the HUPX and the 01.11.2011-12.08.2014 for Nordpool Sweden (4). \par
Polish electricity was mainly traded at three different auctions throughout the day during our chosen time period. As most of the trading, measured in trading volume, takes place at the first auction, we chose to use only the prices of auction I for our modeling approach. Nevertheless, some days are missing within the time series. Those are the 18.03.2011 and the 02.03.2014. The hour values of the 08.01.2008 00:00, 15.02.2008 07:00,  01.04.2008 03:00 and 29.03.2009 01:00 were also missing. Analogous to the missing values of the EXAA, we replaced them with their counterparts 168 hours ago.

Finally all electricity markets except the POLPX trade in EUR or publish the reference price in EUR. 
We use the end-of-day Euro-Z\l{}oty (EUR/PLN) exchange rate of the trading days to make the Polish data comparable to EUR based data. 
%FIRST describe the data (when traded, 
%etc. important feature, e.g. since when negative prices allowd, black november day on EXXA. etc)

Besides the classical approach which uses historical data, we utilize the data of the EXAA for the next day that is available at 10:25 (CET), prior to the submission time point of the other electricity markets. As the EXAA has hourly data this corresponds usually to 24 observations, but when there is a time shift due
to the daylight saving time we have once a year 23 observations in March and 25 observations in October. We denote other electricity markets as $\XX \in \{ \text {EPEX.DE\&AT}, \text{EPEX.CHE}, \ldots , \text{HUPX.HU} \}$. $Y_{t}^{\XX}$ represents the electricity price of market $\XX$ at time $t$. Moreover, we assume that the hourly price $(Y_1^{\XX}, \ldots, Y^{\XX}_{n})$ is observable. In addition we define that $\wtilde{d}(t)$ is
the day that corresponds to the time $t$. So $\wtilde{d}(1)$ is the first observed day, $\wtilde{d}(n)$ the last one and 
$\wtilde{d}(n+1)$ that one which we want to predict.
Given a day $d$ we can denote $H(d)$ as the number of hours that are traded on the corresponding day. 
As mentioned above, $H(d)$ is usually 24 and sometimes 23 or 25.
%If we fix our first time point $Y_0^{\XX}$ it follows that 
For every market $\XX$ at day $\wtilde{d}(n)$ the observable set of prices of the EXAA market is $(Y_1^{\EXAA}, \ldots, Y^{\EXAA}_{n+H(\wtilde{d}(n+1))})$, as there are additional $H(\wtilde{d}(n+1))$ observations available. Further, we denote $\bsY_t = (Y_t^{\EXAA}, Y_t^{\XX}  )'$ as a two-dimensional process. Our focus of interest is now the series of $Y_t^\XX$, especially the values $(Y_{n+1}^{\XX}, \ldots, Y_{n+H(\wtilde{d}(n+1))}^{\XX})$ which we will forecast using the EXAA prices.

\section{Models for electricity prices} \label{models}

In the following we present several models for $Y_t^\XX$. 
These models are easily structured and based on two very common modeling approaches. They are the persistent or na\"ive model and the autoregressive process of order $p$ - AR($p$). Especially the AR($p$) modeling approach is considered as fundamental for an econometric analysis of electricity prices. The process itself or slightly modified versions of it are very often used in the literature, for instance in \cite{weron2008forecasting}, \cite{ferkingstad2011causal} or \cite{kristiansen2012forecasting}. The persistent and the AR($p$) also serve as benchmark models, for instance in \cite{serinaldi2011distributional} and \cite{ziel2015efficient}.
Furthermore, every stationary and invertible ARMA($p$, $q$) process
can be well approximated as an high order AR process,
 as we can rewrite ARMA($p$, $q$) as AR($\infty$) with exponential fast decaying memory, see e.g. \cite{hamilton1994time}. ARMA type models are very popular for electricity price modeling, see e.g. \cite{hickey2012forecasting} or  \cite{liu2013applying}. The relationship holds true also for seasonal ARMA processes, as every seasonal ARMA process is a special 
ARMA($p$, $q$) process, this includes double and triple seasonal ARMA models as used in \cite{taylor2010triple}. 
However, the mentioned seasonal ARMA models are different to those ARMA 
models with a seperate modelling of seasonal trends as e.g. used by \cite{koopman2007periodic} or \cite{keles2013combined}. 
Such ARMA type models can not necessarily be well approximated by an AR($p$).
\par

Below we denote by $\eps_t$ the univariate and by $\bseps_t$ the multivariate error term of the considered univariate or multivariate model.
We assume that the error terms have zero mean with constant and finite variance resp. covariance matrix.
Of course, the assumption of
homoscedasticity is not satisfied, as there is a seasonal structure in the data that also effects
the (conditional) variance of the error term. However, for simplicity reasons we assume homoscedasticity. \par 

% In the following we denote $\eps_t$ and $\bseps_t$ respectively, as model error term that
% is assumed to have zero mean and constant and finite variance. Of course, the assumption of
% homoscedasticity is not satisfied, as there is a seasonal structure in the data that also effects
% the (conditional) variance of the error term. However, for simplicity reasons we assume homoscedasticity. \par 

To point out the effect of the EXAA towards the other electricity markets, we estimate both models in their standard fashion and extend them afterwards by considering the EXAA price in various ways. By providing a comprehensive forecasting study we can test, whether the basic model is significantly outperformed by its counterpart with the EXAA. \par

\subsection{Persistent model}

The first basic model we introduce is the very simple and fast to estimate persistent, or na\"ive model, where 
the electricity price is estimated to be the same as 168 hours ago, which represents usually one week.
It is given by
$Y_t^\XX = Y_{t-168}^\XX + \eps_t$
and can be estimated by $\what{Y}^\XX_{t+h} = Y^\XX_{t+h-168}$ for $1 \leq h \leq H(\wtilde{d}(n+1))$.

\subsection{Univariate AR($p$)}

The second basic model is the well-known autoregressive process of order $p$ (AR($p$)),
which usually provides a high goodness-of-fit and is also estimated in a very short time. 

It is given by
\begin{equation}
Y_t^\XX = \mu + \sum_{k=1}^p \phi_k (Y_{t-k}^\XX - \mu)+ \eps_t
\label{eq_ar} 
\end{equation}
with $\mu$ as mean of the time series and coefficients $\phi_k$ for $1\leq k \leq p$.
For the estimation procedure there are several options available, like the estimation by solving the Yule-Walker equations, 
(conditional) least squares estimation,
or (conditional) likelihood estimation. We estimate the AR($p$) process by solving the Yule-Walker
equations which guarantees a stationary solution{, for more details see e.g. \cite{hamilton1994time}}. The mean $\mu$ is estimated by the sample mean in advance.

The order $p$ of the model is selected via an information criterion. This is also a very common approach in the literature, see for instance \cite{karakatsani2008forecasting} or \cite{liebl2013modeling}. We perform the selection of $p$ by minimizing the Akaike information criterion (AIC) which is given by
$AIC = n \log(\what{\sigma}^2) + 2(p+1)$ with $\what{\sigma}^2$ as estimated mean square error. 
Other criteria like the Bayesian information criterion (BIC) 
could be a reasonable choice, too. To carry out the selection procedure we decide on a maximal possible model order $p_{max}$. Starting with $p=1$, we estimate the model, calculate the AIC, increase $p$ by 1 and repeat this procedure until $p_{max}$ is reached. The model with the minimal AIC is then declared as the final model.  The 
upper bound $p_{max}$ is chosen to be 1400. This is large enough so that all chosen values for $p$ are at least one week, e.g. 168 hours,
smaller than 
$p_{max}$. Note, that the estimated order is usually larger than $336$. \par
Forecasting can be done iteratively by $\what{Y}^\XX_{n+h} = \what{\mu} + \sum_{k=1}^p \what{\phi}_k (\what{Y}^\XX_{n+h-k}- \what{\mu})$ 
for $1 \leq h \leq H(\wtilde{d}(n+1))$ where
$\what{Y}^\XX_t = Y^\XX_t$ for $t\leq n$.

\subsection{Persistent EXAA based model}

This model is the EXAA type equivalent of the persistent model. Here we simply assume 
that the electricity price on market $\XX$ is the same as the EXAA price.
Because the EXAA price is settled at an earlier point in time, the price for the corresponding hour is observable.
Hence, the persistent EXAA based model is given by
$Y_t^\XX = Y_{t}^\EXAA + \eps_t$
and can be estimated 
by $\what{Y}^\XX_{t+h} = Y^\EXAA_{t+h}$ for $1 \leq h \leq H(\wtilde{d}(n+1))$.

%As we want to consider a modification that takes into account the different mean levels of the estimators.
%$Y_t^\XX = \mu + Y_{t}^\EXAA + \eps_t$
%and can be estimated 
%by $\what{Y}^\XX_{t+h} = \what{\mu} + Y^\EXAA_{t+h}$ for $1 \leq h \leq H(\wtilde{d}(n+1))$.
%The mean $\what{\mu}$ should be estimated by the sample mean of $Y_t^\XX - Y_t^\EXAA$.

\subsection{2-dimensional AR($p$)}

Similarly to the univariate AR approach 
discussed above we can model the two dimensional $\bsY_t$, which contains both, the EXAA price and the time series of the investigated exchange.
% is a very popular benchmark in electricity price modelling,
% as it usually gives a quite good estimator and easy to estimate.

In accordance with equation \eqref{eq_ar} it is given by
$$\bsY_t = \bsmu + \sum_{k=1}^p \bsPhi_k (\bsY_{t-k} - \bsmu) + \bseps_t .$$
with mean vector $\bsmu$ and parameter matrices $\bsPhi_k$. 

We estimate the AR($p$) process by solving the multivariate Yule-Walker
equations and determine $p$ using the AIC strategy as described above, where  $p_{max} = 700$.

For the forecasting  we can now exploit the fact that $Y_{n+h}^\EXAA$ is
already observed for $1\leq h\leq H(\wtilde{d}(n+1))$.
Thus the forecast is given by
$\what{\bsY}_{n+h} = (Y^\EXAA_{n+h} , \what{Y}^\XX_{n+h})'$ where
$\what{Y}^\XX_{n+h} =  \what{\mu}^{\XX} + \sum_{k=1}^p (\what{\bsPhi}_k)_2 (\what{\bsY}_{n+h-k} - \what{\bsmu})$ 
for $1 \leq h \leq H(\wtilde{d}(n+1))$ with $(\what{\bsPhi}_k)_2$ as second row of $\what{\bsPhi}_k$ and
$\what{\bsY}_t = \bsY_t $ for $t\leq n$.
Henceforth only the second autoregressive equation that models $Y_t^\XX$ is required to be estimated, as
we want to forecast only one day. 
%The forecasting for $h>n+H(\wtilde{d}(n+1))$ 
%can be obtained iteratively by using  
%$\what{\bsY}_{n+h} - \what{\bsmu} = \sum_{k=1}^p \what{\bsPhi}_k (\what{\bsY}_{n+h-k}-\what{\bsmu})$. However, this is irrelevant for our paper as we are only considering 24 hour ahead forecasts.
%can be estimated by $\what{Y}^\XX_{t+h} = Y^\XX_{t+h-168}$ for $1 \leq h \leq H(\wtilde{d}(n+1))$.

\subsection{Modified 2-dimensional AR($p$)}

In the multivariate AR($p$) model above the observed
EXAA values $Y_{n+h}^\EXAA$ for $1\leq h\leq H(\wtilde{d}(n+1))$ were only considered in the forecasting in order to replace the estimates $\what{Y}_{n+h}^\EXAA$ by its true value $Y_{n+h}^\EXAA$. Hence, we can adjust the model so that this information is also directly used in the model estimation procedure. 
We denote by $\wtilde{\bsY}_t = (Y_{t+H(\wtilde{d}(t+1))}^\EXAA ,Y_{t}^\XX)$ the two dimensional process 
that contains the price for $\XX$ at time $t$ in the second component and the price for $\EXAA$ one day ahead in the first component.
Then the bivariate autoregressive model for $\wtilde{\bsY}_t$ is given by
$$
\wtilde{\bsY}_t = \bsmu + \sum_{k=1}^p \wtilde{\bsPhi}_k 
(\wtilde{\bsY}_{t-k} - \bsmu) + \bseps_t.
$$
In this case the forecasting is as simple as in the univariate case.
It is iteratively done with $\what{\wtilde{\bsY}}_{n+h} = \what{\bsmu} +  \sum_{k=1}^p \what{\wtilde{\bsPhi}}_k ( \what{\wtilde{\bsY}}_{n+h-k} - \what{\bsmu})$ 
for $1 \leq h \leq H(\wtilde{d}(n+1))$ where
$\what{\wtilde{\bsY}}_t = \wtilde{\bsY}_t$ for $t\leq n$.

We want to highlight that both modelling two-dimensional approaches are different, as the additional information is used in another way.
In the first approach the forecast $\what{Y}^\XX_{n+1}$ is independent of the observed EXAA information for future hours,
whereas for the second approach it substantially matters. There $\what{Y}^\XX_{n+1}$ can depend on $\what{Y}^{\EXAA}_{n+1}$, but also on observed information $\what{Y}^{\EXAA}_{n+h}$ for $h\geq2$. For the forecasting, the maximum amount of considered deterministic future EXAA hour values for the first forecasted hour is 23, for the second 22 and so on.

\subsection{Difference based AR($p$)}
The last two models are based on the difference of the target price $Y_t^\XX$ to the EXAA electricity price $Y_t^\EXAA$.
Hence we define $\Delta_t = Y_t^\XX - Y_t^\EXAA$. If $\Delta_t$ would be an i.i.d. noise then
the persistent EXAA based estimator would be a reasonable choice and modeling the difference would not lead to any improvement in the price prediction.
However, if there is some correlation structure left, the assumption that $\Delta_t$ follows an AR($p$) seems to be reasonable.
Therefore, we assume that
\begin{equation}
\Delta_t = \mu + \sum_{k=1}^p \phi_k (\Delta_{t-k} - \mu)+ \eps_t
\label{eq_Delta_ar} 
\end{equation}
holds true for some lags $p$, where $\mu$ represents the mean of the differences. 
As in the univariate case we use the Yule-Walker equations with the AIC for estimation,
where we choose $p_{\max} = 1400$.

Indeed, \eqref{eq_Delta_ar} can be rewritten as 
$$Y_t^\XX  =  Y_t^\EXAA + \mu + \sum_{k=1}^p \phi_k (Y_{t-k}^\XX - Y_{t-k}^\EXAA - \mu)  + \eps_t$$
which shows that this is in fact a special case of an error correction model. So this is basically a 
special case of the 2-dimensional AR($p$) on $\wtilde{\bsY}_t$ considered above.

The forecast of \eqref{eq_Delta_ar} is done iteratively by
$\what{Y}^\XX_{n+h} = Y^\EXAA_{n+h} + \what{\mu} + \sum_{k=1}^p \what{\phi}_k \what{\Delta}_{n+h-k}$ 
for $1\leq h \leq H(\wtilde{d}(n)+1)$ where $\what{\Delta}_{t} = \Delta_{t}$ for $t\leq n$,
 $\what{\Delta}_{t} = \what{Y}^\XX_t - \what{Y}^\EXAA_t$ for $t>n$,
 and $\what{Y}^\EXAA_{n+h} = Y^\EXAA_{n+h}$ for $1\leq h \leq H(\wtilde{d}(n)+1)$. Forecasts further than 
one day-ahead can not be covered directly by this model, as we have to specify a model for $Y_t^\EXAA$ 
to plug-in the corresponding estimates.

% Of course we have now the option to apply the same on the corresponding 24-dimensional setting.
% Thus we have $\bsDDelta_d = \YY^\XX_d - \YY^\EXAA_d$.
% The resulting 24-dimensional AR($p$) is given by
% \begin{equation}
% \bsDDelta_d = \bsmu + \sum_{k=1}^p \bsPhi_k (\bsDDelta_{d-k} - \bsmu)+ \bseps_t
% \label{eq_Delta_ar} 
% \end{equation}
% The forecasting requires
% $\what{\YY}^\XX_{\wtilde{d}(n)+1} = \what{\mu} + Y^\EXAA_{\wtilde{d}(n)+1}
% + \sum_{k=1}^p \what{\bsPhi}_k \what{\bsDDelta}_{\wtilde{d}(n)+1 -k}$ 
%  where $\what{\bsDDelta}_{d} = \bsDDelta_{d}$ for $d\leq \wtilde{d}(n)$.
% %  ,$\what{\bsDDelta}_{d} = \what{\YY}^\XX_d - \what{\YY}^\EXAA_d$ for $d> \wtilde{d}(n)$,
% %  and $\what{\YY}^\EXAA_{\wtilde{d}(n) +1 } = \YY^\EXAA_{\wtilde{d}(n) +1}$.
% 
% 
%  
%  TODO: (-mod. variants???)
% We can also introduce $\wtilde{\Delta}_t = Y_t^\XX - Y_{t+ H(\wtilde{d}(t)+1)}^\EXAA$ resp.
% multivariate $\wtilde{\bsDDelta}_d = \YY_d^\XX - Y_{d+1}^\EXAA$.
% but we check this before we write this because it makes mathemematically perfectly sense, but economically not clear...,

%$\wtilde{\bsY}_t = (Y_{t+H(\wtilde{d}(t+1))}^\EXAA ,Y_{t}^\XX)$

\subsection{Model summary}
 
 A summary table of all considered models with the most relevant information is given in Table \ref{tab_models}. In the following sections we will refer to the models as presented in the abbreviation column of this table.

\begin{table}[tbh]
\centering
\begin{tabular}{p{60mm}lp{25mm}l}
  \hline
  \textbf{Model} & \textbf{Abbreviation} & \textbf{Uses EXAA information} & $p_{\max}$\\ \hline
  persistent model 				& na\"ive			& no 	& -\\
  univariate AR($p$) 				& AR($p$) 			& no 	& 1400 \\
%  \centering 24-dimensional AR($p$) 			& AR24 			& no 	& 15 \\ \hline5
%  \centering 24-dimensional AR($p$) with diagonal matrices & AR24$_diag$ 	& no 	& 70 \\ \hline
  persistent EXAA based model 			& na\"ive-EXAA 		& yes 	& -\\ 
%  \centering EXAA persistent model with mean 		& $\mu$-pers.$^\EXAA$ 	& yes	& -\\ 
  2-dimensional AR($p$) on $\bsY_t$ 		& 2d-AR($p$) 			& yes	& 700 \\
  2-dimensional AR($p$) on $\wtilde{\bsY}_t$ 	& 2d-$\widetilde{\text{AR}}$($p$) 		& yes	& 700 \\
%  \centering 48-dimensional AR($p$) on $\bsY_t$ 		& AR48 			& yes	& 8\\
%  \centering 48-dimensional AR($p$) on $\wtilde{\bsY}_t$	& AR48-mod. 		& yes	& 8\\
  univariate AR($p$) on differences 		& $\Delta$-AR($p$) 		& yes	& 1400\\
%  \centering 24-dimensional AR($p$) on difference 	& $\Delta$-AR24 		& yes	& 15\\ \hline
 \end{tabular}
\caption{Summary table of considered models taken into account for the forecast}
\label{tab_models}
\end{table}

\section{Setup of the forecasting study}

For evaluating the forecasting performance and the desired impact of the EXAA price we carry out a forecasting
study. We face the situation that we sometimes have to forecast 23 or 25 prices instead of 24, which complicates the notation and forecasting. Nevertheless, the occurence of such specific days is considered within the analysis. As mentioned previously the available data covers $7\times 365 = 2555$ days which is about 7 years.

In the forecasting study we use a rolling window of hourly data 
$(\bsY_{1+R(r)}, \ldots, \bsY_{T+R(r)})$ of length $T$ with $R(r)$ as rolling index shift. % that will be explained later on.
The length of the considered sample is $D= 2\times 365 = 730$ days which corresponds to an in-sample period of usually 2 years.
Hence, for the amount of used observations we have $T= \sum_{d=1}^{D} H(\wtilde{d}(M(d)+R(r)))$ with $M(1)=1$ and $M(d)= 1+ \sum_{i=1}^{d-1} H(i)$ for $d>1$. This expression is usually about $24\times 2\times 365 = 17520$.
Given the 2 years of data we do the estimation procedure on the given window. As introduced above we shift
the window by $R(r)$ for $1 \leq r \leq r_{\max} $ with $r_{\max} = 4\times365 + 366-1 = 1825$\footnote{The 
forecasting range contains one leap year (2012)} days. Therefore $r_{\max}$ covers the remaining $5$ years of observations minus
one day. % that we want to forecast.
In detail we have $R(1) =0$ and $R(r) = \sum_{i=1}^{r-1} H(i) $ for $1 < r\leq  r_{\max}$. 
In the latter formula we usually have $R(r) = 24\times (r-1)$.

After the estimation on a given window we do the forecast of the next day traded values
$\left(\what{Y}^\XX_{T+R(r) + 1},\right.$ $ \left. \ldots,   \what{Y}^\XX_{T+R(r) + H( \wtilde{d }(T+R(r) +1)  ) }\right) $ of the electricity time series of interest.
Remember that  $H( \wtilde{d }(T+R(r) +1)  )$ is the amount of traded hours of the proceeding day, 
it is in general 24, but sometimes 23 or 25.
In order to compare our forecasts we compute the mean absolute error ($\MAE$) and
the root mean square error ($\RMSE$) of all forecasted values.
% In addition to the hourly MAE and RMSE we compute the overall mean square error and root mean square
They are given by 
$$\MAE^\XX = \frac{1}{R(r_{\max}+1) -1 } \sum_{t=T+1}^{T+ R(r_{\max}+1) -1} |Y^\XX_{t} - \what{Y}^\XX_{t}|  \text{ and}  $$
$$\RMSE^\XX = \sqrt{  \frac{1}{R(r_{\max}+1) -1 }  \sum_{t=T+1}^{T+ R(r_{\max}+1) -1} |Y^\XX_{t} - \what{Y}^\XX_{t}|^2 } .$$
\section{Results and Discussion}
All results are based on out-of-sample data. The estimated MAE and RMSE are given in Table \ref{tab_mae_rmse}. Every bold print number corresponds to the best model in terms of MAE or RMSE. An underlined value represents a model, which produced a MAE or RMSE which was at least in the confidence interval of the best model. The number in brackets shows the standard deviation, which was estimated via bootstrapping with a sample size of $B=1000$.
First of all we can observe that the standard deviations of the MAE values are smaller in comparison to the RMSE ones.
Thus, their results seem to be more reliable. The reason is likely that most of the electricity prices are heavy tailed
which also affects the model residuals. Hence, squaring the residuals as done in the calculation of the RMSE halves
the corresponding tail index. This automatically leads to more unreliable results. Considering the 
MAE values we see that for all neighboring price regions there is at least one model which involves the EXAA information 
and is superior to the na\"ive and AR($p$) model without EXAA information. Taking into account the 2-sigma range we can obtain that
the performances are significantly better, except for the case of Sweden. Hence based on the the MAE 
we can improve the forecasting results taking into account the EXAA information. 
An interesting feature is that for the EPEX.DE\&AT price the na\"ive-EXAA model seems to be significantly better 
than the other complex AR type models that involve EXAA information. 
As both, the EPEX.DE\&AT price such as the EXAA.DE\&AT price, consider the same region
this result may give an indication that the market efficiently prices the information observed at the EXAA. The relationship between both prices seem not to exhibit recognizable autoregressive patterns, and therefore could not be exploited for an investment strategy. However, this holds only true for our investigated type of AR model. The na\"ive-EXAA model is also superior for the OTE.CZ price, which covers the delivery region of Czech. This indicates that the markets seem to have a strong relationship.

The results considering the RMSE are basically similar. Nevertheless,
the higher confidence regions due to high standard deviations make an interpretation more unstable.

In addition we define hourly versions of the MAE and RMSE to compare the forecasting performance at a specific hour of a day.
The $\MAE^\XX_h$ and $\RMSE^\XX_h$ are given by
$$ \MAE^\XX_h = \frac{1}{ \#(h) } \sum_{r=1}^{r_{\max}} \bsone_{\{h\leq H(D+r) \}}
|Y^\XX_{T+R(r)+h} - \what{Y}^\XX_{T+R(r)+h} | \text{ and} $$
$$ \RMSE^\XX_h = \sqrt{ \frac{1}{ \#(h) }   \sum_{r=1}^{r_{\max}} \bsone_{\{h\leq H(D+r) \}}
|Y^\XX_{T+R(r)+h} - \what{Y}^\XX_{T+R(r)+h} |^2 }  $$
with 
$\#(h) = \sum_{r=1}^{r_{\max}} \bsone_{\{h\leq H(D+r) \} } $ for $1 \leq h \leq 25 $. 
Note that the expression $\#(h)$ is the number of forecasts that correspond to hour $h$. 
For $1\leq h \leq 23$ this is simply $r_{\max}$, whereas for $h=24$ it is a lower amount due to the clock change in March. For $h=25$ this value is not really of interest as it contains only a very few October observations, also due to clock change.
The $\MAE^\XX_h$ and $\RMSE^\XX_h$ are visualized for all markets and models and hours in Figures
\ref{fig_MAE_h} and \ref{fig_RMSE_h} in the Appendix. The figures show clearly, that the evening to night hours from approximately 20:00 to 06:00 face a smaller forecasting error than, for instance, the hours around midday. Comparing both figures the before mentioned issue with the RMSE becomes visible. Due to heavy outliers especially during the midday hours the RMSE is often 10 times higher than at the other hours of the day. This complicates the interpretation of RMSE types of errors, which analyze the 24 hour forecast as a whole. Moreover, it can be obtained, that at least one of the EXAA related models seem to outperform all basic models for each time series in terms of hourly MAE.

% As we have 
% Hence we receive 
% 
% and 

% The models describes in the previous section were all tested against each other.

% latex table generated in R 3.0.2 by xtable 1.7-3 package
% Tue Oct  7 10:30:29 2014
% \begin{table}[ht]
% \centering
% \small
% 
% \caption{Estimated MAE with corresponding standard deviation, bold = best, underlined = not significantly worse than best
% (indicated by the 2-sigma range of the best model, corresponds to one-sided significance level of 2.27\% under normality) }
% \label{tab_mae}
% \end{table}
\begin{table}[ht]
\centering
\small
\resizebox{\textwidth}{!}{
\begin{tabular}{rrrrrrr}
\multicolumn{7}{c}{MAE}\\
  & na\"ive & AR($p$) & na\"ive-EXAA & 2d-AR($p$) & 2d-$\widetilde{\text{AR}}$($p$) & $\Delta$-AR($p$) \\
  \hline
  EPEX.DE\&AT & 8.273(0.056) & 5.980(0.037) & \textbf{3.835(0.029)} & 4.597(0.032) & 4.021(0.030) & 3.899(0.028) \\ 
  EPEX.CHE & 7.839(0.046) & 5.232(0.030) & 6.990(0.046) & 4.676(0.026) & \textbf{4.166(0.025)} & \underline{4.209(0.025)} \\ 
  EPEX.FR & 9.738(0.230) & 8.915(0.157) & 6.887(0.155) & 8.054(0.152) & 7.908(0.156) & \textbf{6.178(0.148)} \\ 
 APX.NL & 6.498(0.033) & 4.801(0.023) & 6.809(0.039) & 4.266(0.020) & \textbf{3.935(0.020)} & 4.540(0.023) \\ 
  %   APX.NL 
 % & 8.335(0.104) & \textbf{6.778(0.075)} & 14.732(0.091) & 6.936(0.079) & 6.929(0.077) & 8.618(0.081) \\ 
  NP.DK.West & 8.076(0.141) & 6.281(0.101) & 7.140(0.103) & 5.624(0.099) & \textbf{5.358(0.099)} & 5.850(0.101) \\ 
  NP.DK.East & 9.669(0.160) & 8.120(0.121) & 8.652(0.124) & \underline{7.267(0.118)} & \textbf{7.158(0.115)} & 7.526(0.124) \\ 
  NP.SW4 & 6.974(0.139) & \underline{5.226(0.101)} & 11.562(0.110) & \underline{5.151(0.103)} & \textbf{5.147(0.102)} & 6.822(0.102) \\ 
  POLPX.PL & 3.421(0.026) & 2.459(0.016) & 8.157(0.039) & \textbf{2.223(0.015)} & \underline{2.246(0.016)} & 4.348(0.021) \\ 
  OTE.CZ & 7.554(0.042) & 5.338(0.027) & \textbf{2.707(0.022)} & 4.135(0.022) & 2.852(0.019) & \underline{2.738(0.018)} \\ 
  HUPX.HU & 10.159(0.160) & 8.030(0.119) & 7.012(0.146) & 7.174(0.109) & \underline{6.717(0.111)} & \textbf{6.579(0.115)} \\ 
   %\hline
  % \vspace{1mm}
\end{tabular}
}

\resizebox{\textwidth}{!}{
\begin{tabular}{rrrrrrr}
\multicolumn{7}{c}{RMSE}\\
  & na\"ive & AR($p$) & na\"ive-EXAA & 2d-AR($p$) & 2d-$\widetilde{\text{AR}}$($p$) & $\Delta$-AR($p$) \\
  \hline
  EPEX.DE\&AT & 14.221(0.395) & 9.659(0.340) & \underline{7.299(0.452)} & \underline{7.965(0.418)} & \underline{7.323(0.445)} & \textbf{7.090(0.436)} \\ 
  EPEX.CHE & 12.607(0.157) & 8.125(0.132) & 11.644(0.102) & 7.279(0.131) & \textbf{6.693(0.125)} & \underline{6.788(0.108)} \\ 
  EPEX.FR & 46.848(6.662) & \underline{34.051(6.092)} & \underline{33.107(6.258)} & \underline{33.187(6.106)} & \underline{33.144(6.153)} & \textbf{32.377(6.196)} \\ 
   APX.NL & 9.516(0.101) & 6.878(0.086) & 10.768(0.073) & 6.100(0.087) & \textbf{5.724(0.090)} & 6.701(0.080) \\ 
%  & \underline{23.574(5.884)} & \underline{17.448(5.479)} & \underline{23.945(4.003)} & \underline{17.440(5.528)} & \textbf{17.429(5.601)} & \underline{18.728(5.179)} \\ 
  NP.DK.West & 31.342(4.111) & \underline{22.357(3.948)} & \underline{22.875(4.135)} & \underline{21.799(4.190)} & \textbf{21.688(4.232)} & \underline{21.889(4.285)} \\ 
  NP.DK.East & 36.254(2.779) & \underline{27.378(2.825)} & \underline{27.667(2.828)} & \underline{26.136(2.935)} & \textbf{25.992(2.841)} & \underline{26.633(2.864)} \\ 
  NP.SW4 & 30.167(2.397) & \underline{22.467(2.223)} & \underline{26.005(2.018)} & \underline{21.731(2.408)} & \textbf{21.657(2.400)} & \underline{22.763(2.191)} \\ 
  POLPX.PL & 6.522(0.110) & 4.286(0.075) & 11.377(0.082) & \textbf{3.934(0.085)} & \underline{3.969(0.084)} & 6.293(0.051) \\ 
  OTE.CZ & 11.526(0.131) & 7.745(0.077) & 5.197(0.127) & 6.221(0.087) & \underline{4.832(0.104)} & \textbf{4.736(0.102)} \\ 
  HUPX.HU & 15.624(0.346) & 11.830(0.256) & 13.018(0.320) & \underline{10.837(0.267)} & \textbf{10.462(0.289)} & \underline{10.467(0.299)} \\ 
  % \hline
\end{tabular}
}
\caption{Estimated MAE and RMSE with corresponding standard deviation, bold = best, underlined = not significantly worse than best
(indicated by the 2-sigma range of the best model, corresponds to one-sided significance level of 2.27\% under normality) 
}
\label{tab_mae_rmse}
\end{table}

Additionally we conduct the popular Diebold-Mariano (DM) test to compare the forecasting performance, as done 
for instance in \cite{hong2012day}, \cite{bordignon2013combining} and \cite{nan2014forecasting}.
For a recent discussion to the use and abuse of the test see \cite{diebold2012comparing}.

The DM-test is based on evaluating loss differences of the forecasting errors of two different models.
Given the forecasting errors $\what{\eps}^\XX_t(m_i) = Y^\XX_{t}(m_i) - \what{Y}^\XX_{t}(m_i)$ this loss differential
for two models $m_1$ and $m_2$
is commonly given by $\delta^{\XX,m_1, m_2}_{p,t} = |\what{\eps}^\XX_t(m_1)|^p - |\what{\eps}^\XX_t(m_2)|^p$.
Often this test is carried out by using squared residuals with $p=2$. In our situation it turned out to
provide unreliable results, as the residuals seem to be to heavy tailed.
Instead we consider the case of absolute residuals with $p=1$ which corresponds to the MAE case.

One crucial assumption on the DM-test is that $\delta^{\XX,m_1, m_2}_{p,t}$ exhibits a homoscedastic and covariance-stationary process.
If we would perform the DM-test on all loss differences $\delta^{\XX,m_1, m_2}_{p,t}$
for $T<t<T+ R(r_{\max}+1)$ the assumptions of the test
must be violated if the underlying process has some linear autoregressive structure that is non-zero.
The reason is that two consecutive forecast errors, e.g. $\what{\eps}^\XX_{t+1}$ and $\what{\eps}^\XX_{t+2}$, have different variances as the estimate for $Y^\XX_{t+2}$ is based on the estimate for $Y^\XX_{t+1}$. The variance for $\what{\eps}^\XX_{t+h}$ increases with the forecasting horizon.

However, if we compare only forecasts that correspond to a specific hour $h$ 
the Diebold-Mariano assumption might be satisfied. 
% Thus we can carry out for each $1\leq h \leq 24$ a DM test on
% the residuals as used in the definition of $\MAE_h$ and $\RMSE_h$. 
Hence, we conduct the DM-test for all prices and hours $1\leq h \leq 24$ (except the 25th) to compare all four 
proposed models that contain EXAA information against the AR($p$) process which does not include the information. 
We consider only the AR($p$) as the MAE and RMSE values suggested that it is the best model considered
that does not use the EXAA information. 
As mentioned above $\delta^{\XX,m_1, m_2}_{p,t}$ must be a covariance-stationary process.
To estimate the order of covariance-stationary we estimate an AR($p$) process for the differential loss series as well and select the order of the best model concerning the AIC.
In our application the optimal order is often chosen to be 7 or 14 which corresponds to weekly cycles.

% More detailed we apply the Diebold-Mariano test twice, first on the absolute residuals. 
% This
% automatically provides analogy to the $\MAE_h$ computed before. We also tried to investigate squared residuals but the
% results for most of the processes does not seems to be reliable in most cases due to the fat tails of the resulting loss differences.
% 
% The DM-test requires the assumption that the loss difference follows a homoscedastic covariance-stationary process. 

\begin{figure}[hbt!]
\centering
\begin{subfigure}[b]{.49\textwidth}
 \includegraphics[width=1\textwidth]{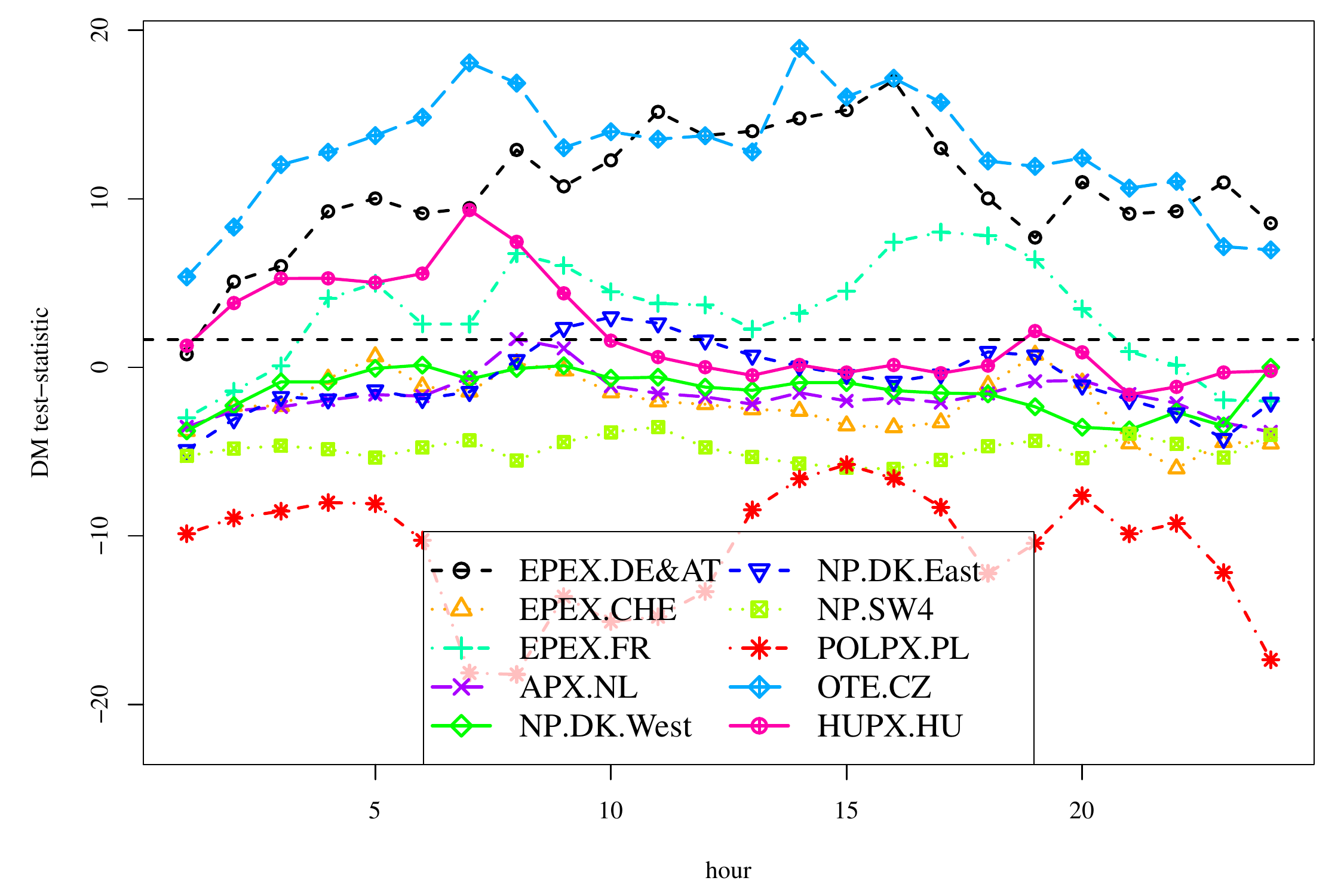} 
  \caption{AR($p$) vs. na\"ive-EXAA}
  \label{fig_dm_1}
\end{subfigure}
\begin{subfigure}[b]{.49\textwidth}
 \includegraphics[width=1\textwidth]{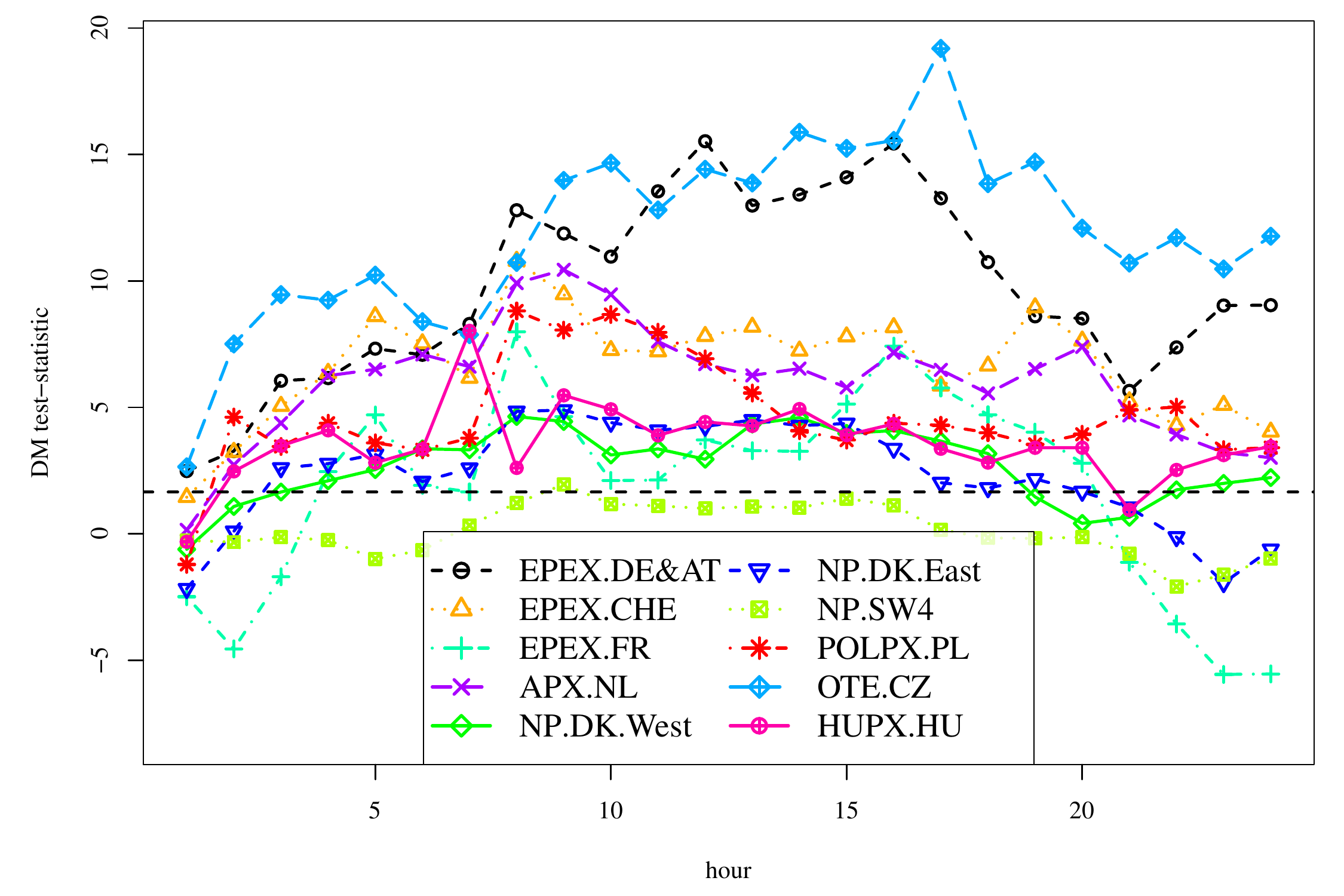} 
  \caption{AR($p$) vs. 2d-AR($p$)}
  \label{fig_dm_2}
\end{subfigure}
\begin{subfigure}[b]{.49\textwidth}
 \includegraphics[width=1\textwidth]{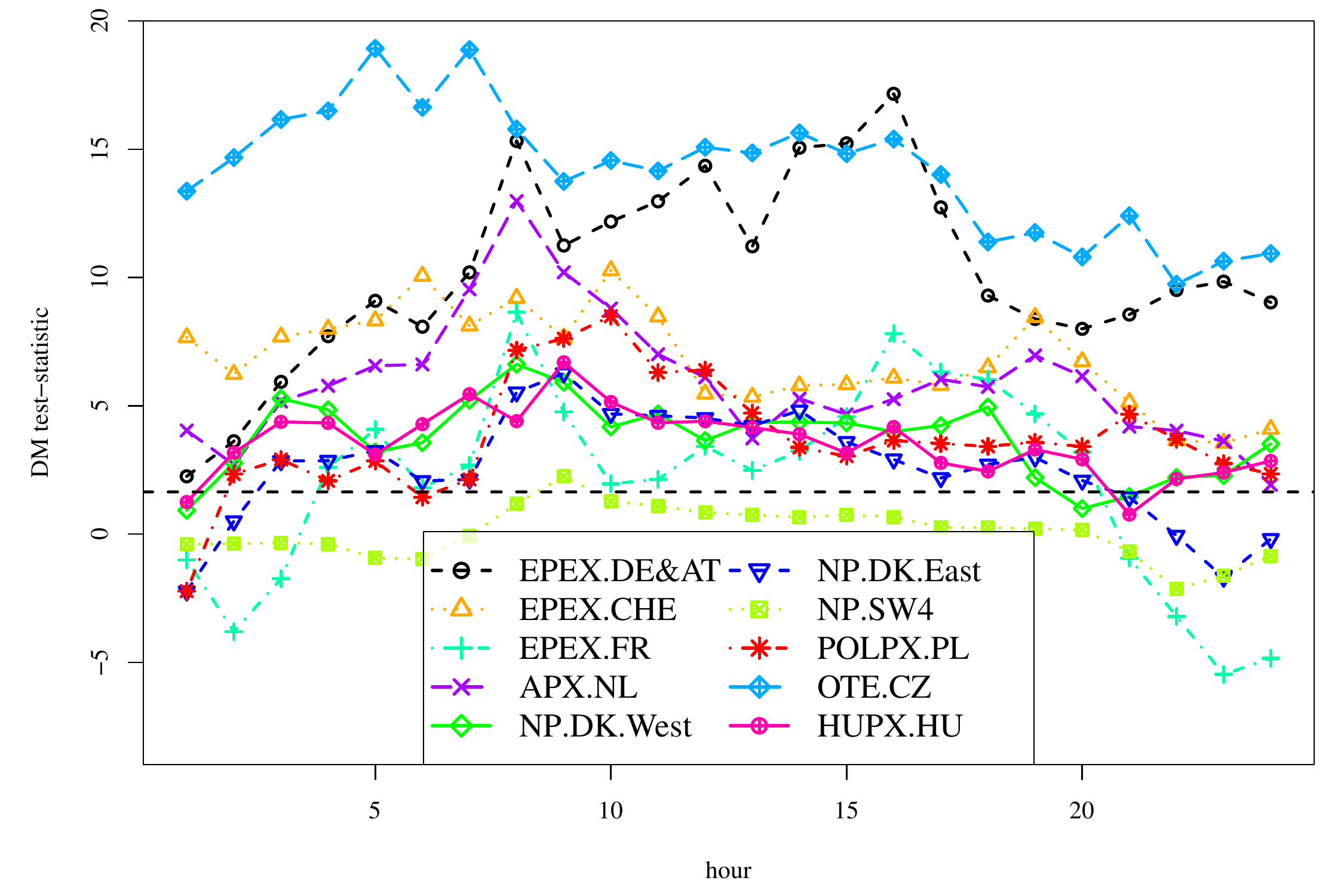} 
  \caption{AR($p$) vs. 2d-$\widetilde{\text{AR}}$($p$)}
  \label{fig_dm_3}
\end{subfigure}
\begin{subfigure}[b]{.49\textwidth}
 \includegraphics[width=1\textwidth]{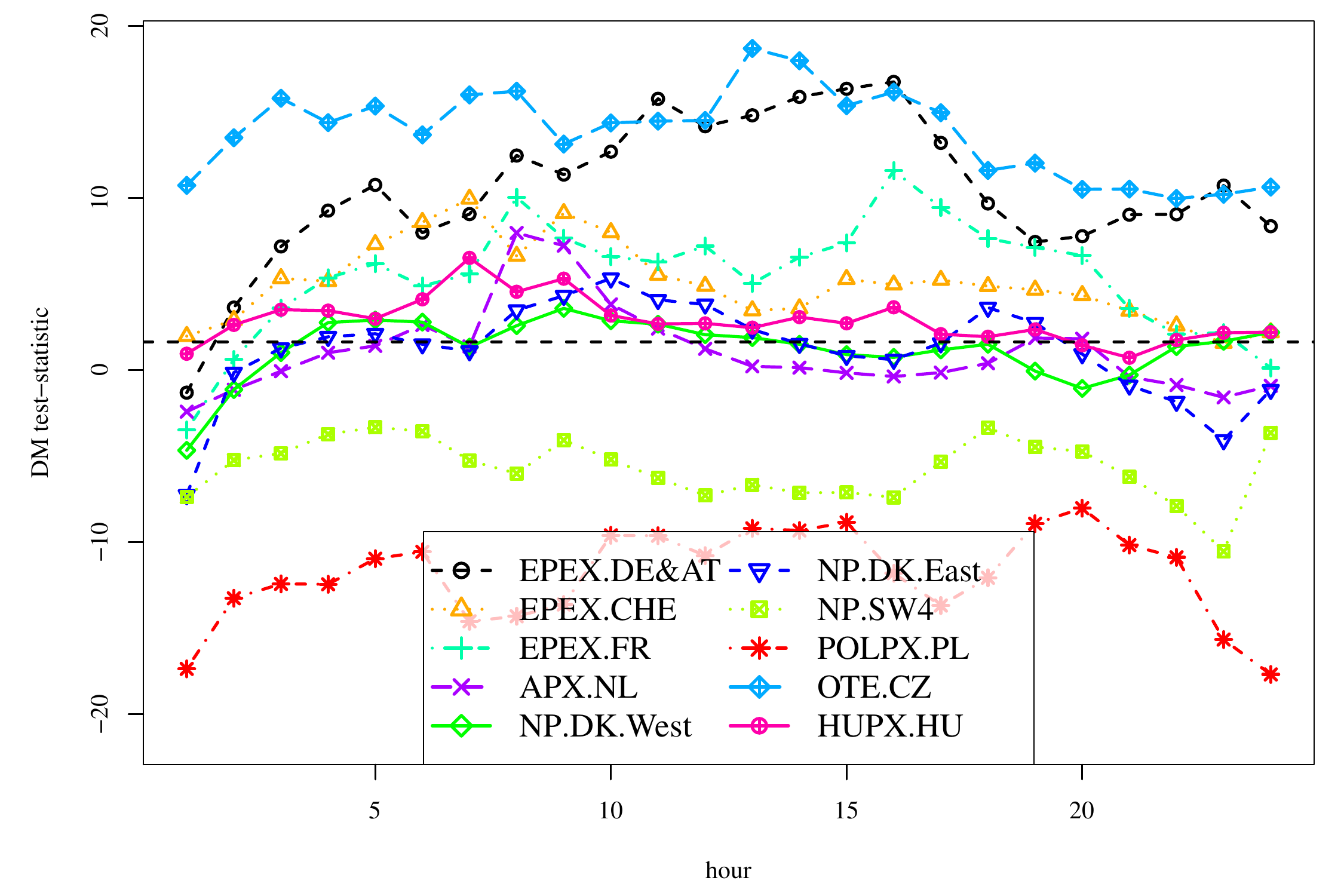} 
  \caption{AR($p$) vs. $\Delta$-AR($p$)}
  \label{fig_dm_4}
\end{subfigure}
\caption{DM-test statistics including the 95\% confidence threshold line (dashed).}
 \label{fig_dm}
\end{figure}

The resulting test statistics for the conducted tests are given in Figure \ref{fig_dm}. A higher DM test-statistic corresponds to a higher model error of the AR($p$) in comparison to the EXAA-based model. All 
values above the dashed 95\% confidence line indicate that the EXAA-based model provides significantly better forecasts for this specific hour than 
the AR($p$).  Interestingly the DM-test results are in strong accordance with the conclusions drawn from the analysis of the MAE. 
So except for Sweden there is at least one model that contains the EXAA information that is significantly 
superior to the AR($p$) in the majority of hours. The significance turns out to be very high especially for 
the German-Austrian EPEX price and the Czech OTE.CZ price. In accordance to our MAE study there is no model which dominates every other model for every hour.
Regarding the hourly test statistics over the day it is difficult to establish any clear pattern. But 
In general we observe that the significance of the first hours, especially the first one is relatively small. 
The reason is that for this hour the AR($p$) performs relatively well, as can be obtained from Figure \ref{fig_MAE_h} in the Appendix.
Moreover, we can see that for the EPEX prices of Germany and Austria, Switzerland and France it seems that the significants is clearer
during the hours from about 5am to 8pm.

 \subsection{Temporal Structure Analysis}

Moreover, we want to analyze the temporal structure of the impact of the EXAA. In detail we want to discuss the daily, monthly and annual seasonal dependence of the impact. The latter might give hints for the progress of the European electricity market integration. 
However, a temporal structure analysis has to be interpreted carefully, as all of our assumed models 
have constant dependence structure over time, especially no seasonal variation in the parameters. Hence we restrict ourselves
only on evaluating point estimates to get indication for the direction of the behavior.
We omit uncertainty measures like confidence intervals as they might not be valid.

As the previous study showed, the MAE provides more robust results. Therefore, we perform the temporal structure evaluation only for the absolute
residuals
\begin{align*}
 \MAE^\XX_{g, \T} &=  \frac{1}{\#(\T)} \sum_{t\in \T} |Y^\XX_{t} - \what{Y}^\XX_{t} | \ \text{ for } \  \T \in \TT_{g}  \ \text{ and } \ g\in \GG %\\
%  \MAE_{\text{weekly}, \T} &=  \frac{1}{\#(\T)} \sum_{t\in \T} |Y^\XX_{t} - \what{Y}^\XX_{t} | \ \text{ for } \  \T \in \TT_{\text{weekly}} \\
%  \MAE_{\text{time}, \T} &=  \frac{1}{\#(\T)} \sum_{t\in \T} |Y^\XX_{t} - \what{Y}^\XX_{t} | \ \text{ for } \  \T \in \TT_{\text{time}} \\ 
\end{align*}
where $\GG = \{\text{monthly}, \text{daily}, \text{annual}\}$ is the set of time partitions,
$\TT_{g}$ is the set of all time points of partition $g \in \GG$ and $\#(\T)$ gives the number of elements of $\T$.
For the monthly partition we use the twelve months of a year, for the daily we use the 7 weekdays and for 
the annual partition we divide our 5 year out-of-sample range into 5 equidistant time groups, each about a year.
All computed $\MAE^\XX_{g, \T}$-values are given in Figures  \ref{fig_MAE_mon}, \ref{fig_MAE_week} and \ref{fig_MAE_cp} in the appendix. Please keep in mind that for the HUPX we did not have the same data period as for the other markets.

Regarding the curves of the of the monthly seasonality which are presented in Figure \ref{fig_MAE_mon}, we can observe that
in general the absolute forecasting error is less over the summer than over the winter. This is a commonly observed phenomenon, which can be mainly explained by the usually higher amount of holidays during winter time and the increasing uncertainty over the renewable energy, especially wind energy, due to more intense weather conditions. Both events tend to create price movements which are more difficult to forecast.

Concerning the impact of EXAA to the other markets it is hard to make general statements. In most cases the superiority of an EXAA-related model over a univariate model seems to maintain over the months. However, for the EPEX Switzerland price we observe that from October to March there is almost no difference 
between the univariate AR process that required no EXAA information to the processes that require EXAA information. Only in the winter months, where forecasting performance in general lacks accuracy, the EXAA inclusion leads to better results.
As we have shown above, the EXAA based models
for the Swiss price were significantly better than the models without EXAA information. So we can conclude that this 
significance is mainly driven by a better performance over the summer. We also observe tremendously high peaks, e.g. in France or Denmark West. But a closer look into the data showed, that they are all due to heavy outliers within a specific month.

For the daily behavior as presented in Figure \ref{fig_MAE_week} we observe no clear behavior that seems to hold true for every country. In 
some countries the weekend are easier to predict or benefit more from the impact of the EXAA than working days from Monday to Friday,
but in some countries this does not hold. In most cases the superiority of an EXAA based model towards univariate model seem to maintain over the days. Only for Sweden 4 it seems that the univariate model sometimes leads to better results for some of the days, but the difference is typically very small. This uncertainty about whether the EXAA inclusion does provide better forecasting results for this market goes along with our previous findings. For the German and Austrian EPEX price we see that
the absolute forecasting error is the largest on Sunday and Monday. The improvement from the na\"ive EXXA-model
to the univariate AR($p$) seem to be quite stable through the days. This is remarkable if we have the fact in mind, that at 
EXAA there are auctions only on working days, whereas the EPEX day-ahead market auction is every day. 

The last temporal structure concerns the variation of the process over the years, as can be seen in figure \ref{fig_MAE_cp}.
If the markets become more and more integrated we would expect that the difference of absolute forecasting error of the univariate model in comparison to an EXAA-based model becomes relatively larger over time. Also, on average, the forecasting performance of EXAA-based model should increase, leading to lower MAE over time.
% If the markets become more
%and more integrated we would expect that the absolute forecasting error with respect to EXAA became
%smaller over time. 
%Similarly we would expect that the impact of EXAA increases as the markets should behave more and more similar.
In fact we observe for some countries that the absolute forecasting error decreases as expected. This seems most 
clear for France, Denmark East, Sweden and Czech Republic. Interestingly for the Netherlands and Denmark West this seems to be the inverse.
It seems as if they are become less integrated with respect to the EXAA. We only have supposition for 
this observation. For the APX Netherlands the reason might be that the Netherlands integrates more towards the APX UK market, which 
results in a lower coupling to the EXAA price. 
% Similarly for Denmark West, we guess that Denmark West was stronger 
% coupled to the German and Austrian price. Now it turns more towards the Nordpool market, especially to Denmark West.
% This suggestion can be underlined by the fact that the forecasting error of Denmark West increases whereas the error in Denmark increased.
%\RS{The difference in univariate to EXAA-based models seem to increase at least slightly for the markets of Germany and Austria, DK West and the Czech Republic. This also provides evidence for a closer relationship of these markets over the years. Again, the APX Netherlands seem to experience a reversed relationship, as the difference of both model types decreases over time, which may support our before-mentioned supposition.}

Furthermore, we can observe that for some countries, most clearly for the Netherlands
the difference between the MAE of the AR($p$) and the best model with EXAA-impact is decreasing. So in general it seems 
that the impact of the EXAA is reduced over time. In contrast for the German and Austrian EPEX price and the Czech price 
we observe the opposite. Interestingly, these are the both markets where the na\"ive EXXA-model performed best. So on markets
that behave very similar to the EXAA the impact of the EXAA seems to be increased over the years of our study.

\subsection{Implication for Market Participants} 

Our results provide implications for several decision makers in the field of electricity trading. First and foremost every active participant of an exchange can consider the market results of connected exchanges as early price information of their own exchange. If their trading strategy is at least partly based on an econometric approach, including these prices will yield a decreased uncertainty about upcoming day-ahead prices. In the special case of Germany and Austria, where the EXAA and the EPEX cover the same regions, investors seem to get a very accurate early snapshot of the market and can therefore conduct valuable adjustments of their investment strategy. As using the early price settlement of exchanges is often accompanied by becoming an active member of this exchange, exchange companies could also utilize our results. If they decide to set their price settlement time point before the price submission of relevant connected exchanges, they may encourage participants of these exchanges to become a 
member, simply to get access to this early price information. Lastly, our findings provide also implications for portfolio managers of electricity companies mainly in Germany and Austria. Analyzing the price patterns of both, the EXAA and EPEX exchange, may lead to an optimal distribution of trading among these two exchanges. Orders, which were not settled at the EXAA, can be placed again at the EPEX -- adjusted by the updated price information coming from the EXAA settlement.

% TODO write more...

%  $\RMSE$ & na\"ive & AR(p) & na\"ive-EXAA & 2d-AR(p) & 2d-$\widetilde{\text{AR}}$(p) & $\Delta$-AR(p) \\

\section{Conclusion}

We investigated several models to show the impact of the EXAA day-ahead price on electricity day-ahead spot prices of regions, which are directly connected to Germany and Austria. To conduct our study we introduced a unique investigation setting, where traders can utilize different price settlement time points of exchanges to get a snapshot of other markets. By analyzing different error metrics and test setups we were also able to provide insights in the relatedness of those markets. It turned out that including the EXAA information in standard and robust time series approaches increased the performance of those models for every examined market. In some cases, e.g. APX Netherlands or EPEX France, considering the early EXAA information resulted in a vast improvement. 

Interestingly, the na\"ive EXAA model, which simply uses the price of the EXAA as predictor, turned out to be the best model for the EPEX Germany and Austria as well as the OTE Czech Republic. This in turn means that common and robust autoregression techniques for this relationship were not able to filter out additional information in the price relation of the EXAA and those exchanges. Especially in the case of the EXAA and the EPEX, which both trade for the same region, this may provide evidence for the reasoning of \citep{viehmann2011risk}, who considered the EXAA price as an approximation for the OTC prices for Germany and Austria.

Our findings provide important implications for traders and decision makers of electricity exchanges. Utilizing the EXAA price disclosure as an early price information for the upcoming prices of the EPEX will lead to a vast improvement of forecasting accuracy. Electricity companies which are also engaged in trading should therefore always incorporate the EXAA price, if they decide to make predictions for the EPEX price. If properly implemented, the EXAA price can also be exploited by market participants of other exchanges, even though they may not trade for the same region. Meeting the requirement of connected markets is enough to improve the trading strategy by using the EXAA price. 

An analysis of the temporal structure of the relationship revealed that the impact of the EXAA towards other markets develops over the years. For some markets, e.g. the Czech Republic, it seems that the relationship becomes stronger, whereas e.g. for the APX Netherlands it becomes weaker. We also observe that for some markets, like Switzerland, the improvement in forecasting accuracy is only very high at some months of the year. An analysis of the daily structure revealed that the improvement by using EXAA-related models is relatively stable across all weekdays and for most of the markets.

Nevertheless, this paper still may function as an early introduction in a new perspective on the investigation of related markets. For our analysis we used very basic but robust modeling approaches for every time series. As explained in detail in section \ref{models}, our autoregressive modeling setup nests many models frequently used in the literature. However, we explicitly avoided higher tier models like seasonal ARIMA models, models with heteroscedasticity or models with regressors in general. This helped us to focus on the analysis of the relationship of the market prices but may also have biased our results, as those extended model may negate the improvement of forecasting we have found using EXAA-based approaches.

Therefore we still leave several issues for future research.

One feature that was not taken into account are the trading days of the spot markets. 
All considers electricity spot markets trade every day except the EXAA.
The EXAA trades only on working days and not on weekends and Austrian public holidays. Therefore on a Friday of a 
common week there will be traded three days, Saturday, Sunday and Monday. Having this fact in mind
it is even more remarkable that the EXAA results are that impressive. Hence, further modeling approaches
could also investigate the trading days, so that the flow of information can be captured better. % of the EXAA more detailed.
% , even though they are 
% trading sometimes far in the future.

Our analysis of the temporal structure of the relationship also gives clear indication for the usage of change points within the time series. Estimating and modeling those change points will very likely result in an overall better forecasting performance, as the markets have obviously changed over the past years. Such change points may also account for the fact that the performance of all models typically suffers during the winter months. The detection of such change points could be driven by real events like milestones in the ongoing coupling of electricity markets in Europe or statistical techniques.

Further research could also go into the direction of constructing a cascade model.
As the Table \ref{tab_exchanges} suggests we can push forward the idea of using certain available information.
So we could e.g. incorporate the EXAA information to estimate the POLPX price, then using the POLPX and EXAA price to
forecast the Czech, Hungary and Swiss electricity price. All these prices could be used to forecast all the other ones that trade later on.
This cascade type model can be extend to a complex net of electricity prices for even greater regions.

Obviously, another room for improvement lays in the types of models we used. We could therefore extend the models to more complex ones. One way in this sense could be to take into account non-linearities, interactions or other regressors such as load, renewable energy feed-in, etc. Another way to improve the models could be the relaxation of some of our assumptions, e.g. allowing the variance to vary periodically over time.
 
% -----------------------------------
% 
% We could do more in the direction, "standard":
% 
% only processes with linear autoregressive structure considered, so far.
% 
% parameter selection, so other information criteria, lasso method
% or dimensional reduction approaches, like pca on the high-dimensional time series first.
% we could consider non-linearities, interactions, other regressors such as load, renewable energy feed-in, etc.
% further all parameters are assumed to be constant over time, this can be relaxed, so e.g. periodically varying or
% some dummy variable approaches or change points.
% 
% 
% "cascade" models...
% 
% "exaa weekend/ph" only traded on working days, others every day...

\section{References}

\bibliographystyle{apalike}
\bibliography{Bibliothek}

\begin{thebibliography}{}

\bibitem[Bollino et~al., 2013]{bollino2013integration}
Bollino, C.~A., Ciferri, D., and Polinori, P. (2013).
\newblock Integration and convergence in european electricity markets.
\newblock {\em Available at SSRN 2227541}.

\bibitem[Bordignon et~al., 2013]{bordignon2013combining}
Bordignon, S., Bunn, D.~W., Lisi, F., and Nan, F. (2013).
\newblock {Combining day-ahead forecasts for British electricity prices}.
\newblock {\em Energy Economics}, 35(0):88--103.

\bibitem[Bosco et~al., 2010]{bosco2010long}
Bosco, B., Parisio, L., Pelagatti, M., and Baldi, F. (2010).
\newblock Long-run relations in european electricity prices.
\newblock {\em Journal of applied econometrics}, 25(5):805--832.

\bibitem[Bunn and Gianfreda, 2010]{bunn2010integration}
Bunn, D.~W. and Gianfreda, A. (2010).
\newblock Integration and shock transmissions across european electricity
  forward markets.
\newblock {\em Energy Economics}, 32(2):278--291.

\bibitem[Diebold, 2012]{diebold2012comparing}
Diebold, F.~X. (2012).
\newblock Comparing predictive accuracy, twenty years later: A personal
  perspective on the use and abuse of diebold-mariano tests.
\newblock Technical report, National Bureau of Economic Research.

\bibitem[Erni, 2012]{erni2012day}
Erni, D. (2012).
\newblock {\em Day-Ahead Electricity Spot Prices-Fundamental Modelling and the
  Role of Expected Wind Electricity Infeed at the European Energy Exchange}.
\newblock PhD thesis, University of St. Gallen.

\bibitem[Ferkingstad et~al., 2011]{ferkingstad2011causal}
Ferkingstad, E., L{\o}land, A., and Wilhelmsen, M. (2011).
\newblock Causal modeling and inference for electricity markets.
\newblock {\em Energy Economics}, 33(3):404--412.

\bibitem[Hamilton, 1994]{hamilton1994time}
Hamilton, J.~D. (1994).
\newblock {\em Time series analysis}, volume~2.
\newblock Princeton university press Princeton.

\bibitem[Hickey et~al., 2012]{hickey2012forecasting}
Hickey, E., Loomis, D.~G., and Mohammadi, H. (2012).
\newblock Forecasting hourly electricity prices using armax--garch models: An
  application to miso hubs.
\newblock {\em Energy Economics}, 34(1):307--315.

\bibitem[Hong and Wu, 2012]{hong2012day}
Hong, Y.-Y. and Wu, C.-P. (2012).
\newblock Day-ahead electricity price forecasting using a hybrid principal
  component analysis network.
\newblock {\em Energies}, 5(11):4711--4725.

\bibitem[Houllier and de~Menezes, 2012]{houllier2012fractional}
Houllier, M.~A. and de~Menezes, L.~M. (2012).
\newblock A fractional cointegration analysis of european electricity spot
  prices.
\newblock In {\em European Energy Market (EEM), 2012 9th International
  Conference on the European Energy Market}, pages 1--6. IEEE.

\bibitem[Huisman and Kili{\c{c}}, 2013]{huisman2013history}
Huisman, R. and Kili{\c{c}}, M. (2013).
\newblock A history of european electricity day-ahead prices.
\newblock {\em Applied Economics}, 45(18):2683--2693.

\bibitem[Kalantzis and Milonas, 2010]{kalantzis2010market}
Kalantzis, F. and Milonas, N.~T. (2010).
\newblock Market integration and price dispersion in the european electricity
  market.
\newblock In {\em European Energy Market (EEM), 2010 7th International
  Conference on the European Energy Market}, pages 1--6. IEEE.

\bibitem[Karakatsani and Bunn, 2008]{karakatsani2008forecasting}
Karakatsani, N.~V. and Bunn, D.~W. (2008).
\newblock {Forecasting electricity prices: The impact of fundamentals and
  time-varying coefficients}.
\newblock {\em International Journal of Forecasting}, 24(4):764--785.

\bibitem[Keles et~al., 2013]{keles2013combined}
Keles, D., Genoese, M., M{\"o}st, D., Ortlieb, S., and Fichtner, W. (2013).
\newblock {A combined modeling approach for wind power feed-in and electricity
  spot prices}.
\newblock {\em Energy Policy}, 59(0):213--225.

\bibitem[Koopman et~al., 2007]{koopman2007periodic}
Koopman, S.~J., Ooms, M., and Carnero, M.~A. (2007).
\newblock {Periodic seasonal Reg-ARFIMA--GARCH models for daily electricity
  spot prices}.
\newblock {\em Journal of the American Statistical Association},
  102(477):16--27.

\bibitem[Kristiansen, 2012]{kristiansen2012forecasting}
Kristiansen, T. (2012).
\newblock {Forecasting Nord Pool day-ahead prices with an autoregressive
  model}.
\newblock {\em Energy Policy}, 49(0):328--332.

\bibitem[Liebl, 2013]{liebl2013modeling}
Liebl, D. (2013).
\newblock {Modeling and forecasting electricity spot prices: A functional data
  perspective}.
\newblock {\em The Annals of Applied Statistics}, 7(3):1562--1592.

\bibitem[Liu and Shi, 2013]{liu2013applying}
Liu, H. and Shi, J. (2013).
\newblock {Applying ARMA--GARCH approaches to forecasting short-term
  electricity prices}.
\newblock {\em Energy Economics}, 37(0):152--166.

\bibitem[Nan et~al., 2014]{nan2014forecasting}
Nan, F., Bordignon, S., Bunn, D.~W., and Lisi, F. (2014).
\newblock The forecasting accuracy of electricity price formation models.
\newblock {\em International Journal of Energy and Statistics}, 2(01):1--26.

\bibitem[Ronn and Wimschulte, 2009]{ronn2009intra}
Ronn, E.~I. and Wimschulte, J. (2009).
\newblock Intra-day risk premia in european electricity forward markets.
\newblock {\em Journal of Energy Markets}, 2(4):71--98.

\bibitem[Serinaldi, 2011]{serinaldi2011distributional}
Serinaldi, F. (2011).
\newblock Distributional modeling and short-term forecasting of electricity
  prices by generalized additive models for location, scale and shape.
\newblock {\em Energy Economics}, 33(6):1216--1226.

\bibitem[Taylor, 2010]{taylor2010triple}
Taylor, J.~W. (2010).
\newblock Triple seasonal methods for short-term electricity demand
  forecasting.
\newblock {\em European Journal of Operational Research}, 204(1):139--152.

\bibitem[Viehmann, 2011]{viehmann2011risk}
Viehmann, J. (2011).
\newblock Risk premiums in the german day-ahead electricity market.
\newblock {\em Energy policy}, 39(1):386--394.

\bibitem[Weron and Misiorek, 2008]{weron2008forecasting}
Weron, R. and Misiorek, A. (2008).
\newblock Forecasting spot electricity prices: A comparison of parametric and
  semiparametric time series models.
\newblock {\em International Journal of Forecasting}, 24(4):744--763.

\bibitem[Zachmann, 2008]{zachmann2008electricity}
Zachmann, G. (2008).
\newblock Electricity wholesale market prices in europe: Convergence?
\newblock {\em Energy Economics}, 30(4):1659--1671.

\bibitem[Ziel et~al., 2015]{ziel2015efficient}
Ziel, F., Steinert, R., and Husmann, S. (2015).
\newblock Efficient modeling and forecasting of electricity spot prices.
\newblock {\em Energy Economics}, 47:98--111.

\end{thebibliography}

\section{Appendix}
 
\begin{figure}[hbt!]
\centering
\begin{subfigure}[b]{.49\textwidth}
 \includegraphics[width=1\textwidth, height=.17\textheight]{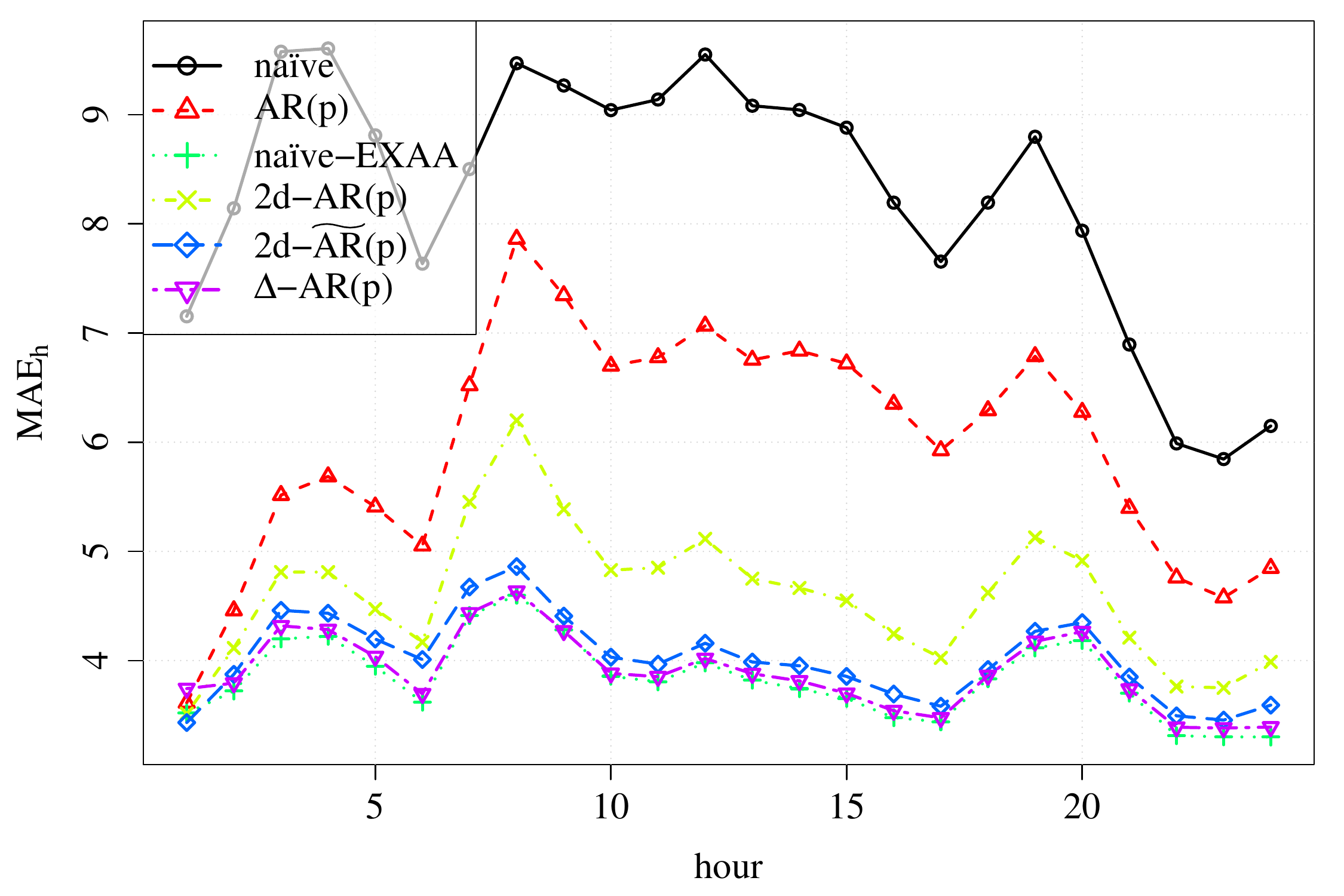} 
  \caption{EPEX DE\&AT}
%   \label{fig_dm_1}
\end{subfigure}
\begin{subfigure}[b]{.49\textwidth}
 \includegraphics[width=1\textwidth, height=.17\textheight]{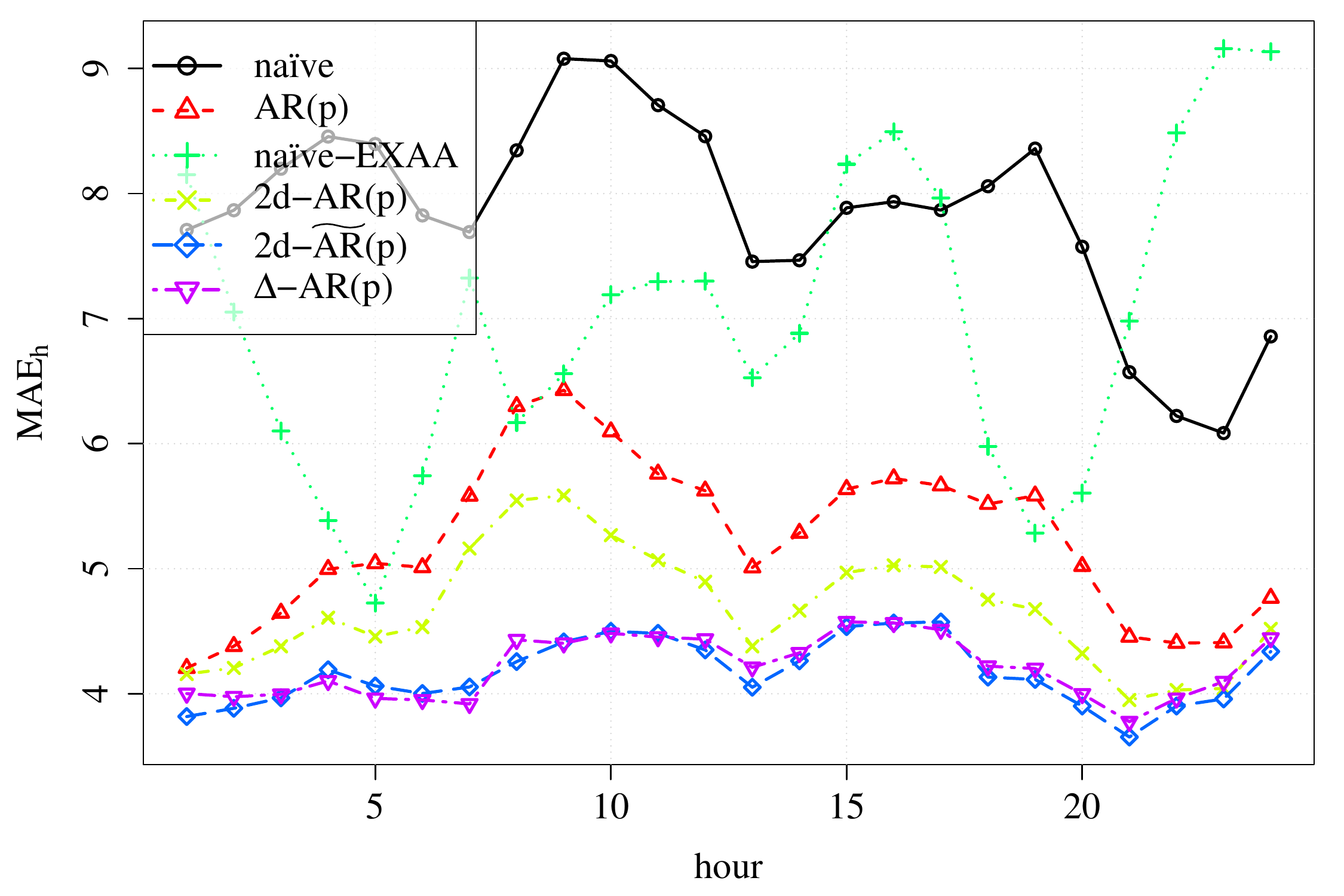} 
  \caption{EPEX CHE}
%   \label{fig_dm_2}
\end{subfigure}
\begin{subfigure}[b]{.49\textwidth}
 \includegraphics[width=1\textwidth, height=.17\textheight]{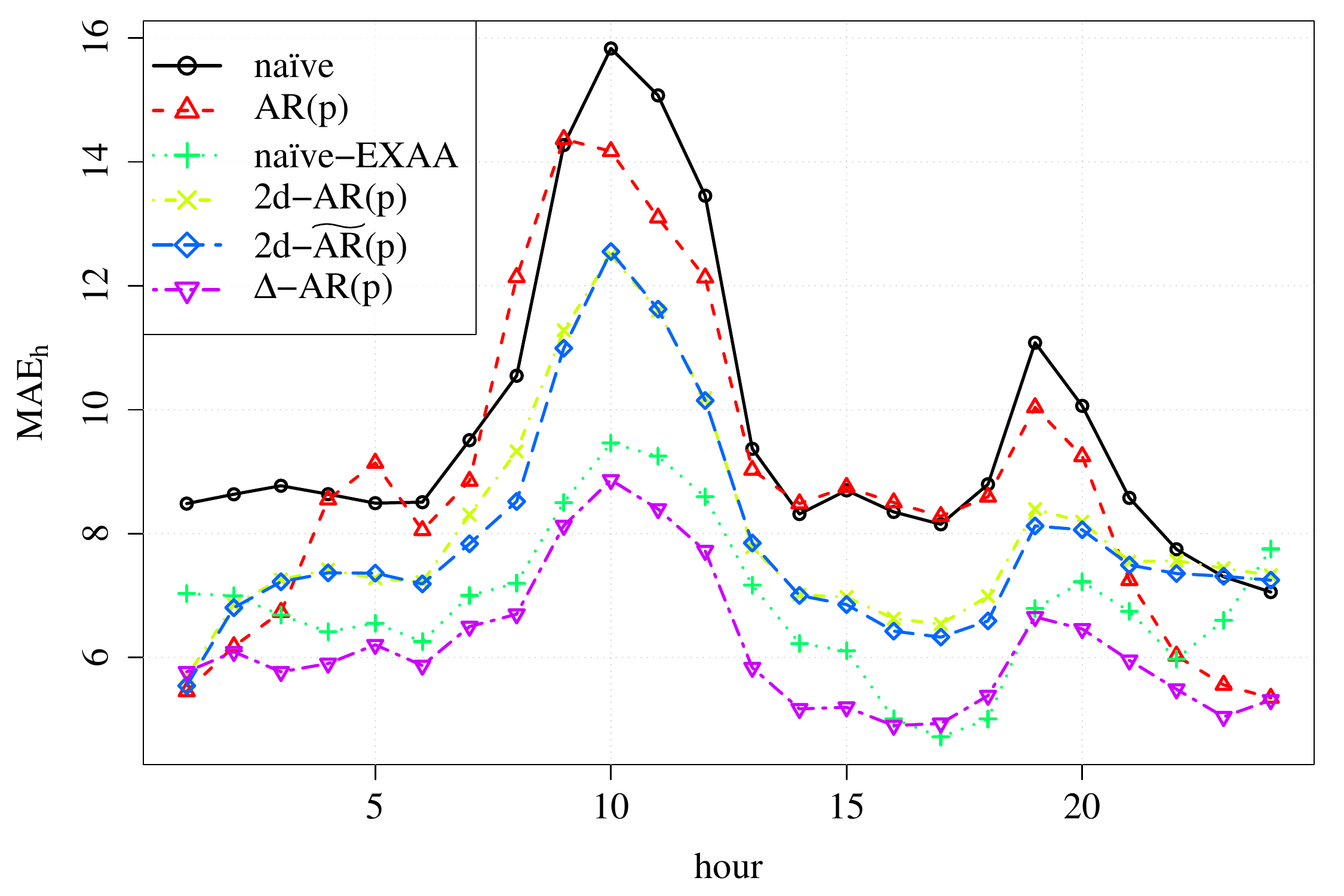} 
  \caption{EPEX FR}
%   \label{fig_dm_3}
\end{subfigure}
\begin{subfigure}[b]{.49\textwidth}
 \includegraphics[width=1\textwidth, height=.17\textheight]{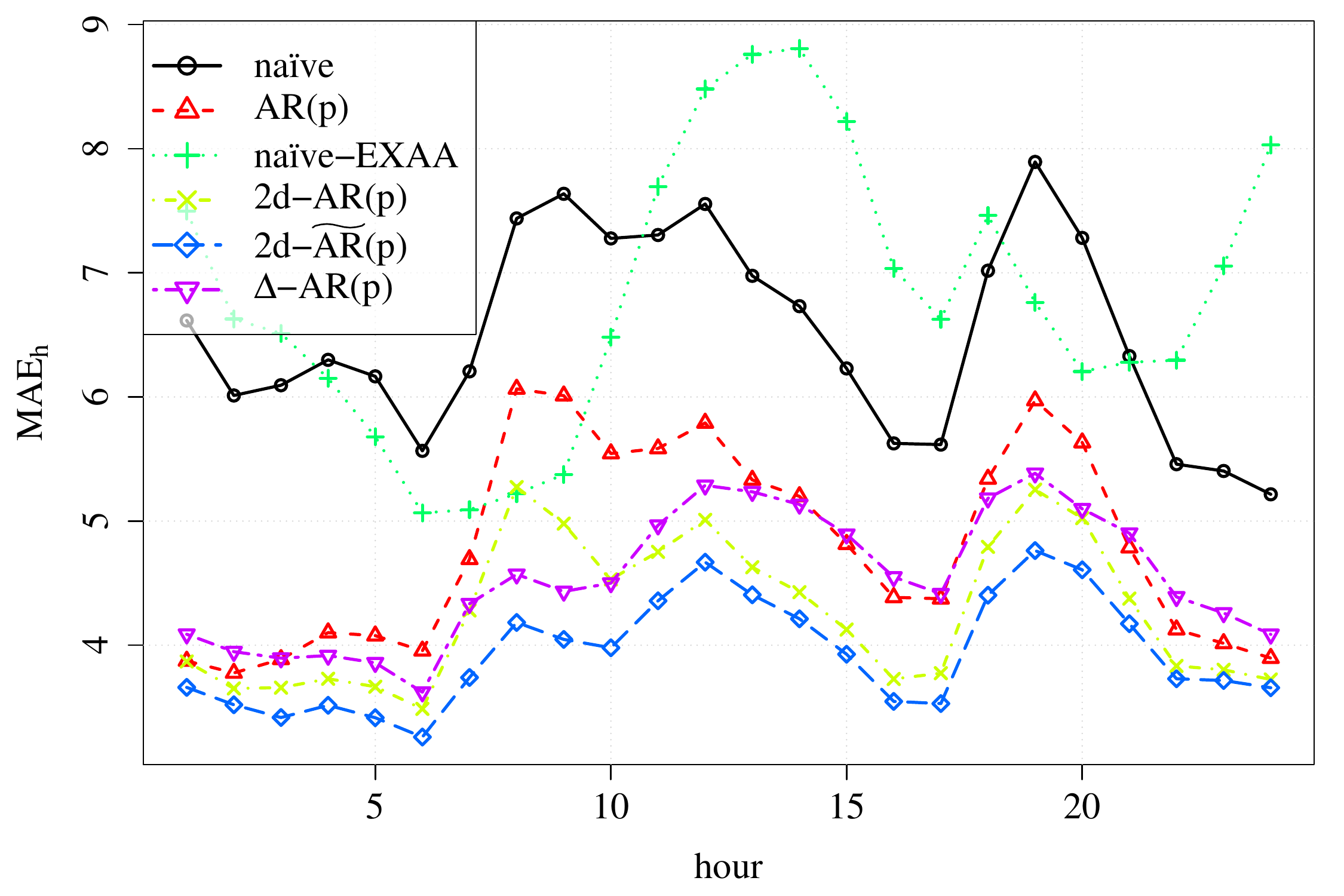} 
  \caption{APX NL}
%   \label{fig_dm_4}
\end{subfigure}
\begin{subfigure}[b]{.49\textwidth}
 \includegraphics[width=1\textwidth, height=.17\textheight]{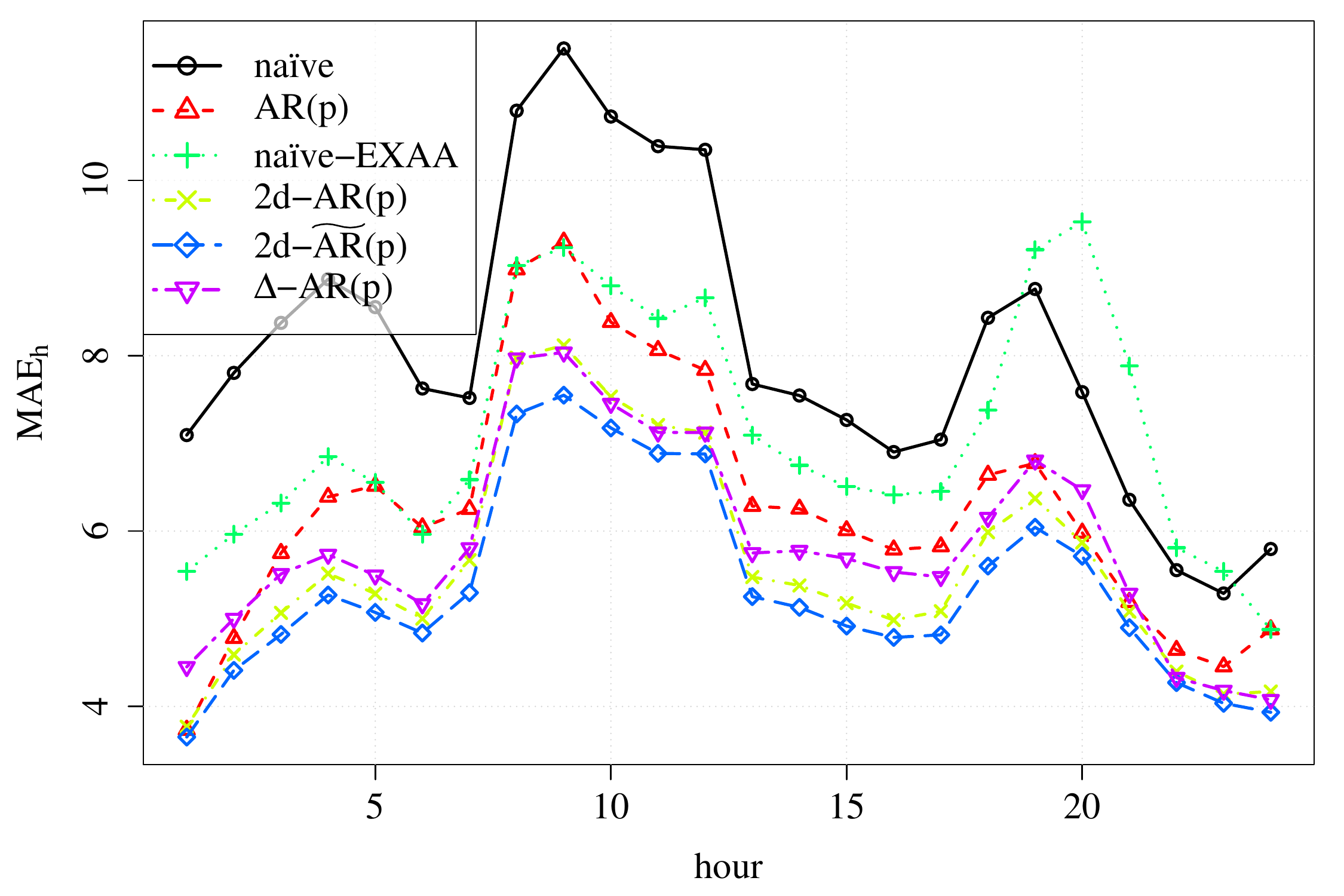} 
  \caption{Nordpool DK West}
%   \label{fig_dm_5}
\end{subfigure}
\begin{subfigure}[b]{.49\textwidth}
 \includegraphics[width=1\textwidth, height=.17\textheight]{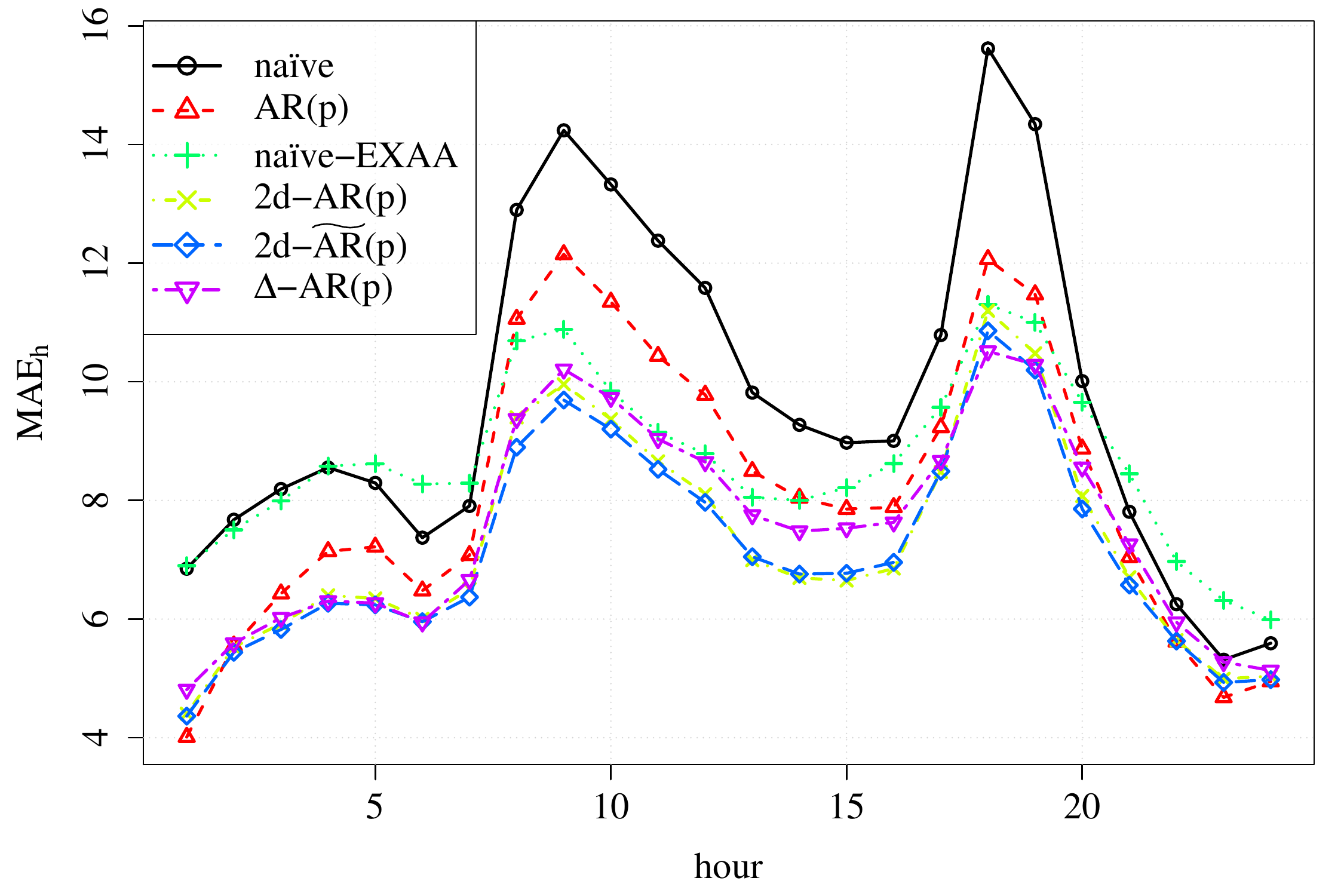} 
  \caption{Nordpool DK East}
%   \label{fig_dm_6}
\end{subfigure}
\begin{subfigure}[b]{.49\textwidth}
 \includegraphics[width=1\textwidth, height=.17\textheight]{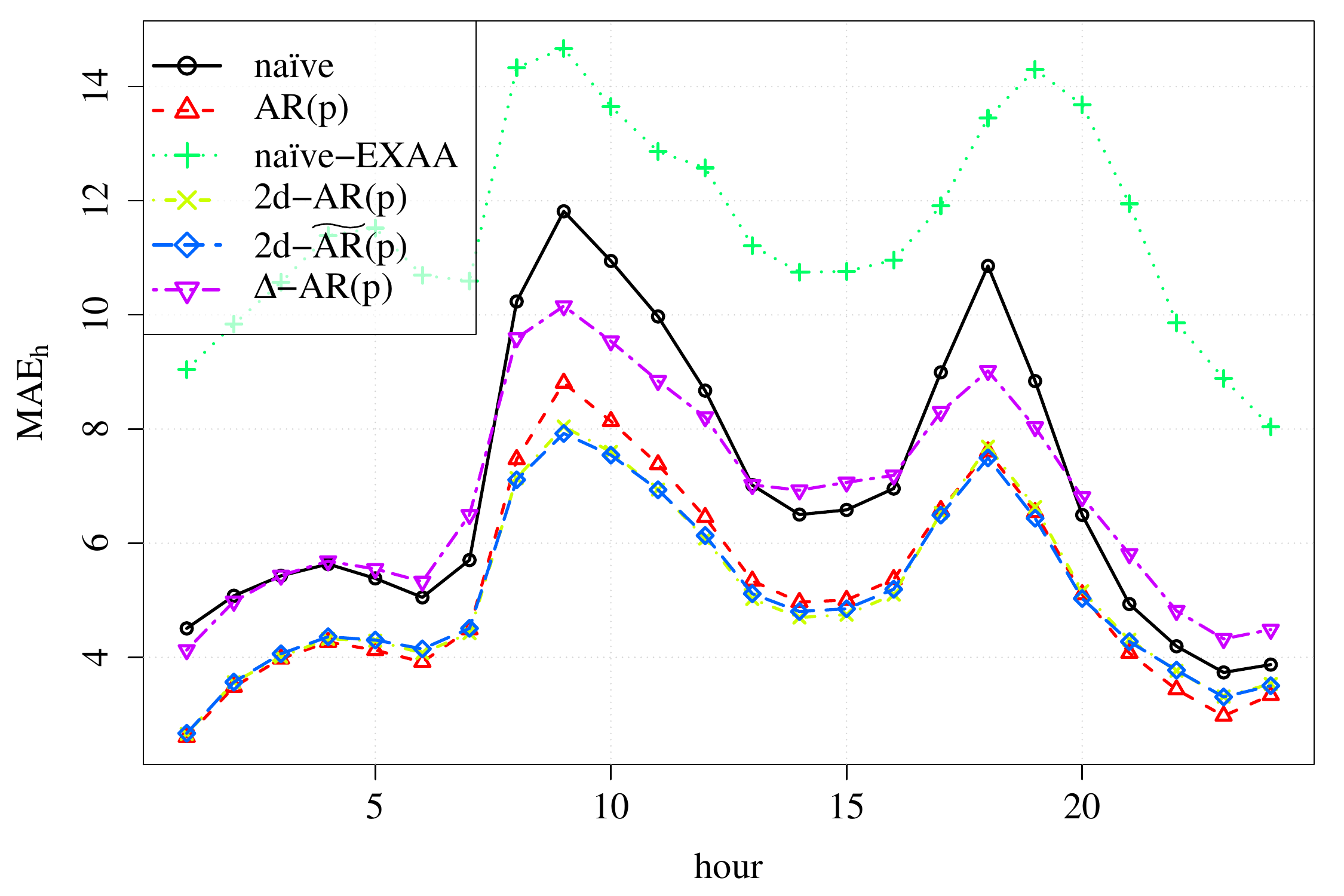} 
  \caption{Nordpool Sweden4}
%   \label{fig_dm_7}
\end{subfigure}
\begin{subfigure}[b]{.49\textwidth}
 \includegraphics[width=1\textwidth, height=.17\textheight]{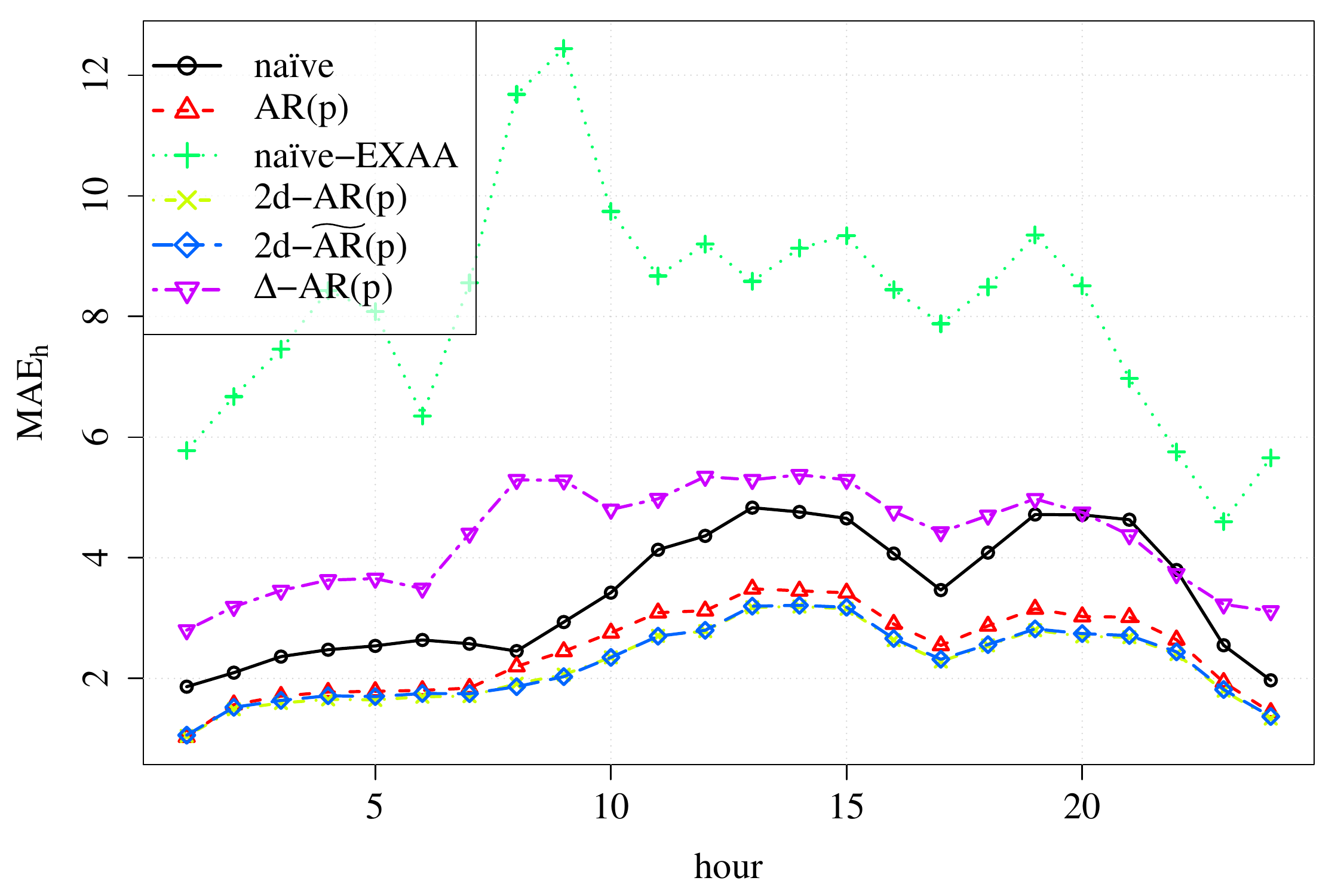} 
  \caption{POLPX PL}
%   \label{fig_dm_8}
\end{subfigure}
\begin{subfigure}[b]{.49\textwidth}
 \includegraphics[width=1\textwidth, height=.17\textheight]{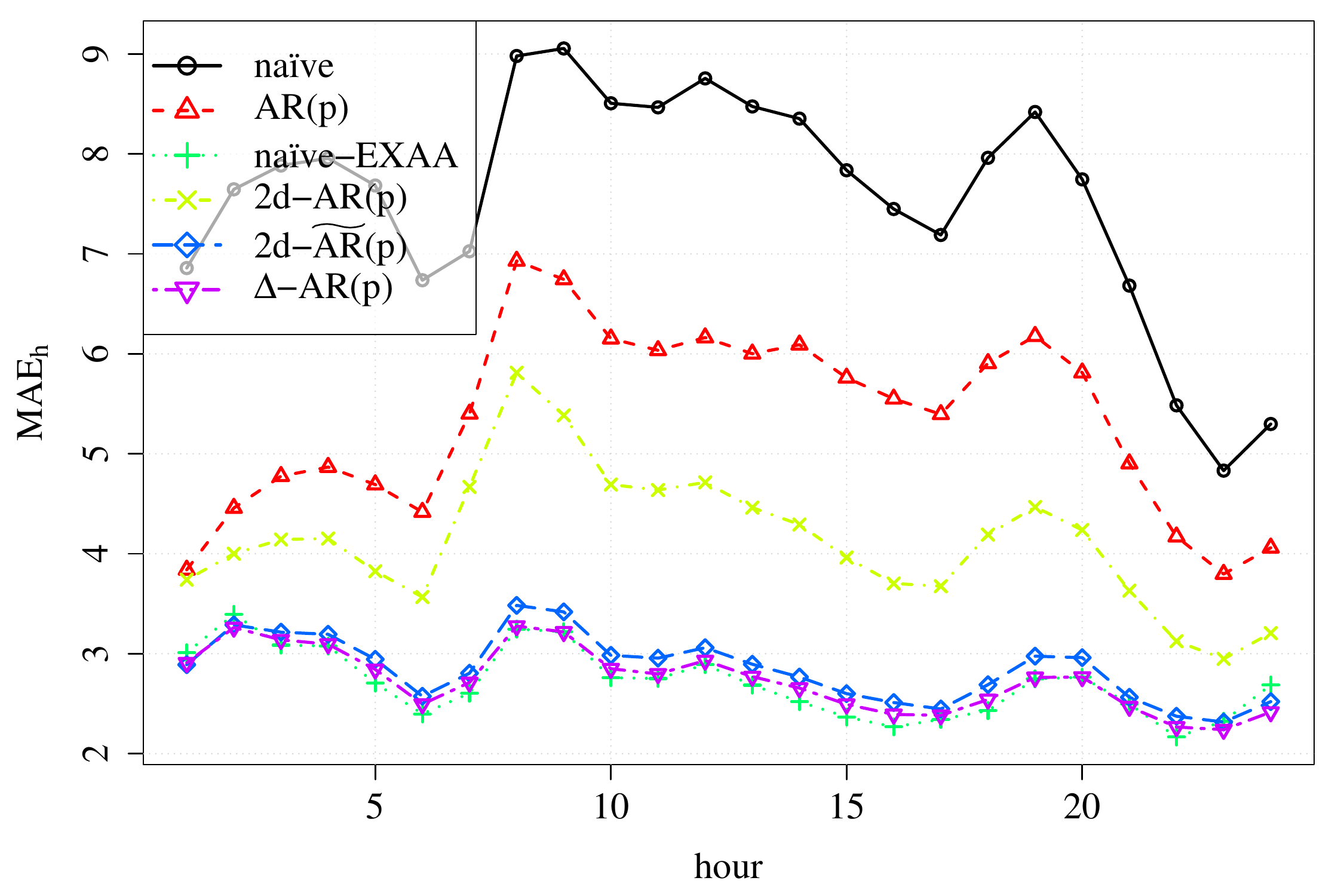} 
  \caption{OTE CZ}
%   \label{fig_dm_9}
\end{subfigure}
\begin{subfigure}[b]{.49\textwidth}
 \includegraphics[width=1\textwidth, height=.17\textheight]{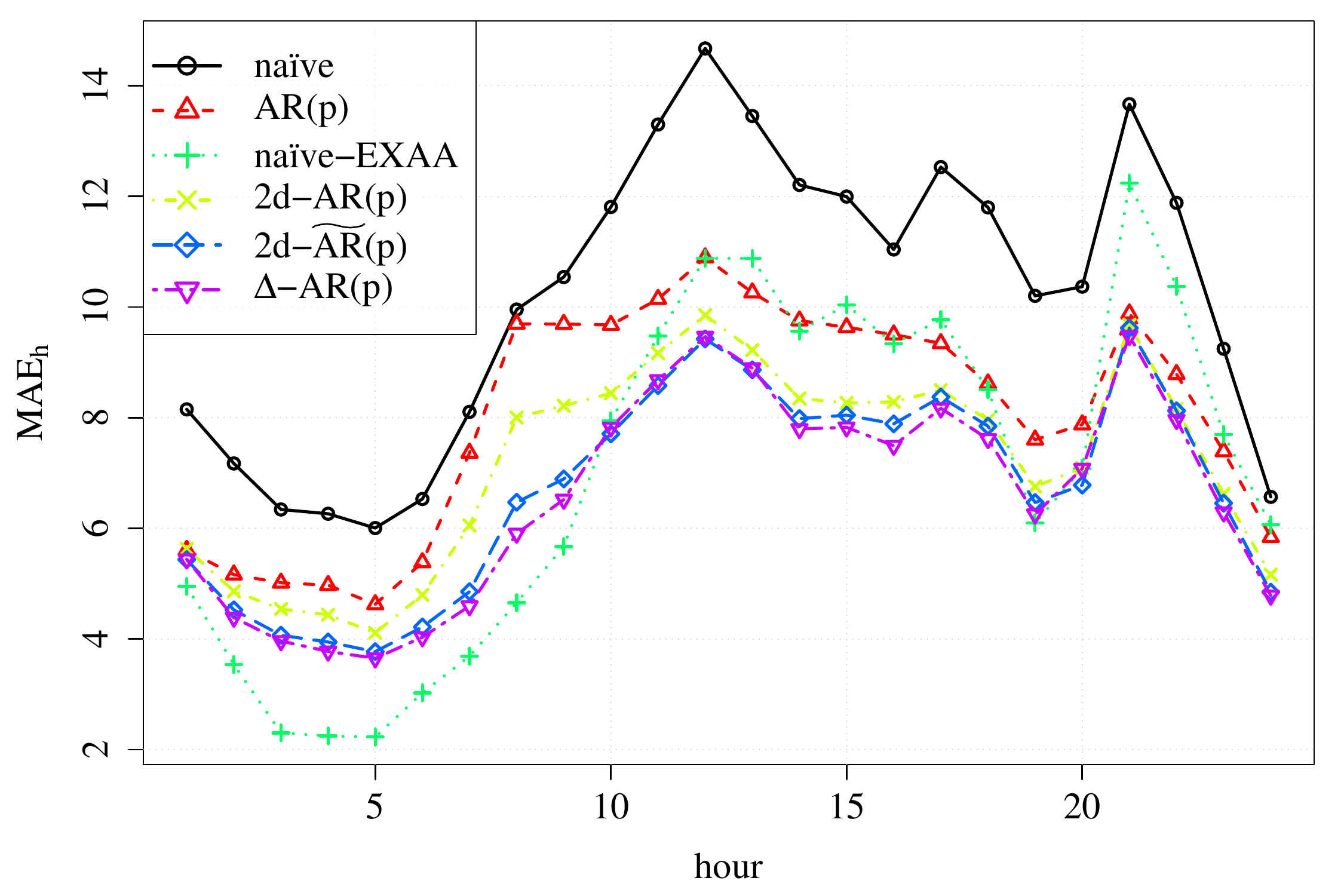} 
  \caption{HUPX HU}
%   \label{fig_dm_10}
\end{subfigure}
\caption{$\MAE^\XX_h$.}
 \label{fig_MAE_h}
\end{figure}

%%%RMSE
\begin{figure}[hbt!]
\centering
\begin{subfigure}[b]{.49\textwidth}
 \includegraphics[width=1\textwidth, height=.17\textheight]{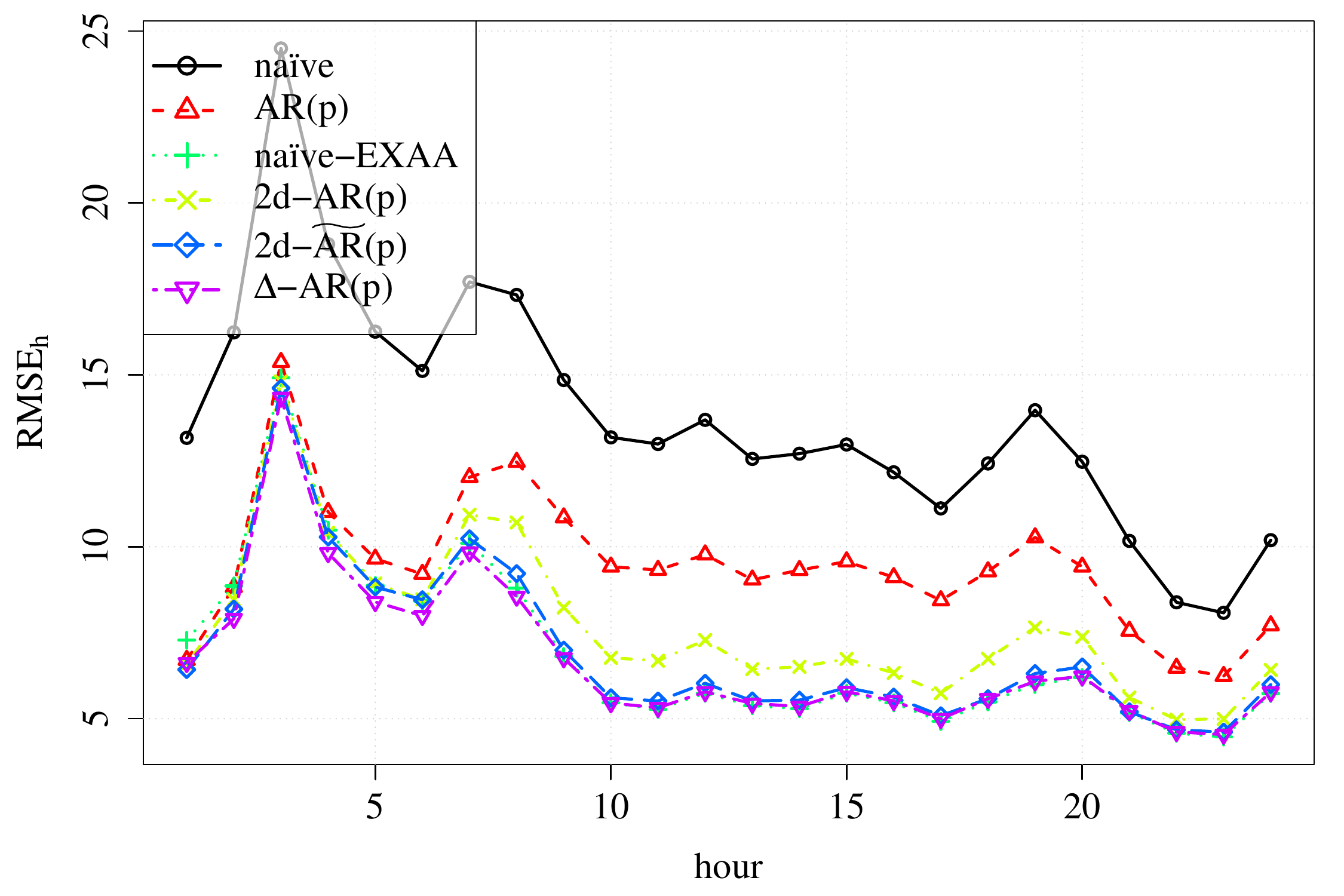} 
  \caption{EPEX DE\&AT}
%   \label{fig_dm_1}
\end{subfigure}
\begin{subfigure}[b]{.49\textwidth}
 \includegraphics[width=1\textwidth, height=.17\textheight]{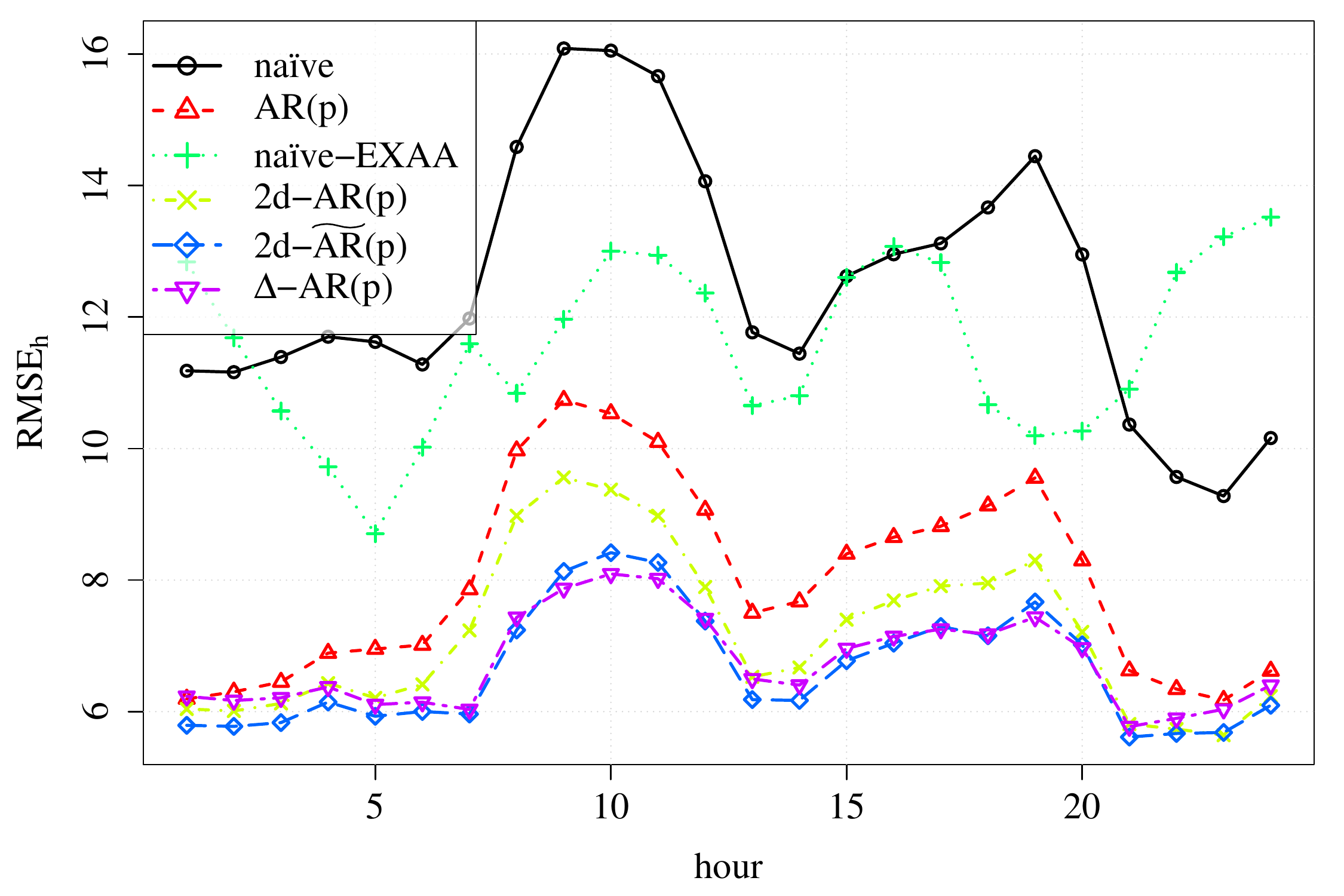} 
  \caption{EPEX CHE}
%   \label{fig_dm_2}
\end{subfigure}
\begin{subfigure}[b]{.49\textwidth}
 \includegraphics[width=1\textwidth, height=.17\textheight]{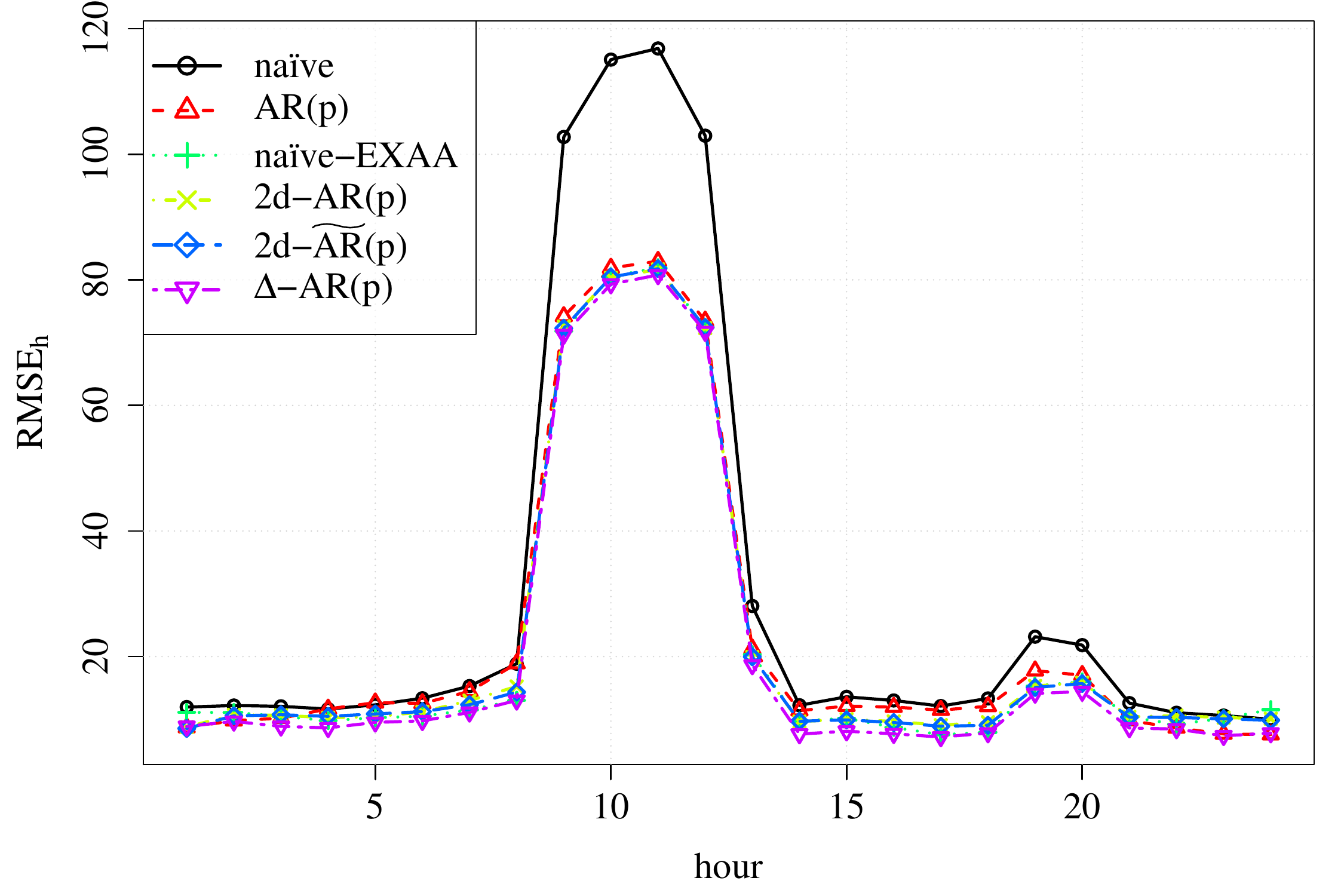} 
  \caption{EPEX FR}
%   \label{fig_dm_3}
\end{subfigure}
\begin{subfigure}[b]{.49\textwidth}
 \includegraphics[width=1\textwidth, height=.17\textheight]{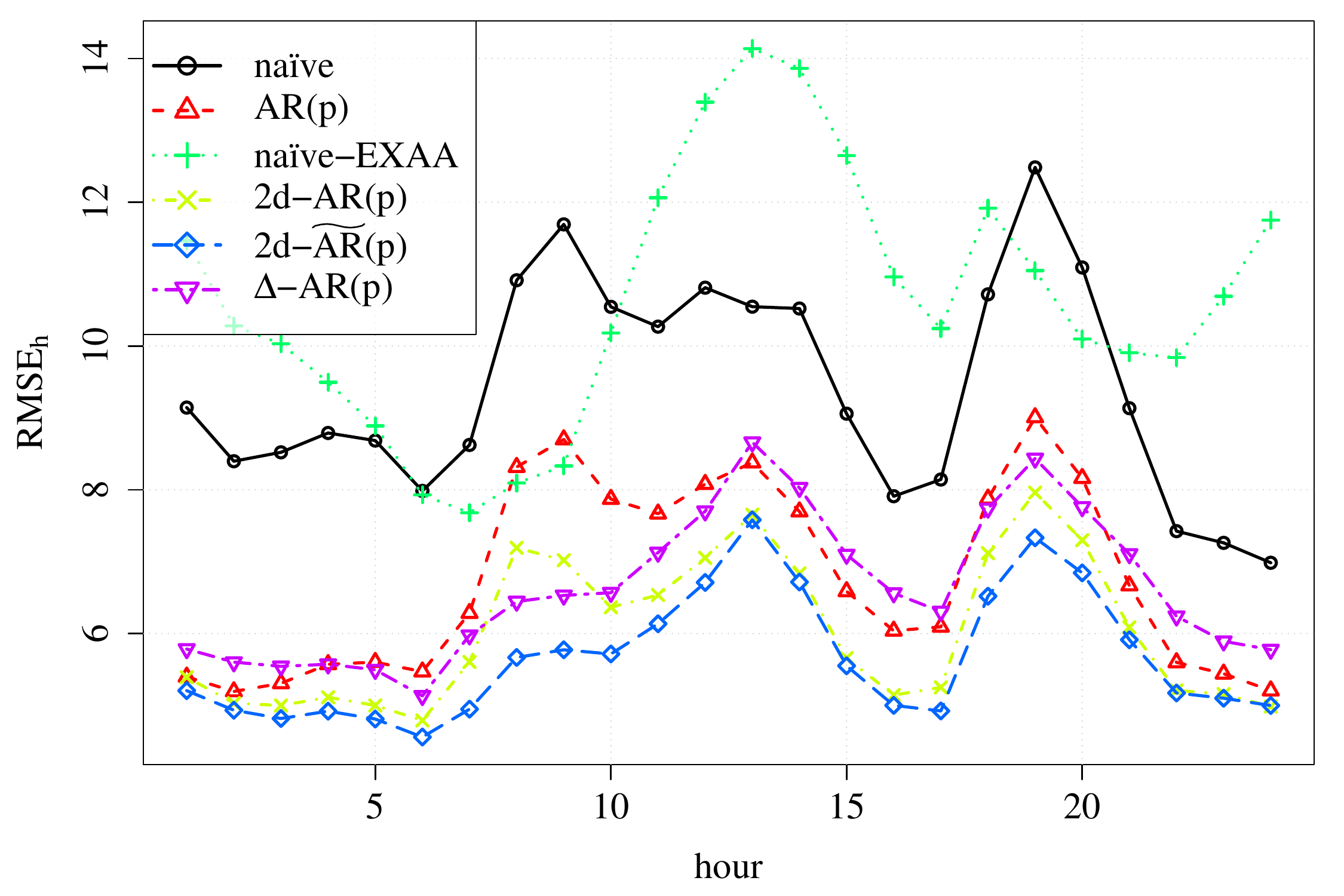} 
  \caption{APX NL}
%   \label{fig_dm_4}
\end{subfigure}
\begin{subfigure}[b]{.49\textwidth}
 \includegraphics[width=1\textwidth, height=.17\textheight]{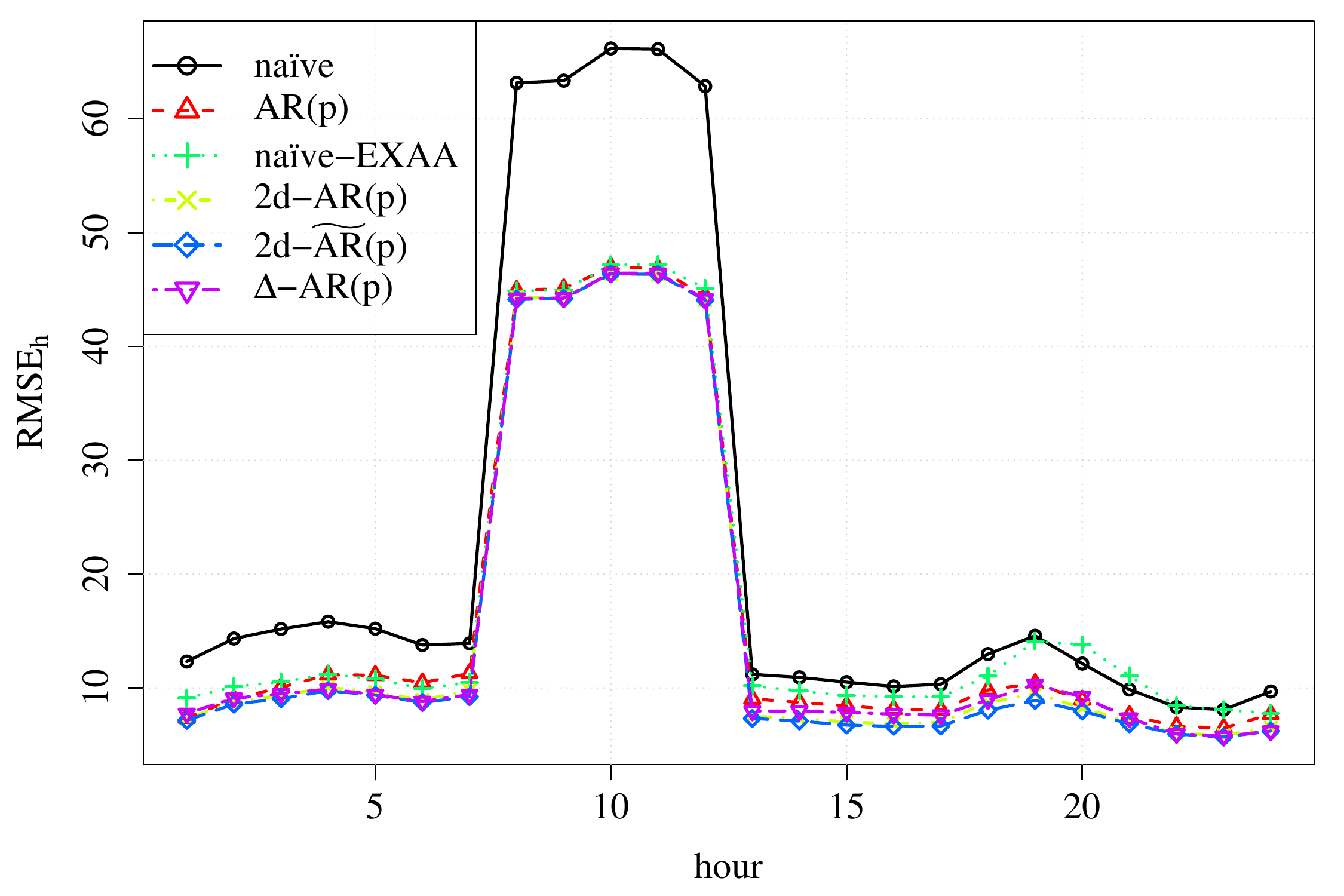} 
  \caption{Nordpool DK West}
%   \label{fig_dm_5}
\end{subfigure}
\begin{subfigure}[b]{.49\textwidth}
 \includegraphics[width=1\textwidth, height=.17\textheight]{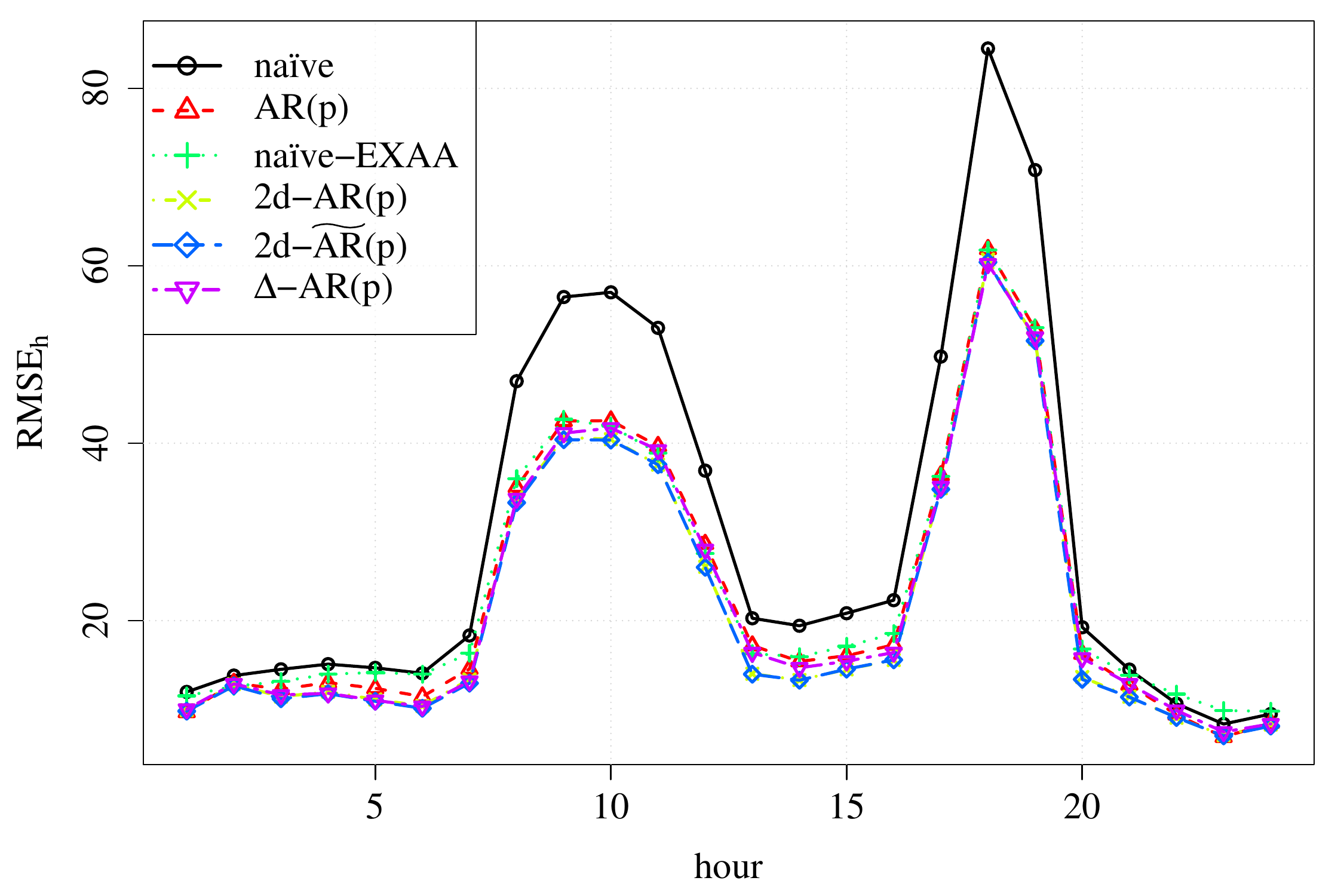} 
  \caption{Nordpool DK East}
%   \label{fig_dm_6}
\end{subfigure}
\begin{subfigure}[b]{.49\textwidth}
 \includegraphics[width=1\textwidth, height=.17\textheight]{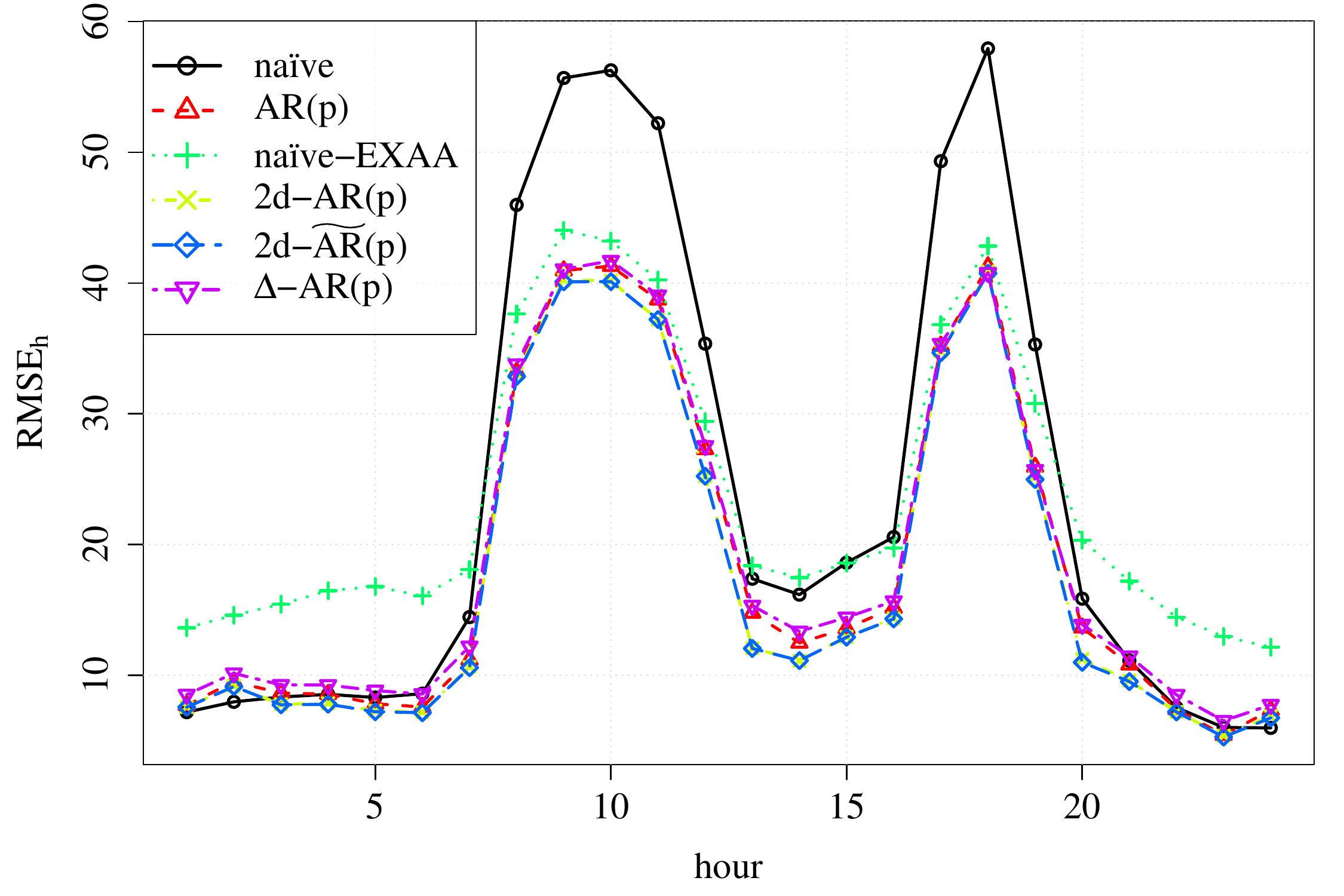} 
  \caption{Nordpool Sweden4}
%   \label{fig_dm_7}
\end{subfigure}
\begin{subfigure}[b]{.49\textwidth}
 \includegraphics[width=1\textwidth, height=.17\textheight]{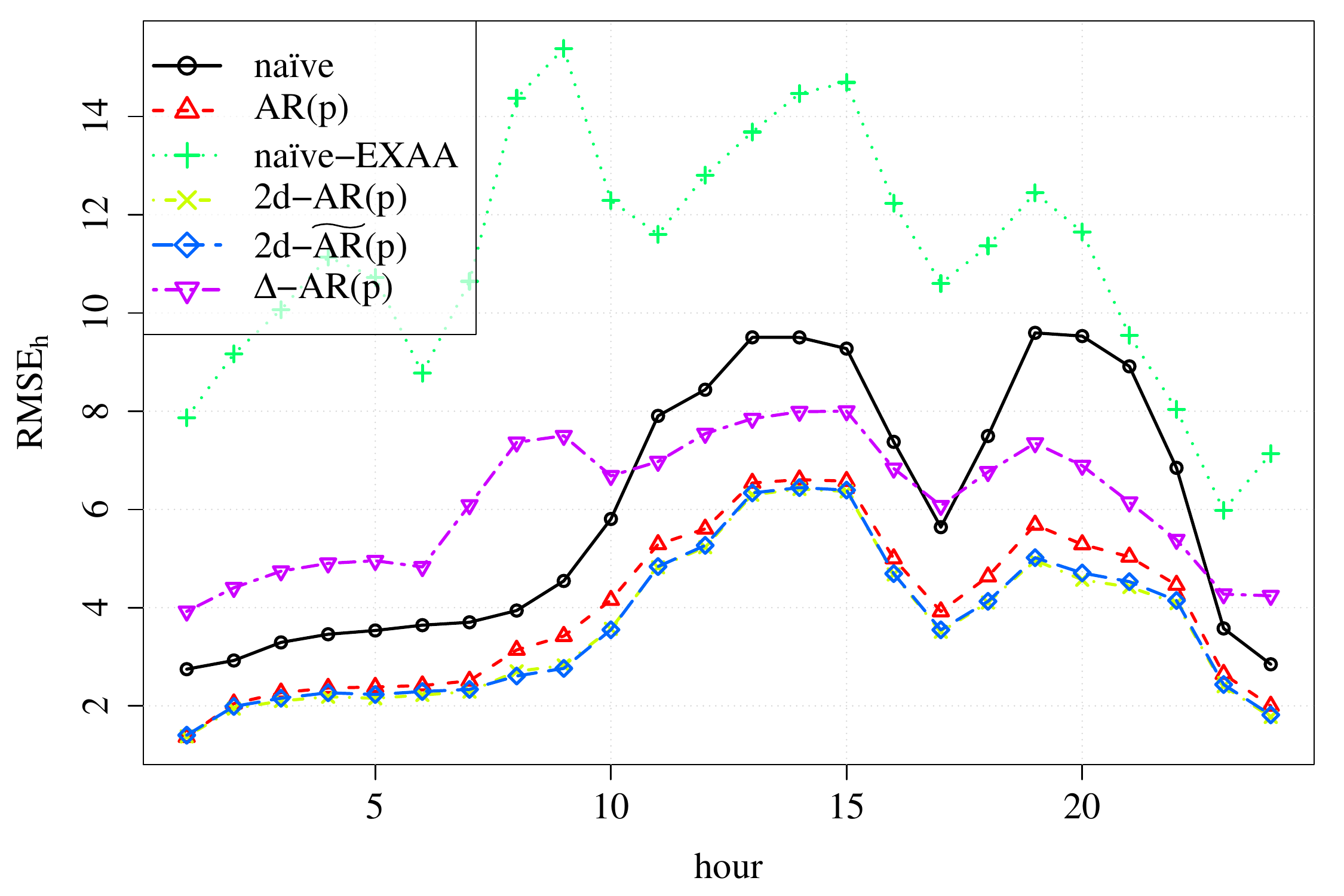} 
  \caption{POLPX PL}
%   \label{fig_dm_8}
\end{subfigure}
\begin{subfigure}[b]{.49\textwidth}
 \includegraphics[width=1\textwidth, height=.17\textheight]{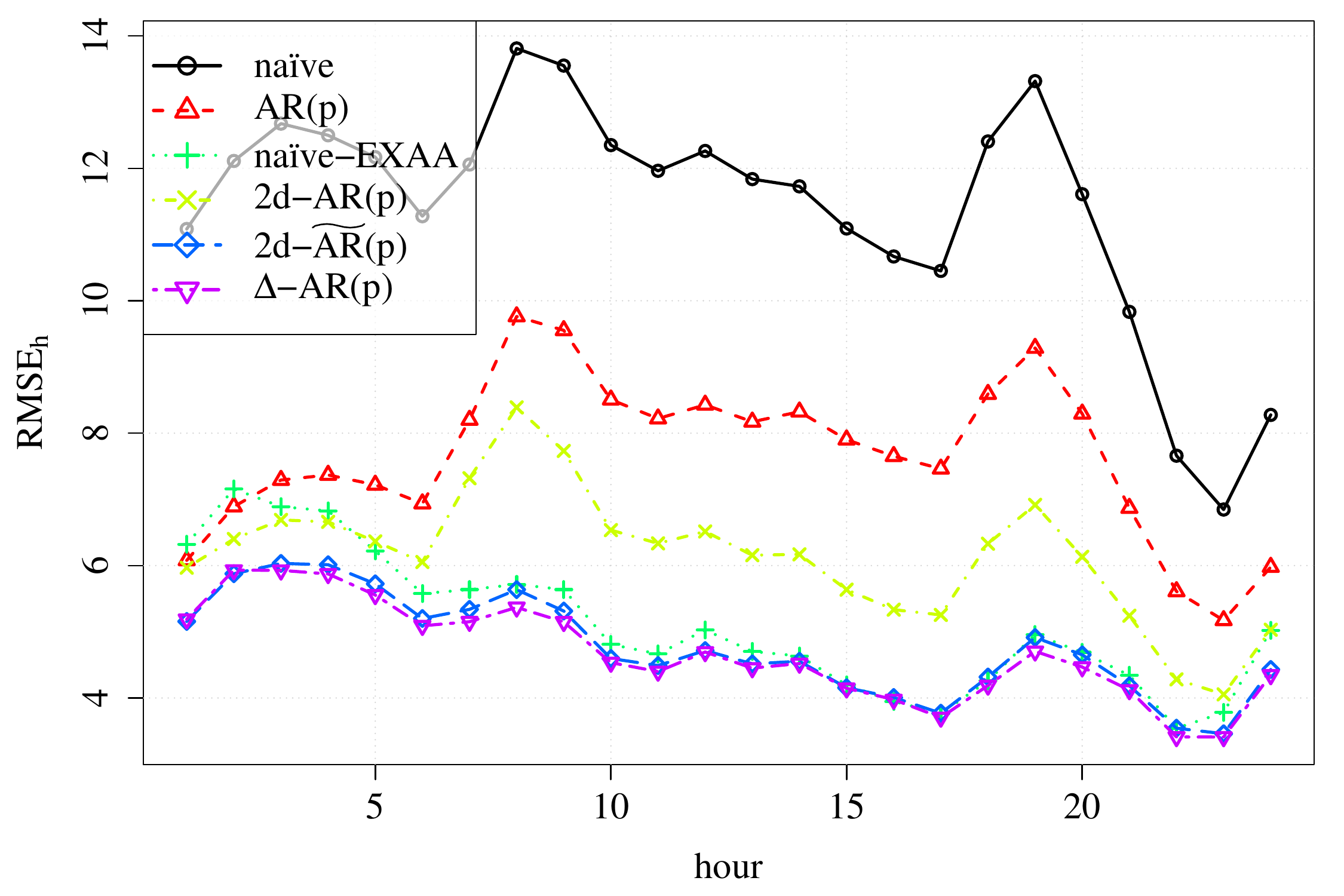} 
  \caption{OTE CZ}
%   \label{fig_dm_9}
\end{subfigure}
\begin{subfigure}[b]{.49\textwidth}
 \includegraphics[width=1\textwidth, height=.17\textheight]{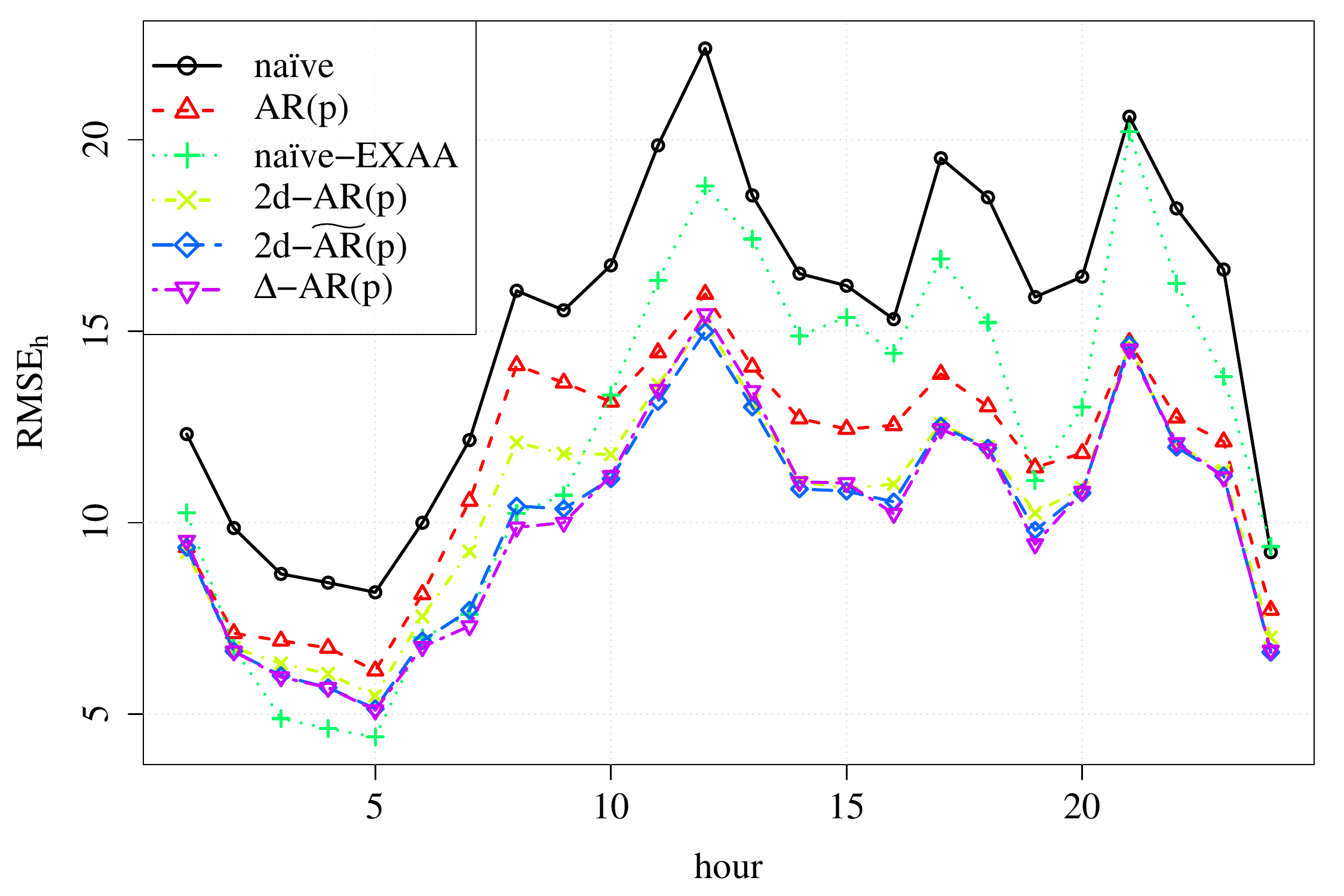} 
  \caption{HUPX HU}
%   \label{fig_dm_10}
\end{subfigure}
%  [1] "EXAA.DE&AT" "EPEX.DE&AT" "EPEX.CHE"   "EPEX.FR"    "BELPEX.BE" 
%  [6] "NP.DK.West" "NP.DK.East" "NP.SW4"     "POLPX.PL"   "OTE.CZ"    
% [11] "HUPX.HU" 

\caption{$\RMSE^\XX_h$.}
 \label{fig_RMSE_h}
\end{figure}

\begin{figure}[hbt!]
\centering
\begin{subfigure}[b]{.49\textwidth}
 \includegraphics[width=1\textwidth, height=.17\textheight]{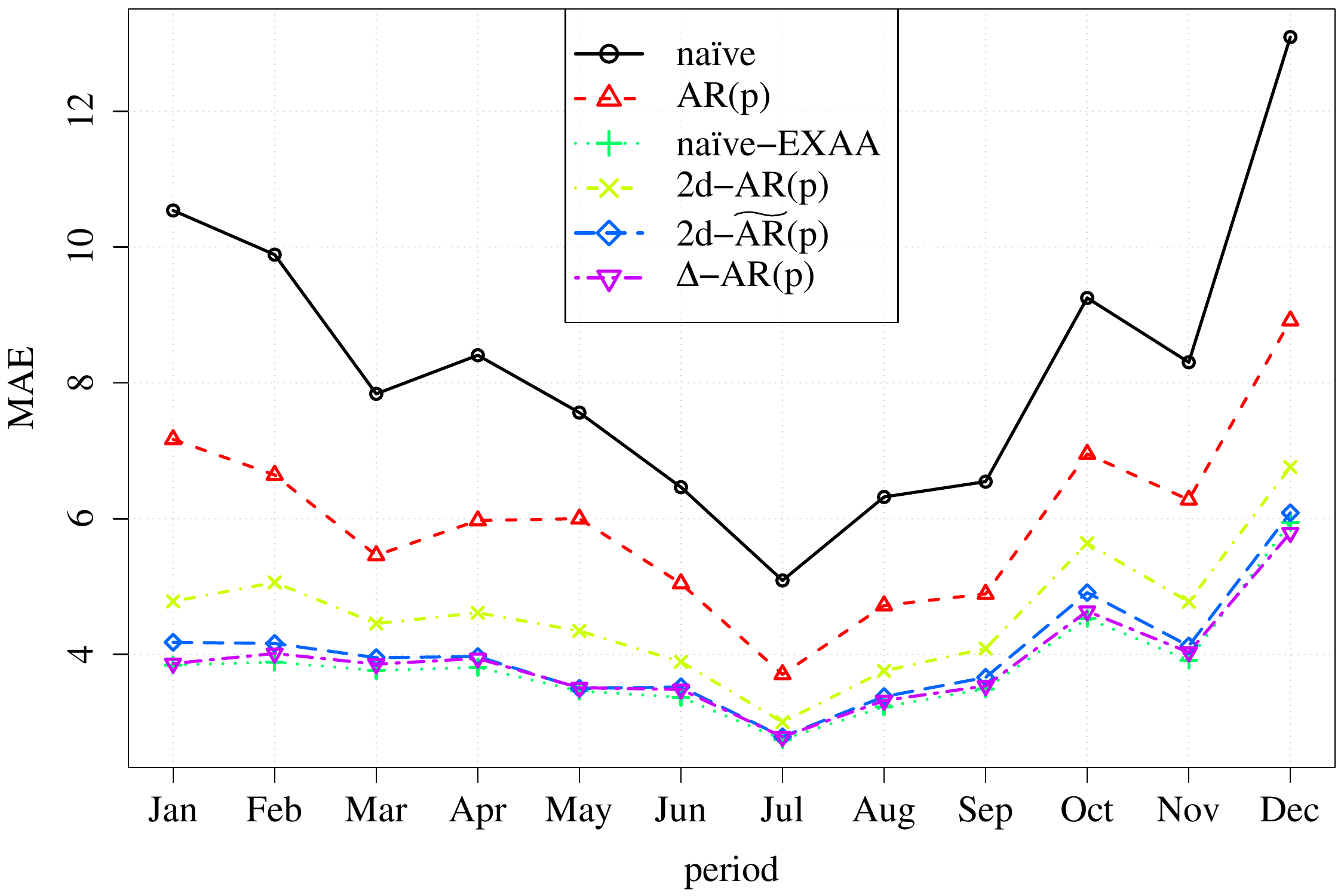} 
  \caption{EPEX DE\&AT}
\end{subfigure}
\begin{subfigure}[b]{.49\textwidth}
 \includegraphics[width=1\textwidth, height=.17\textheight]{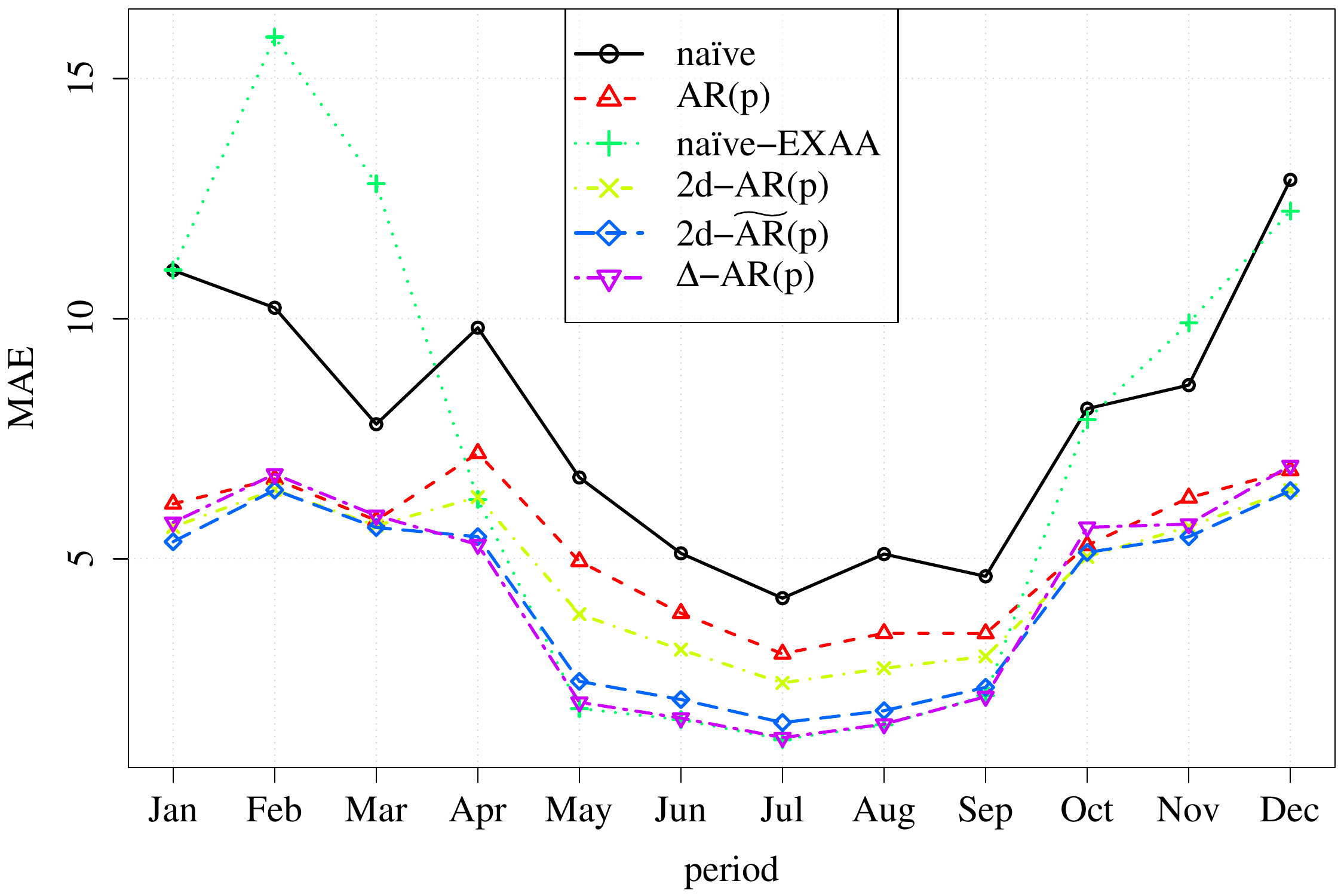} 
  \caption{EPEX CHE}
\end{subfigure}
\begin{subfigure}[b]{.49\textwidth}
 \includegraphics[width=1\textwidth, height=.17\textheight]{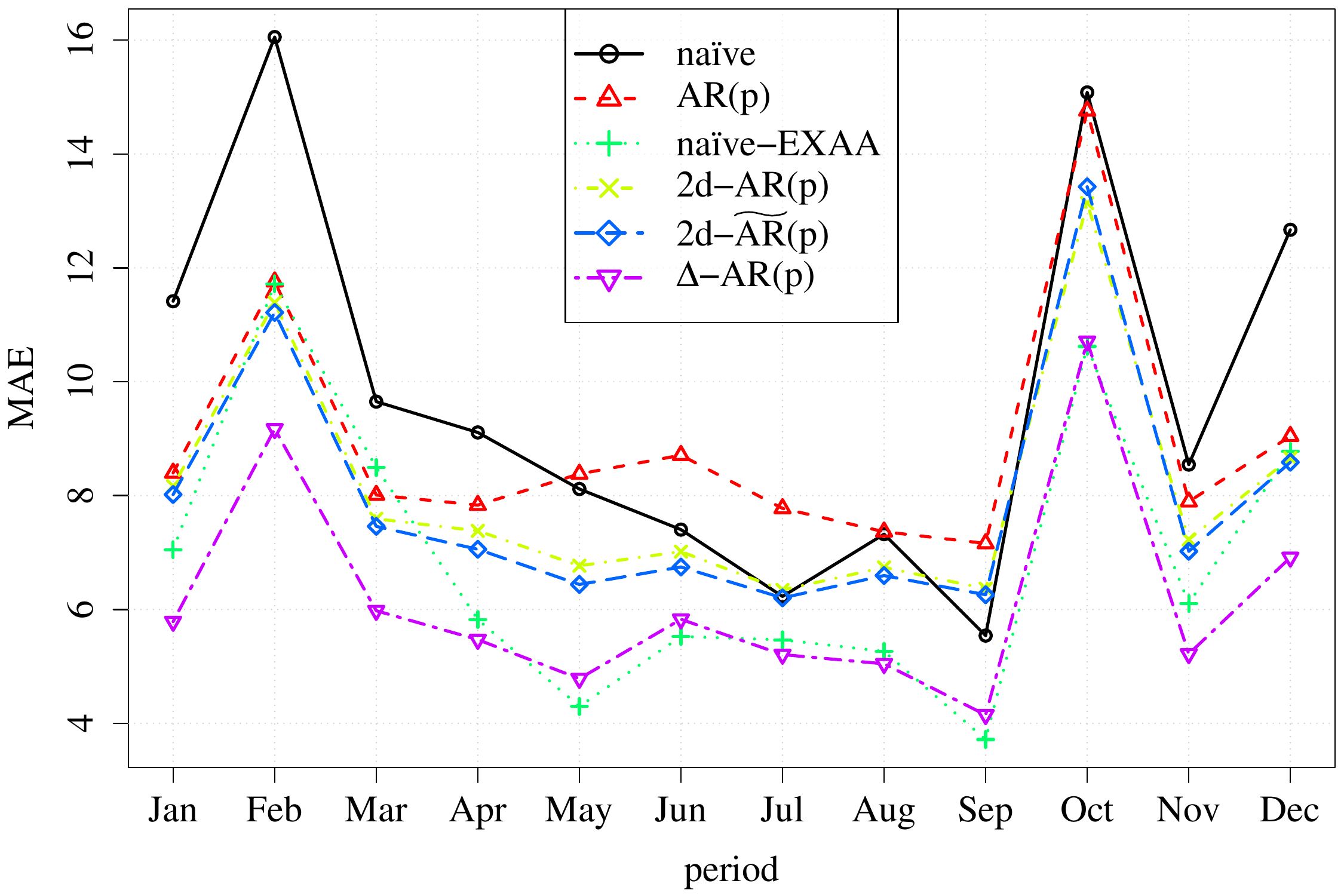} 
  \caption{EPEX FR}
\end{subfigure}
\begin{subfigure}[b]{.49\textwidth}
 \includegraphics[width=1\textwidth, height=.17\textheight]{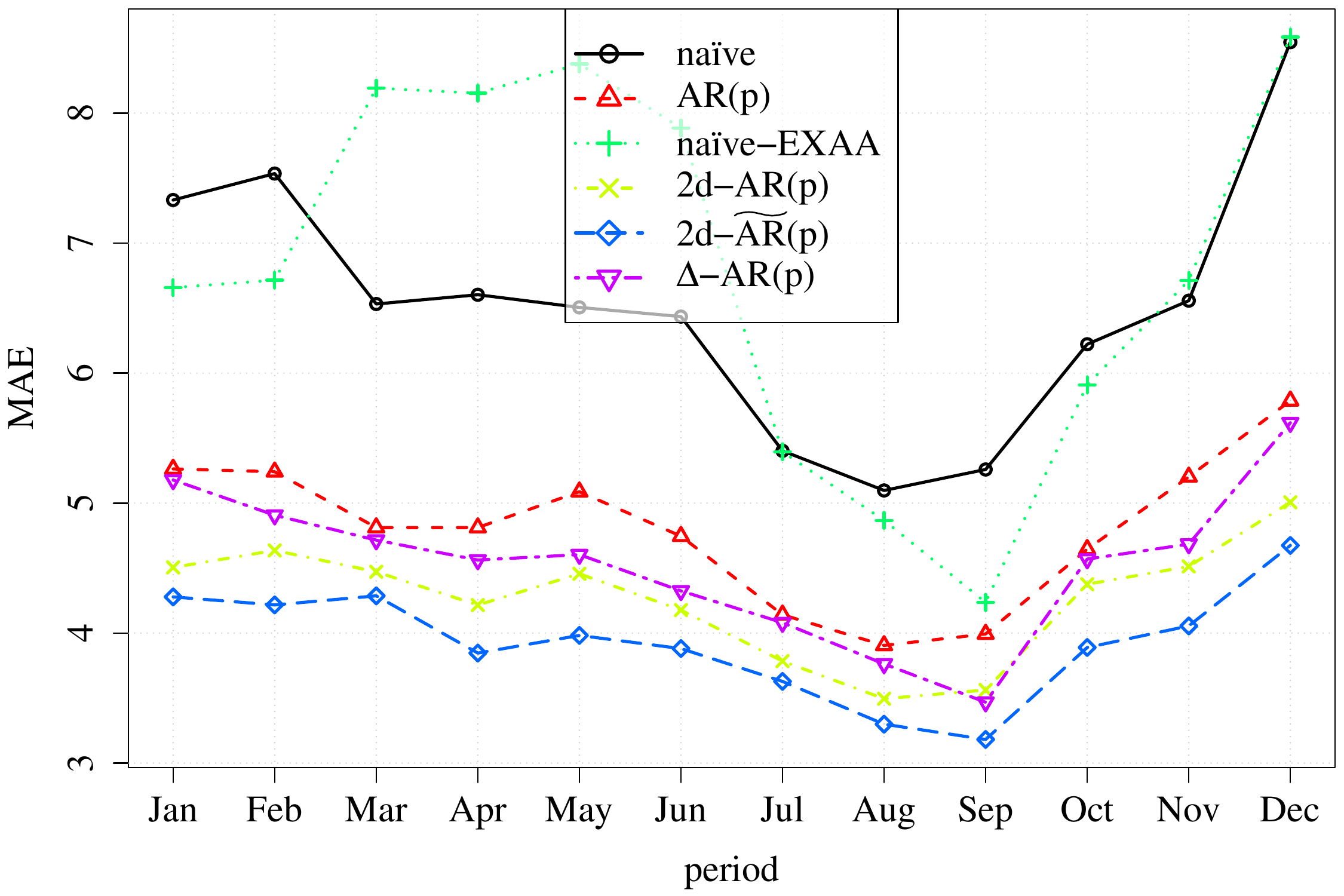} 
  \caption{APX NL}
\end{subfigure}
\begin{subfigure}[b]{.49\textwidth}
 \includegraphics[width=1\textwidth, height=.17\textheight]{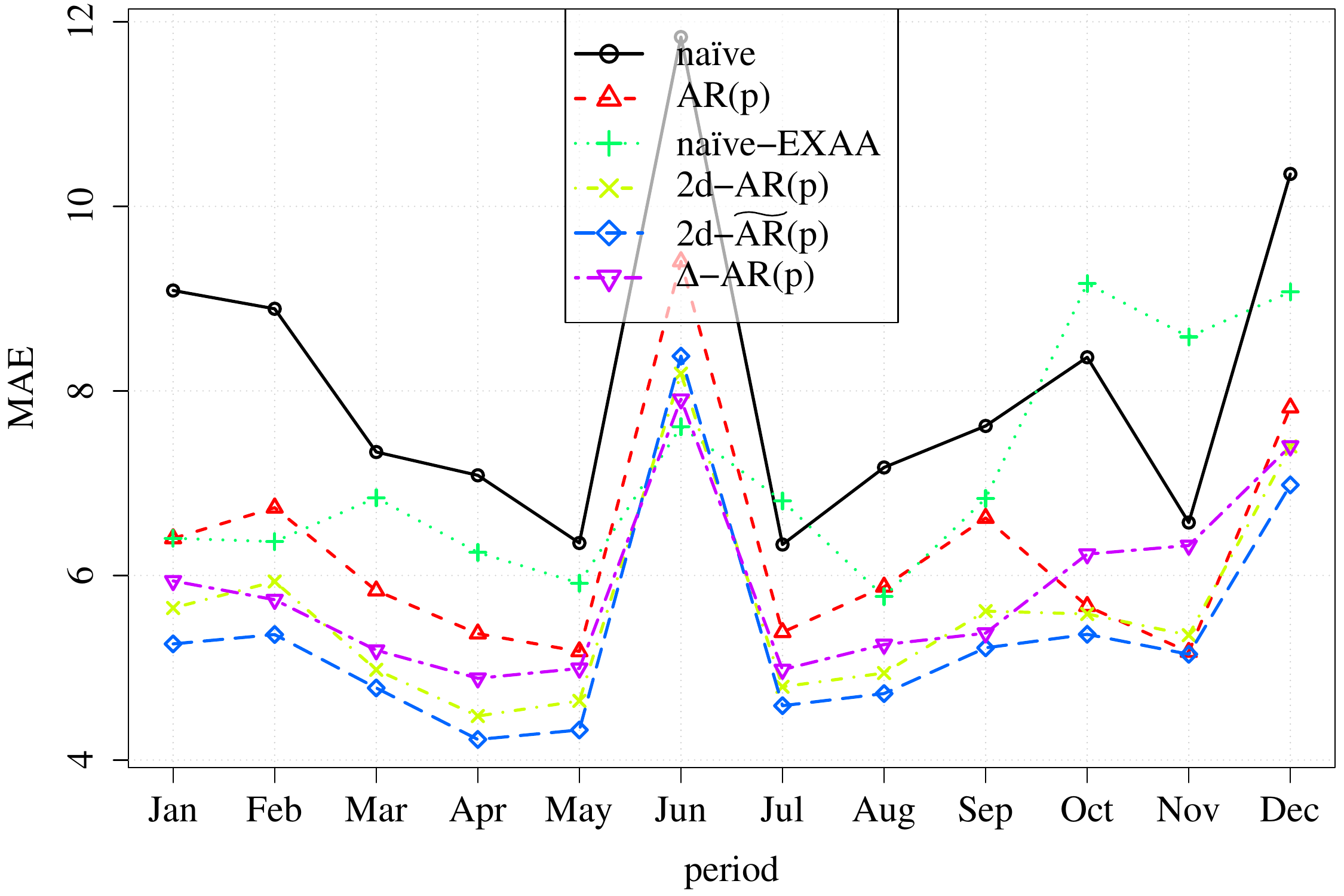} 
  \caption{Nordpool DK West}
\end{subfigure}
\begin{subfigure}[b]{.49\textwidth}
 \includegraphics[width=1\textwidth, height=.17\textheight]{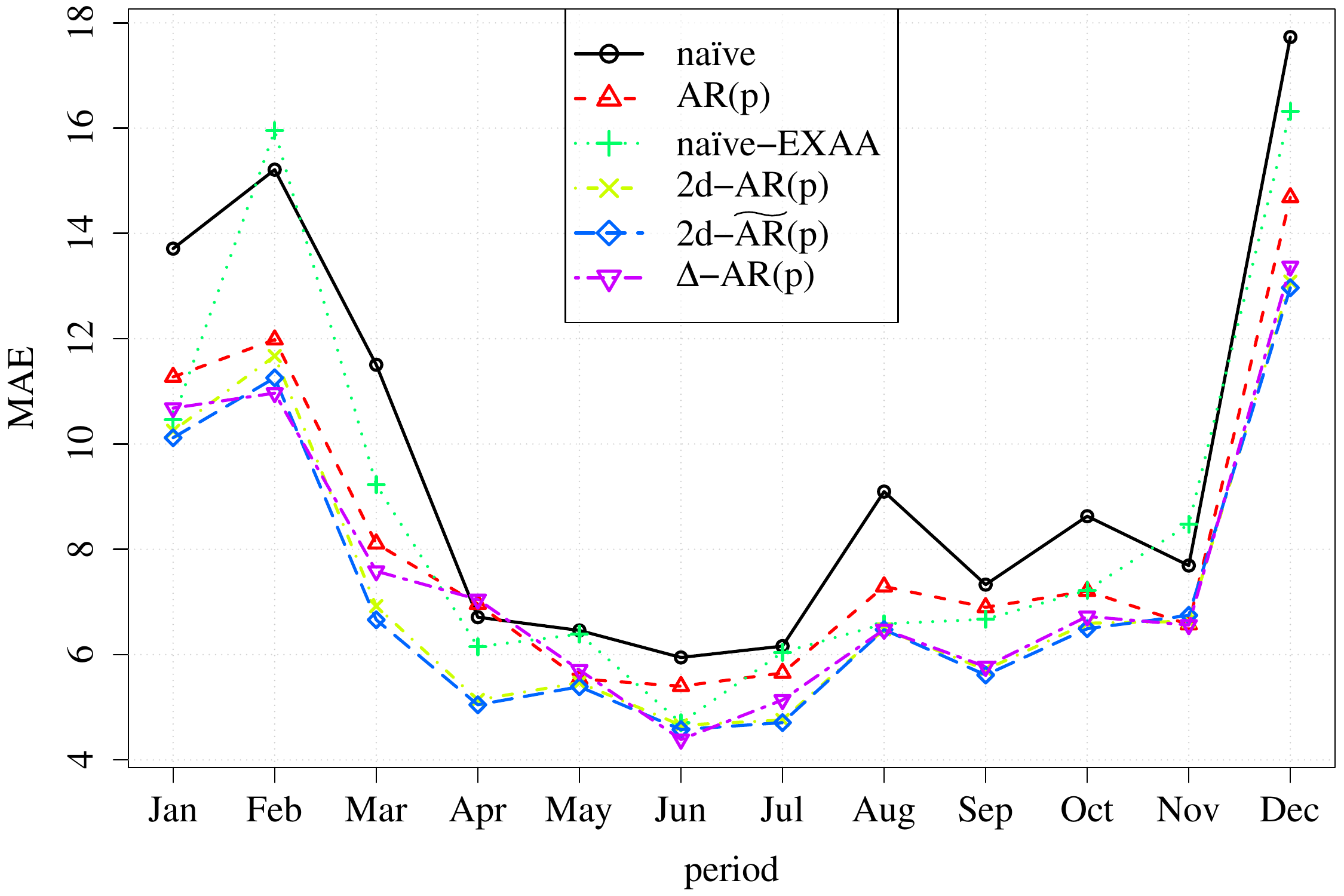} 
  \caption{Nordpool DK East}
\end{subfigure}
\begin{subfigure}[b]{.49\textwidth}
 \includegraphics[width=1\textwidth, height=.17\textheight]{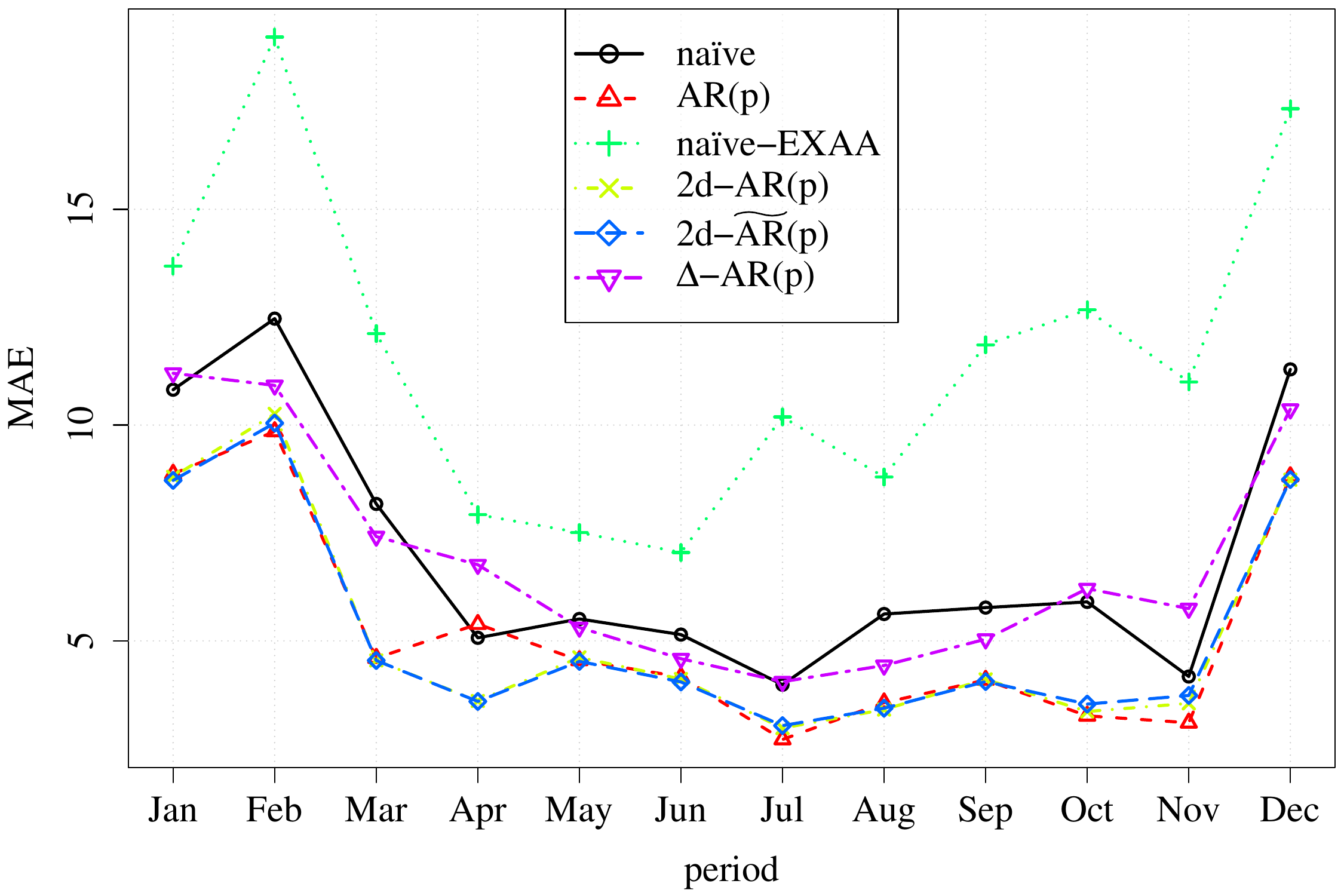} 
  \caption{Nordpool Sweden4}
\end{subfigure}
\begin{subfigure}[b]{.49\textwidth}
 \includegraphics[width=1\textwidth, height=.17\textheight]{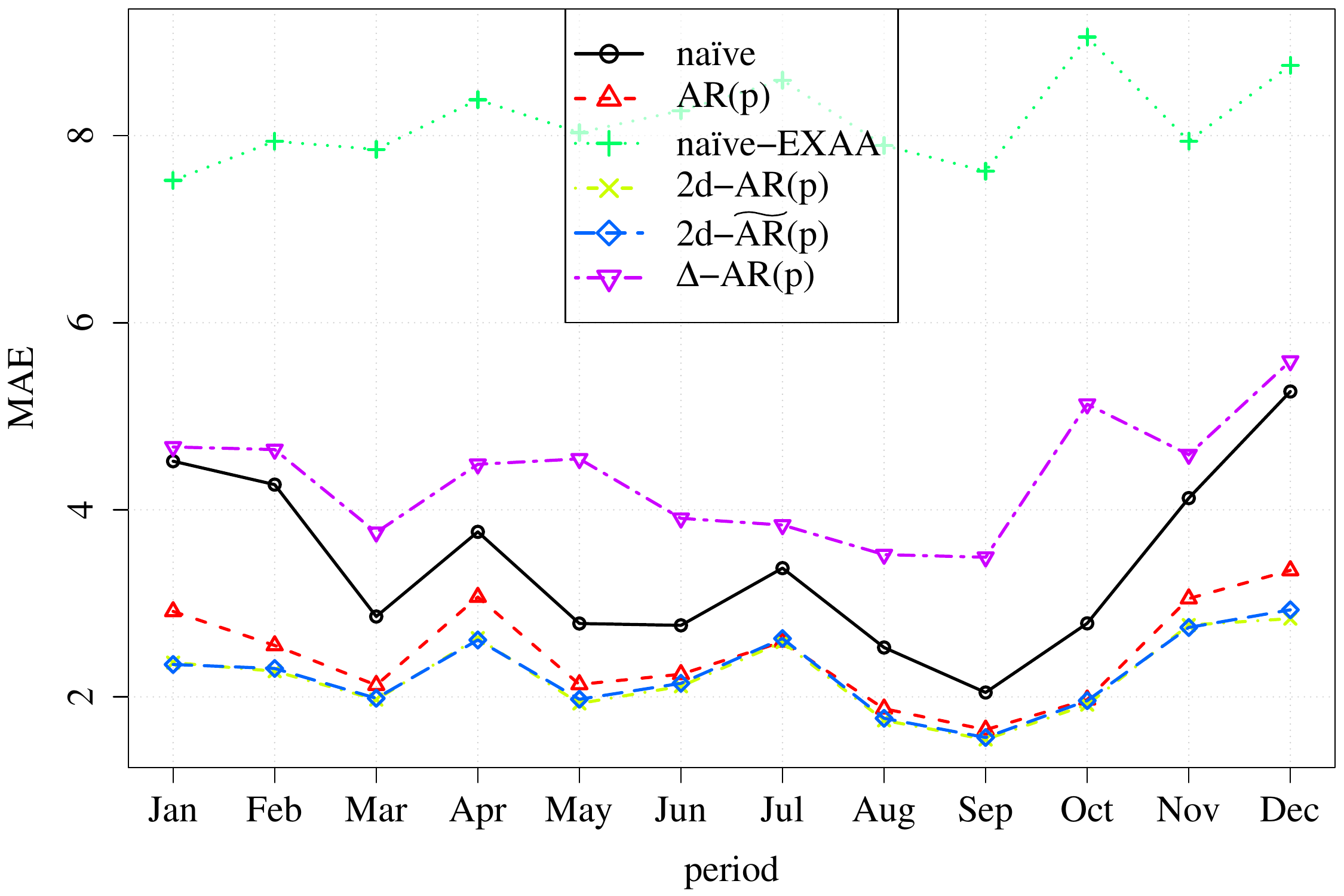} 
  \caption{POLPX PL}
\end{subfigure}
\begin{subfigure}[b]{.49\textwidth}
 \includegraphics[width=1\textwidth, height=.17\textheight]{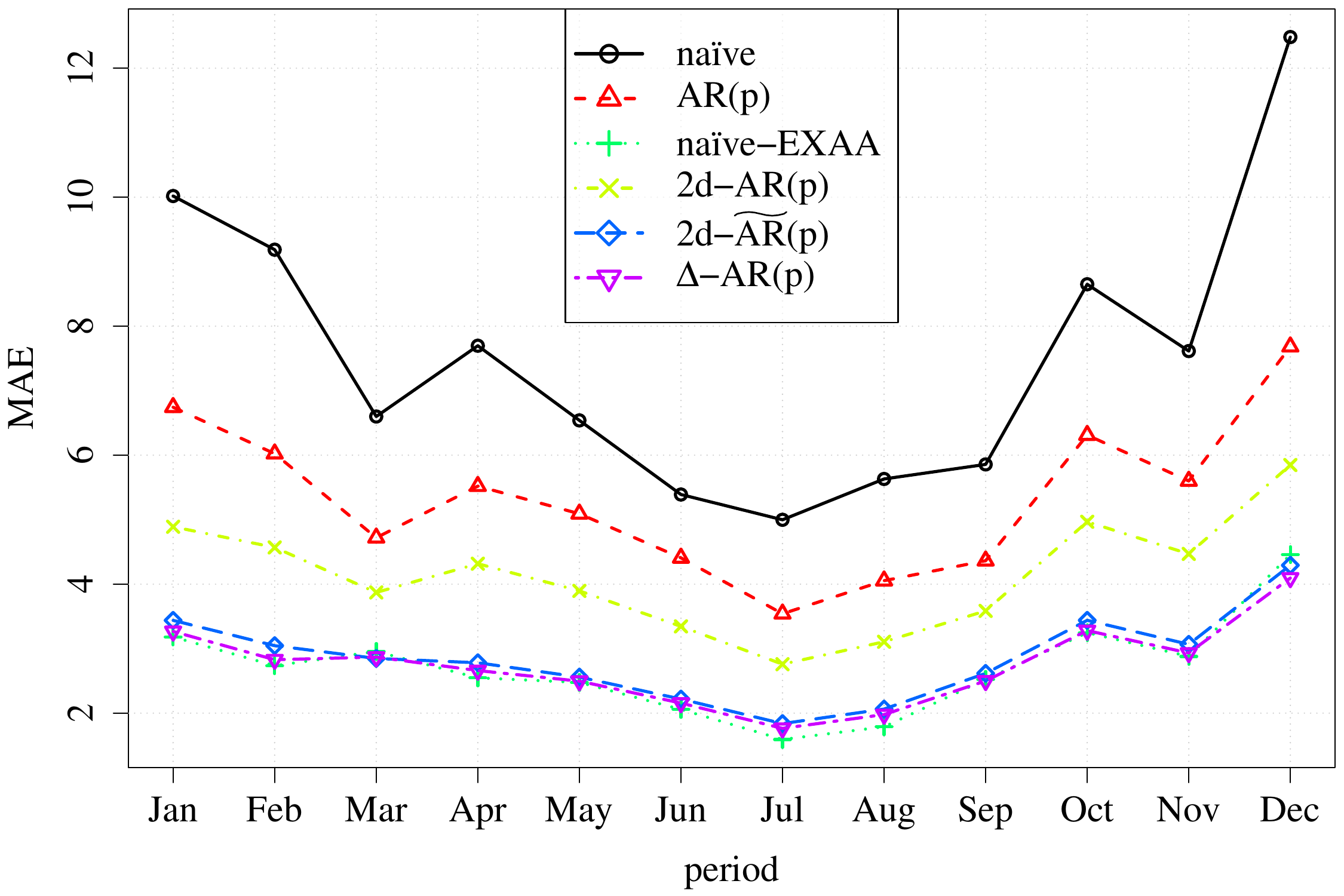} 
  \caption{OTE CZ}
\end{subfigure}
\begin{subfigure}[b]{.49\textwidth}
 \includegraphics[width=1\textwidth, height=.17\textheight]{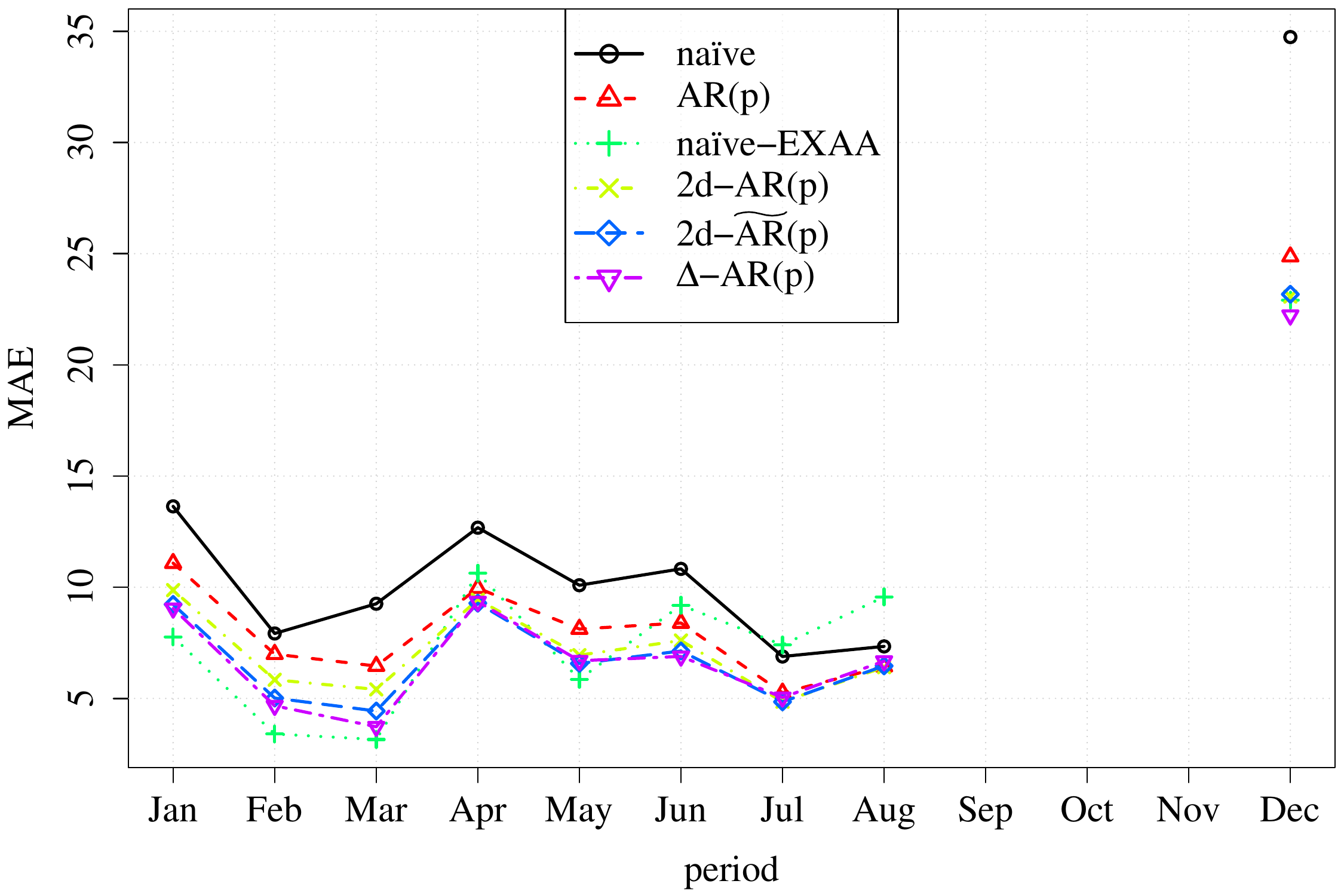} 
  \caption{HUPX HU}
\end{subfigure}
\caption{$\MAE^\XX_{\text{monthly},\T}$ for $\T \in \TT_{\text{monthly}}$ (the 12 months of a year).}
  \label{fig_MAE_mon}
\end{figure}

\begin{figure}[hbt!]
\centering
\begin{subfigure}[b]{.49\textwidth}
 \includegraphics[width=1\textwidth, height=.17\textheight]{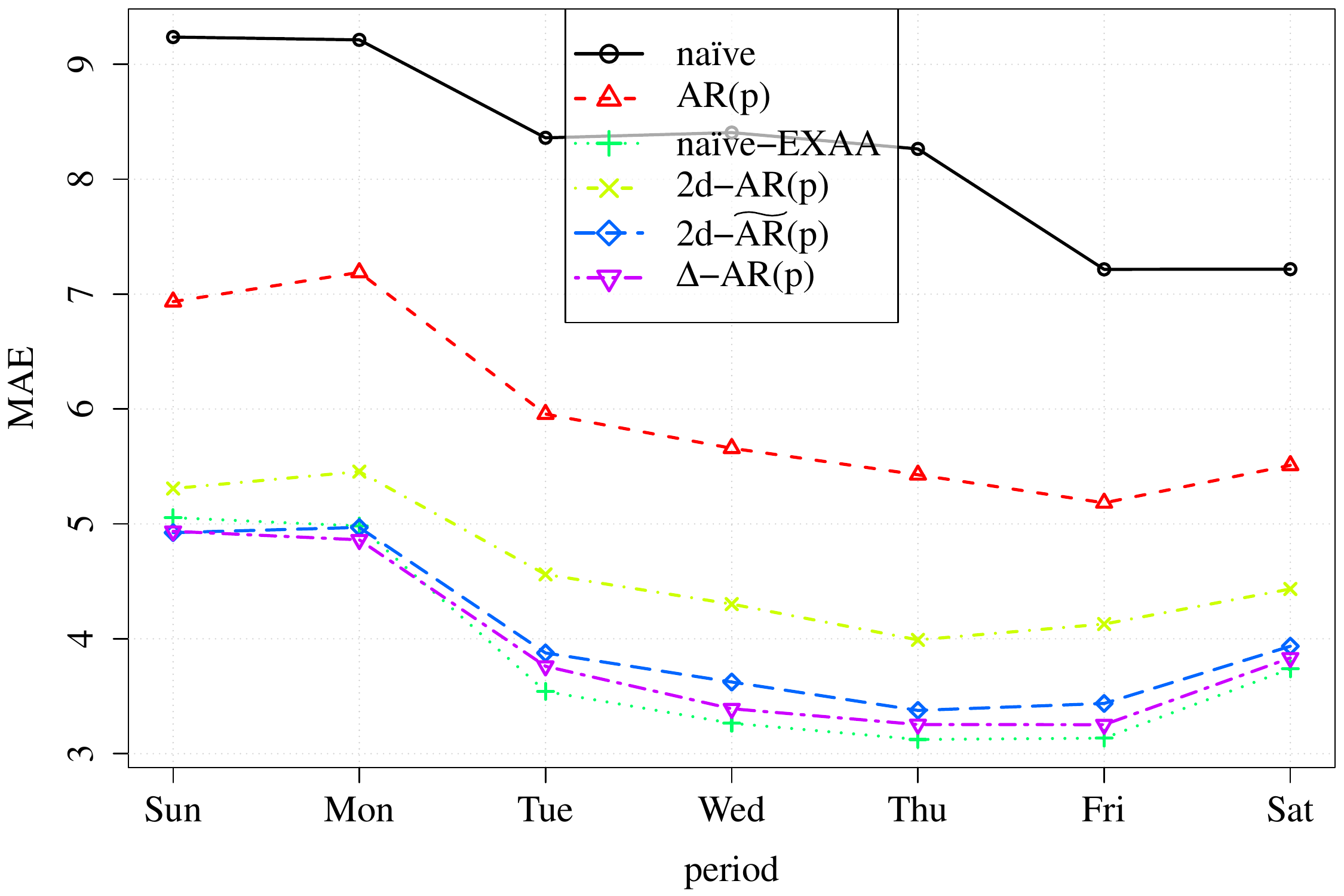} 
  \caption{EPEX DE\&AT}
\end{subfigure}
\begin{subfigure}[b]{.49\textwidth}
 \includegraphics[width=1\textwidth, height=.17\textheight]{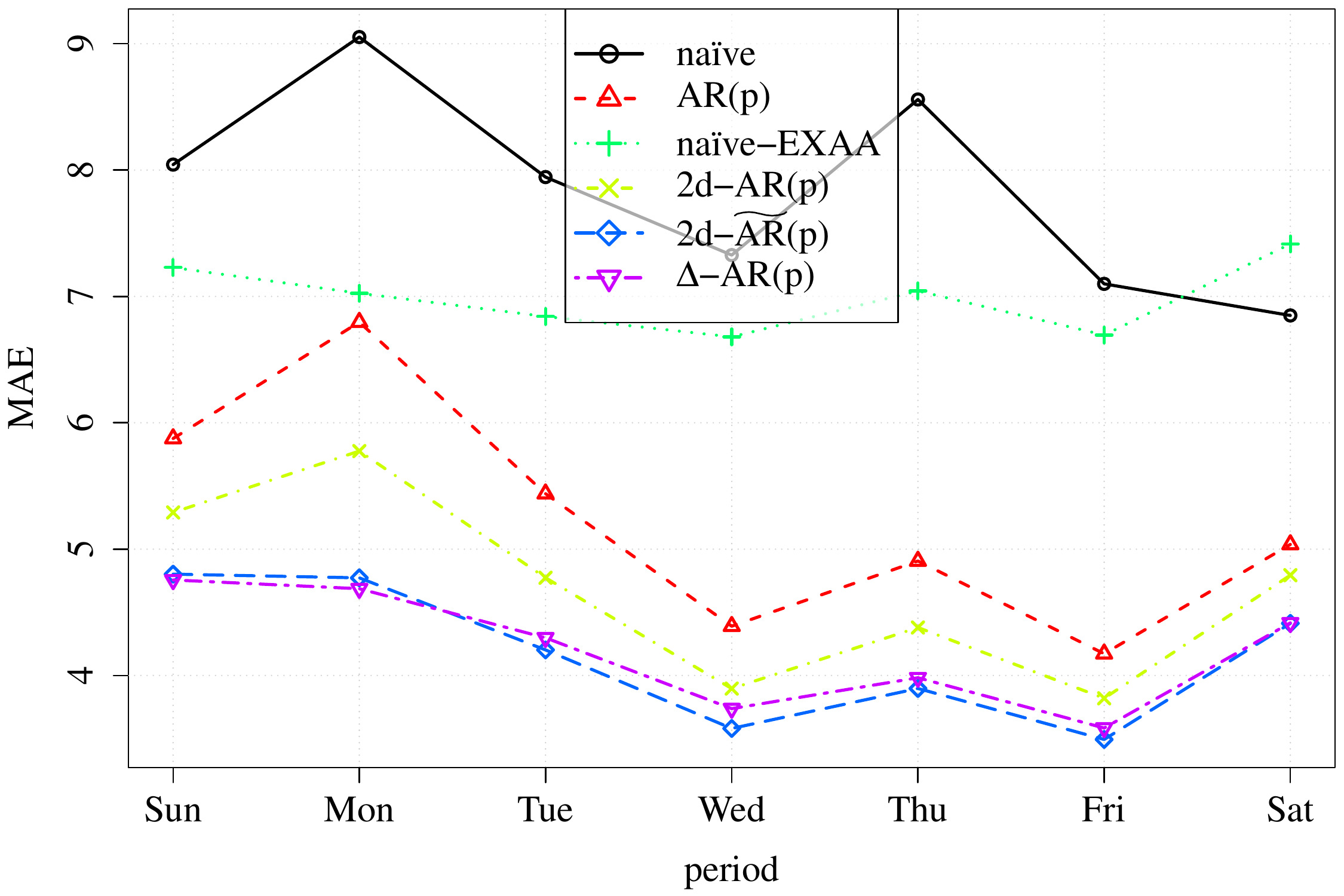} 
  \caption{EPEX CHE}
\end{subfigure}
\begin{subfigure}[b]{.49\textwidth}
 \includegraphics[width=1\textwidth, height=.17\textheight]{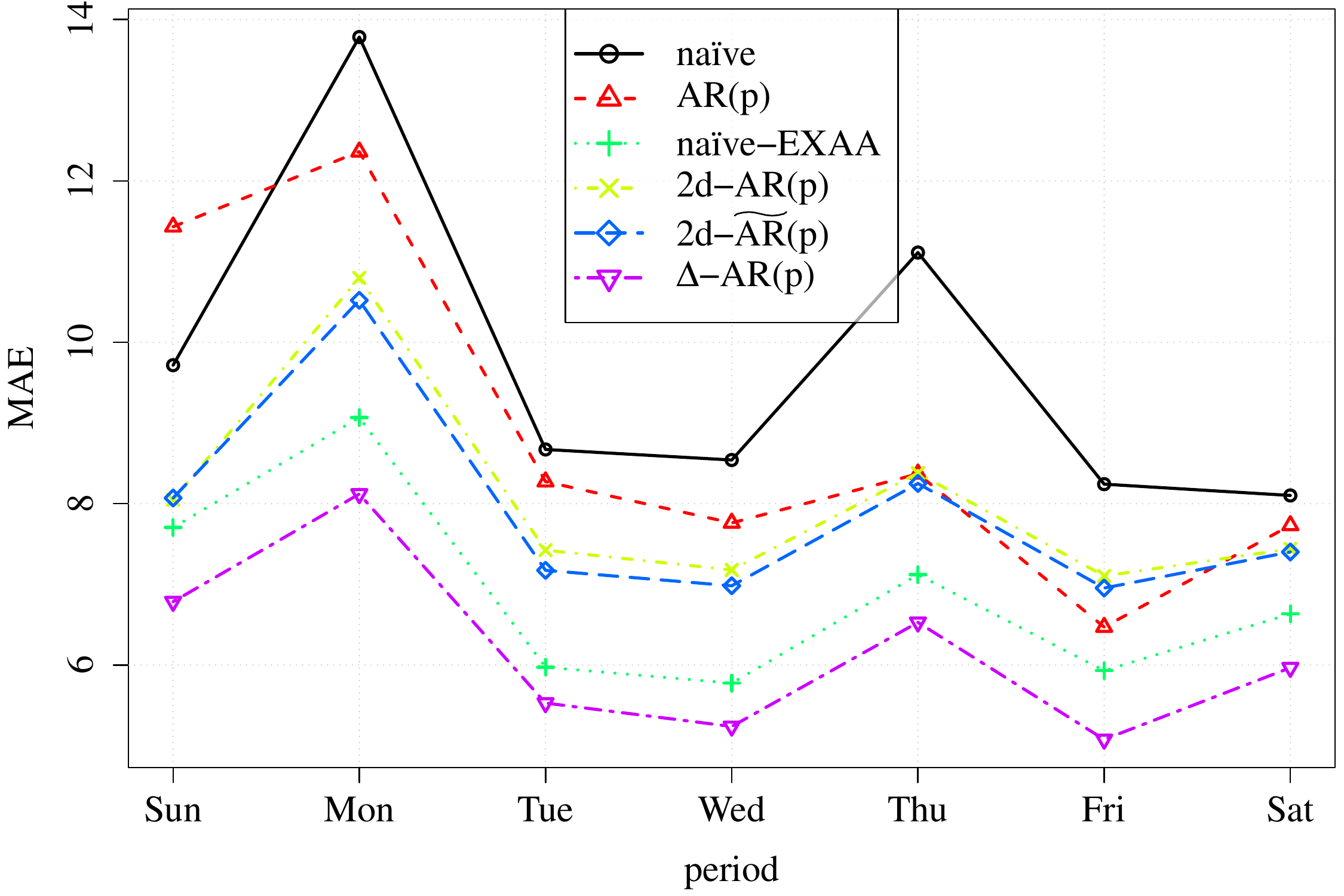} 
  \caption{EPEX FR}
\end{subfigure}
\begin{subfigure}[b]{.49\textwidth}
 \includegraphics[width=1\textwidth, height=.17\textheight]{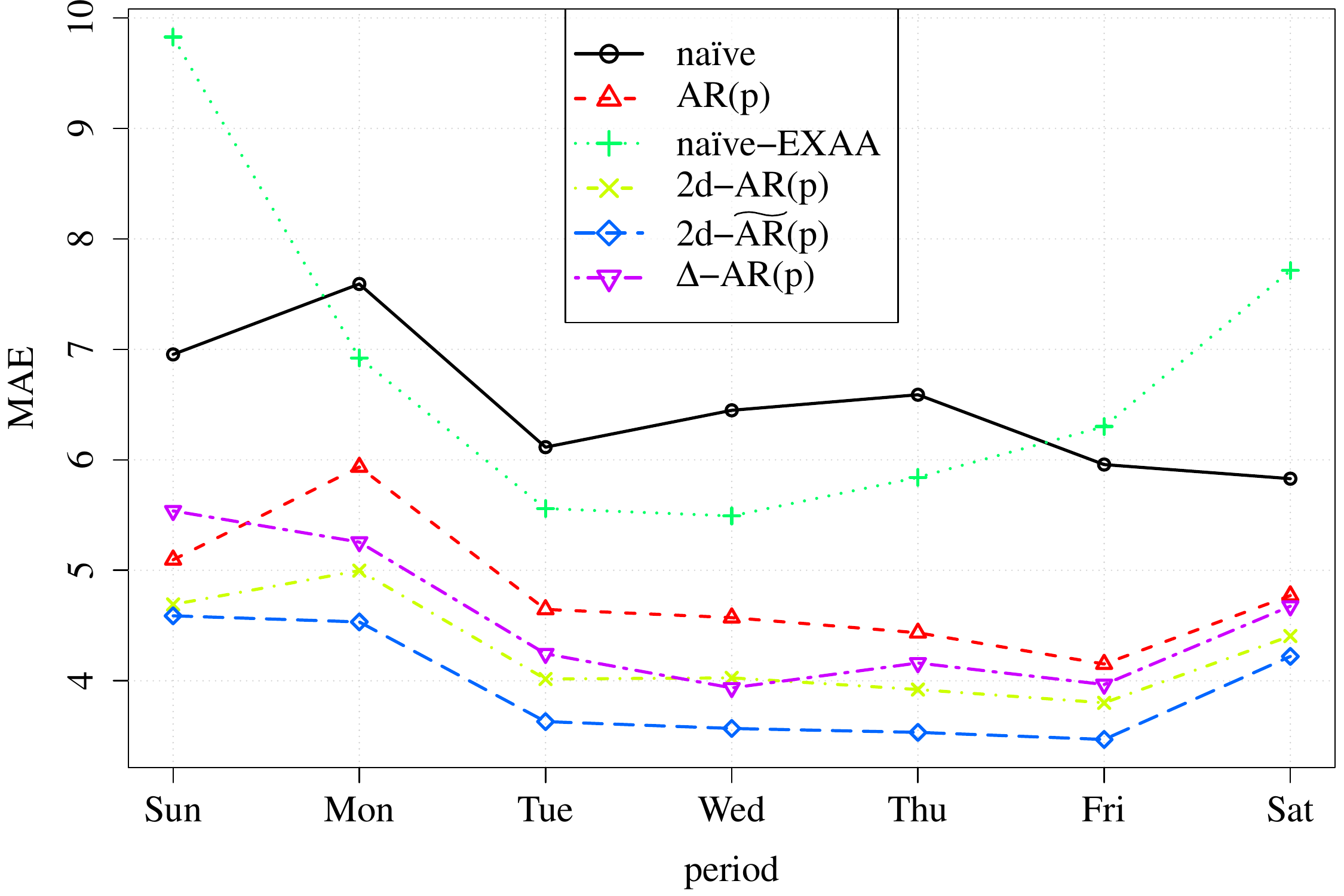} 
  \caption{APX NL}
\end{subfigure}
\begin{subfigure}[b]{.49\textwidth}
 \includegraphics[width=1\textwidth, height=.17\textheight]{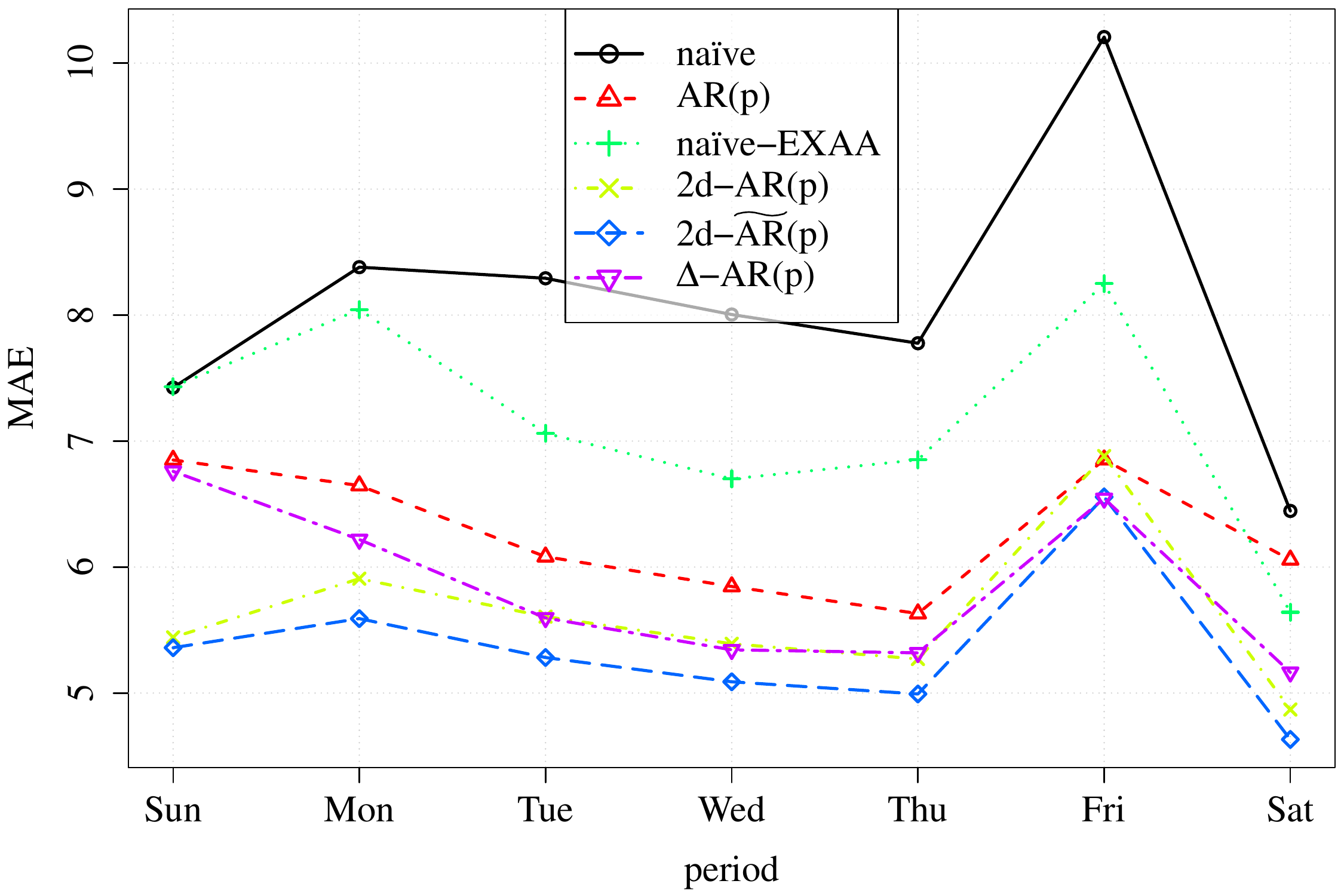} 
  \caption{Nordpool DK West}
\end{subfigure}
\begin{subfigure}[b]{.49\textwidth}
 \includegraphics[width=1\textwidth, height=.17\textheight]{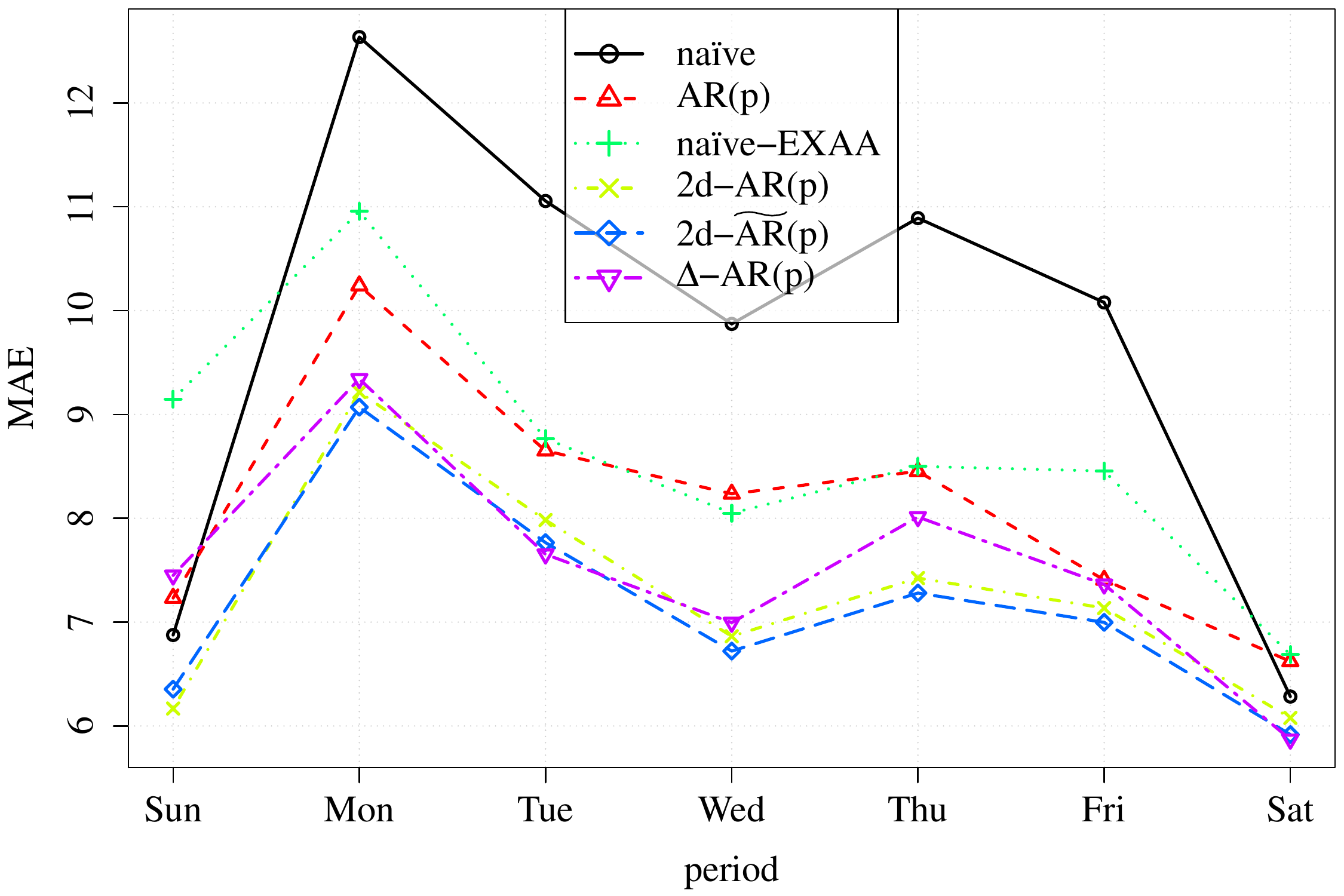} 
  \caption{Nordpool DK East}
\end{subfigure}
\begin{subfigure}[b]{.49\textwidth}
 \includegraphics[width=1\textwidth, height=.17\textheight]{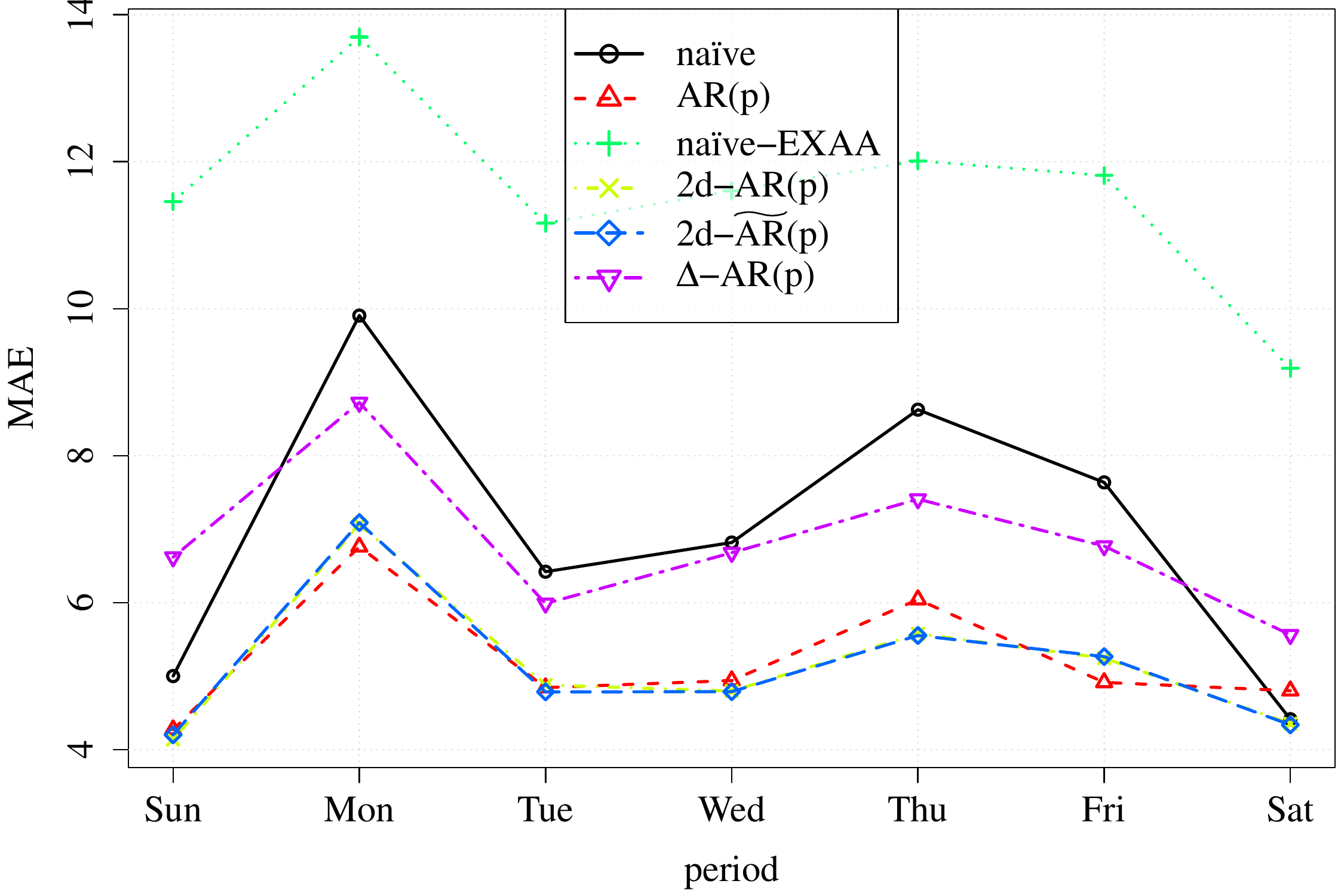} 
  \caption{Nordpool Sweden4}
\end{subfigure}
\begin{subfigure}[b]{.49\textwidth}
 \includegraphics[width=1\textwidth, height=.17\textheight]{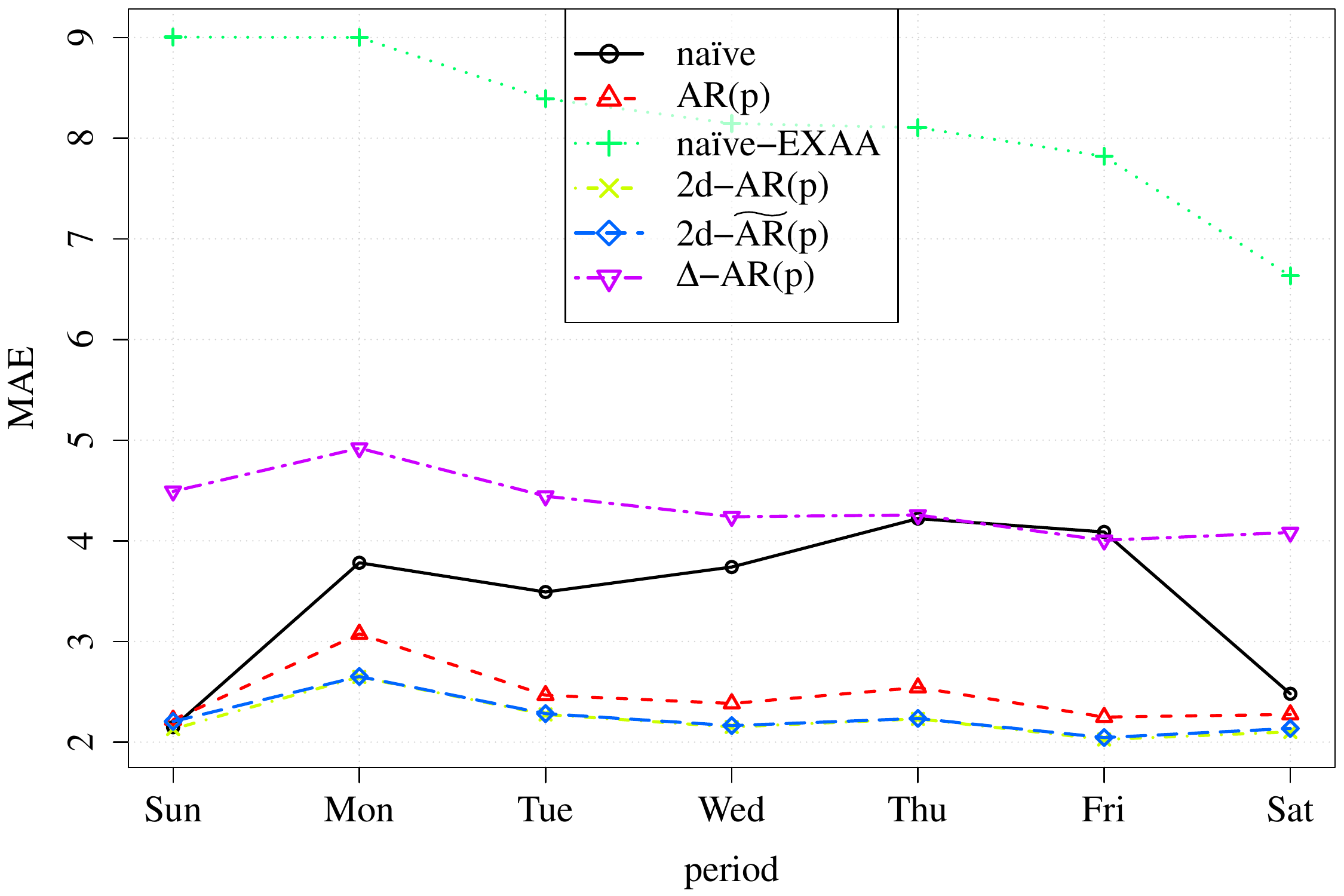} 
  \caption{POLPX PL}
\end{subfigure}
\begin{subfigure}[b]{.49\textwidth}
 \includegraphics[width=1\textwidth, height=.17\textheight]{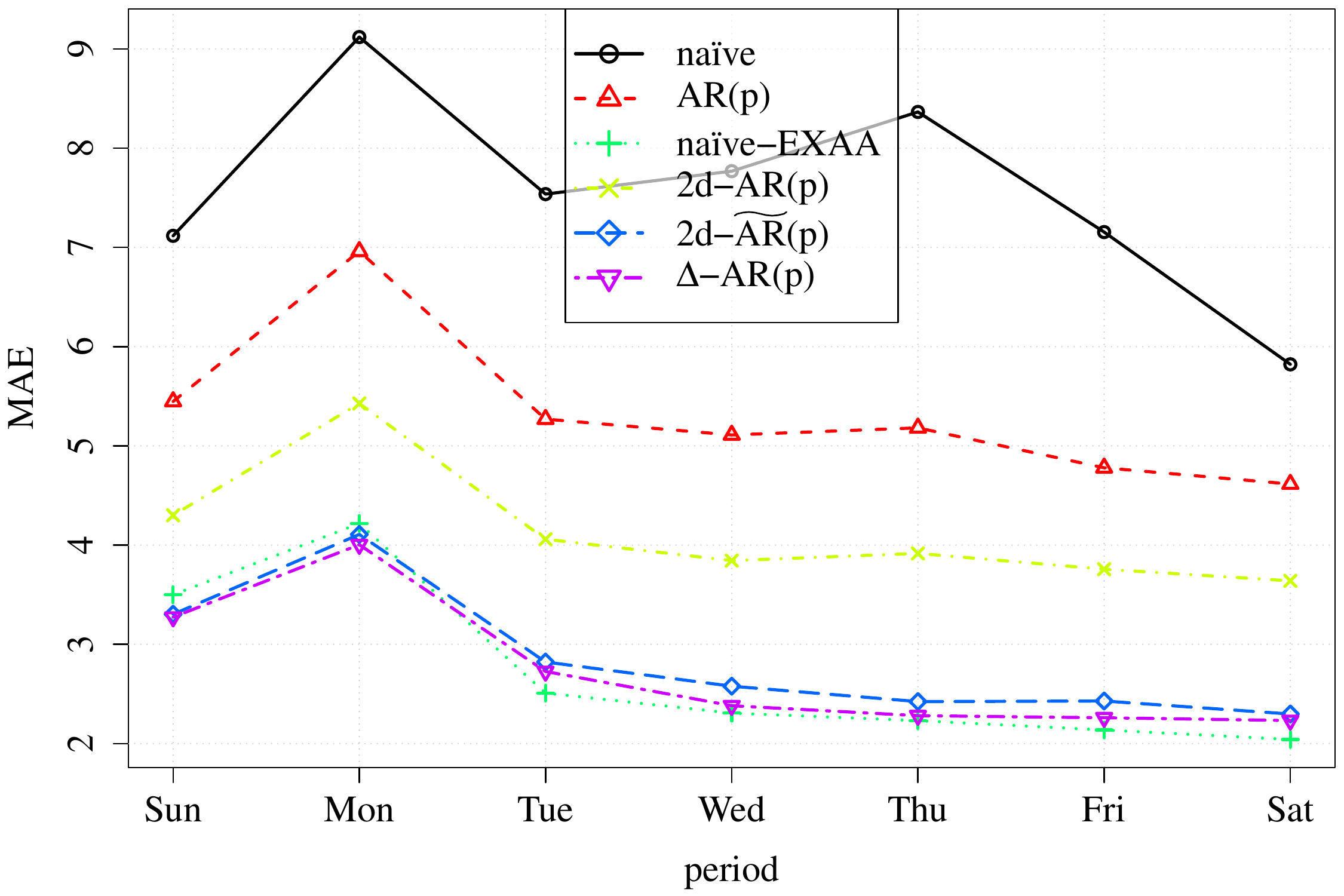} 
  \caption{OTE CZ}
\end{subfigure}
\begin{subfigure}[b]{.49\textwidth}
 \includegraphics[width=1\textwidth, height=.17\textheight]{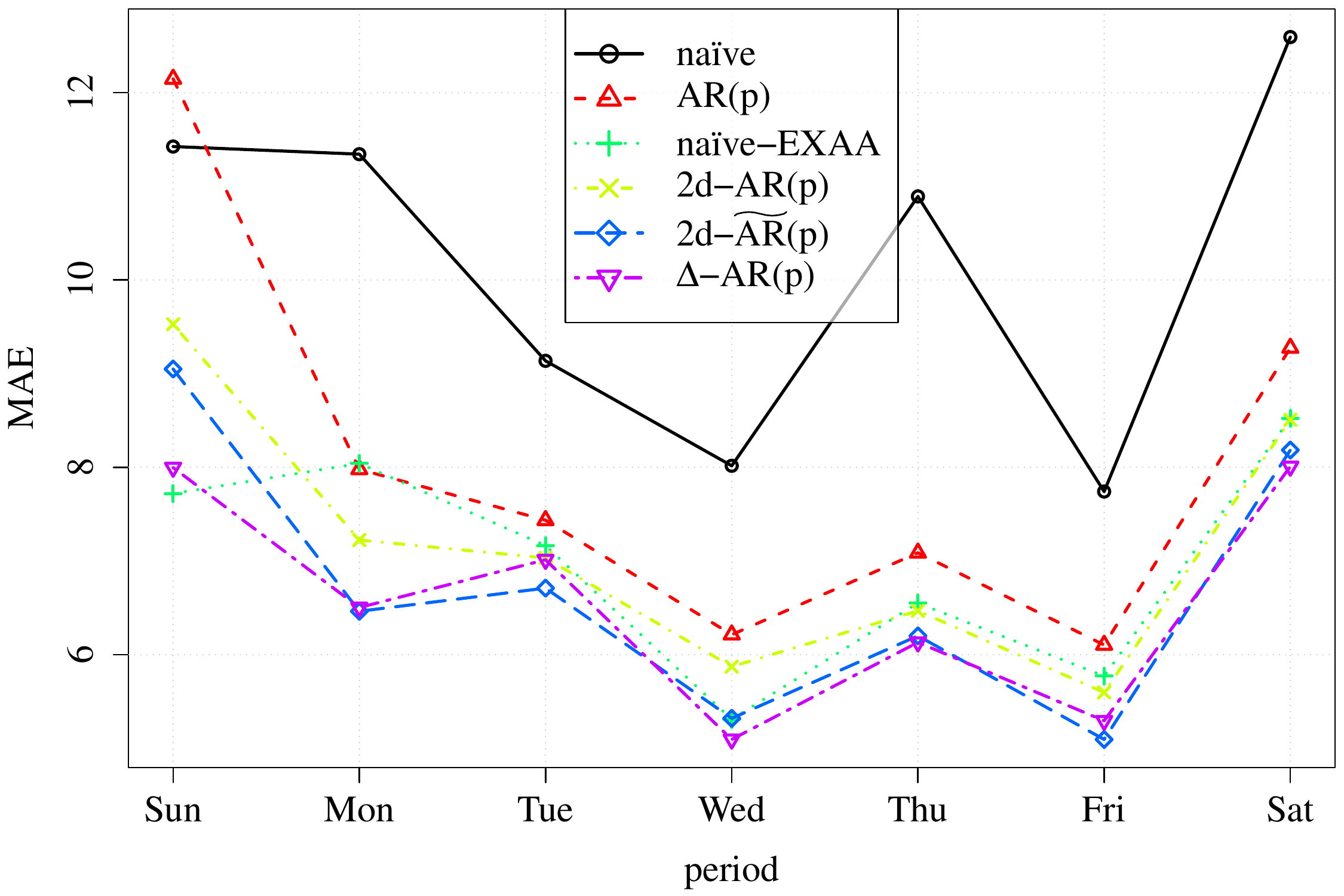} 
  \caption{HUPX HU}
\end{subfigure}
\caption{$\MAE^\XX_{\text{daily},\T}$ for $\T \in \TT_{\text{daily}}$ (the 7 days of a week).}
 \label{fig_MAE_week}
\end{figure}

\begin{figure}[hbt!]
\centering
\begin{subfigure}[b]{.49\textwidth}
 \includegraphics[width=1\textwidth, height=.17\textheight]{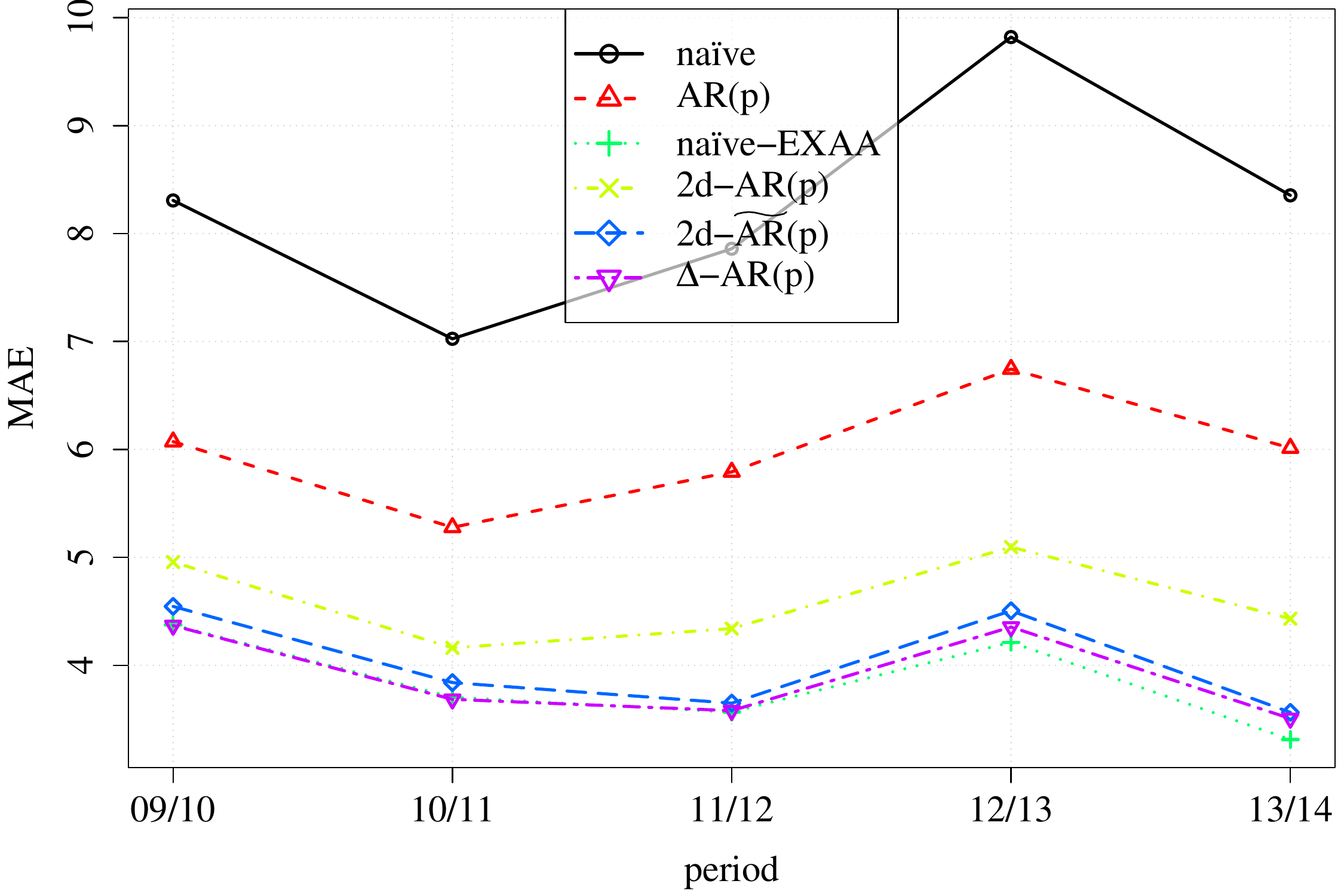} 
  \caption{EPEX DE\&AT}
\end{subfigure}
\begin{subfigure}[b]{.49\textwidth}
 \includegraphics[width=1\textwidth, height=.17\textheight]{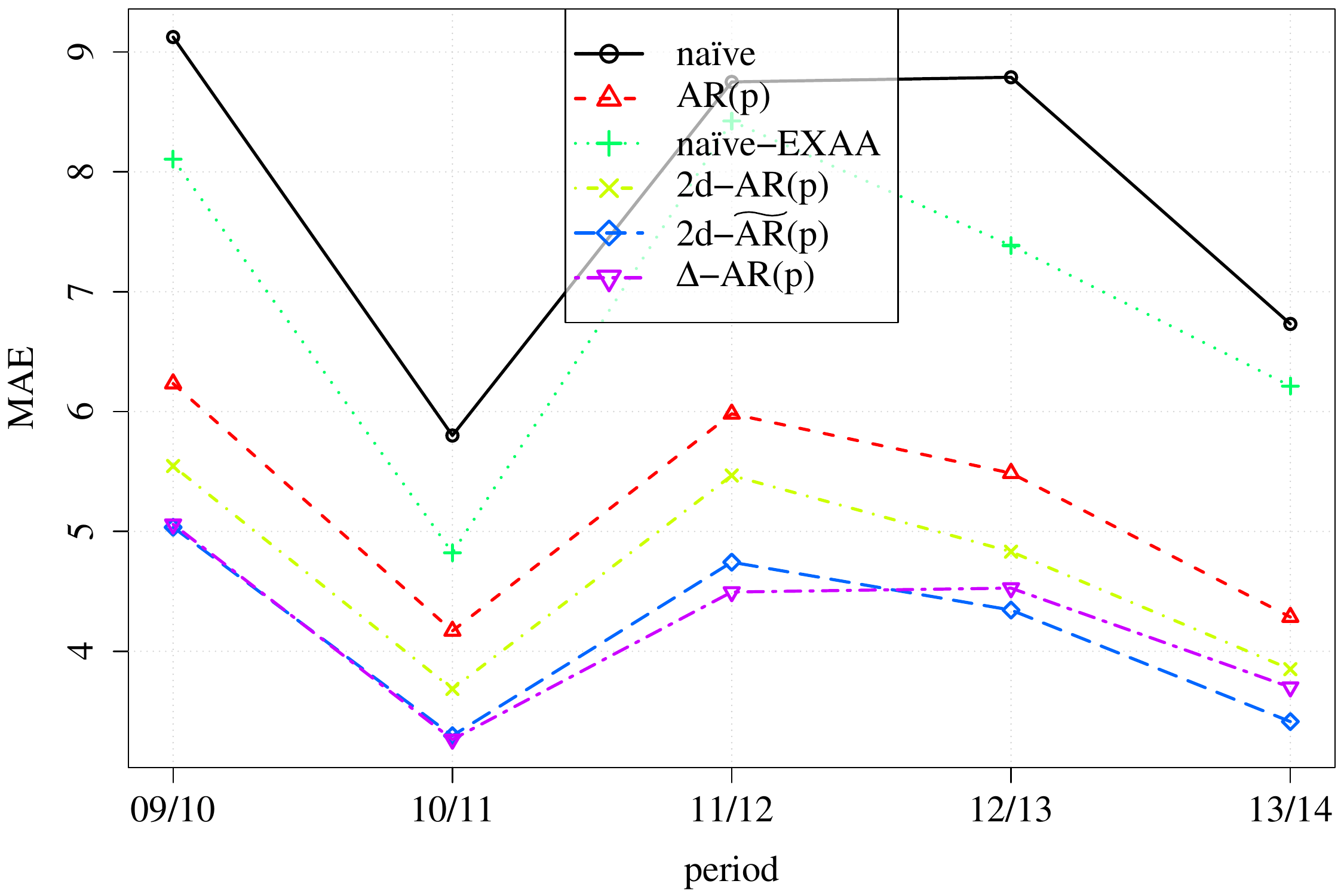} 
  \caption{EPEX CHE}
\end{subfigure}
\begin{subfigure}[b]{.49\textwidth}
 \includegraphics[width=1\textwidth, height=.17\textheight]{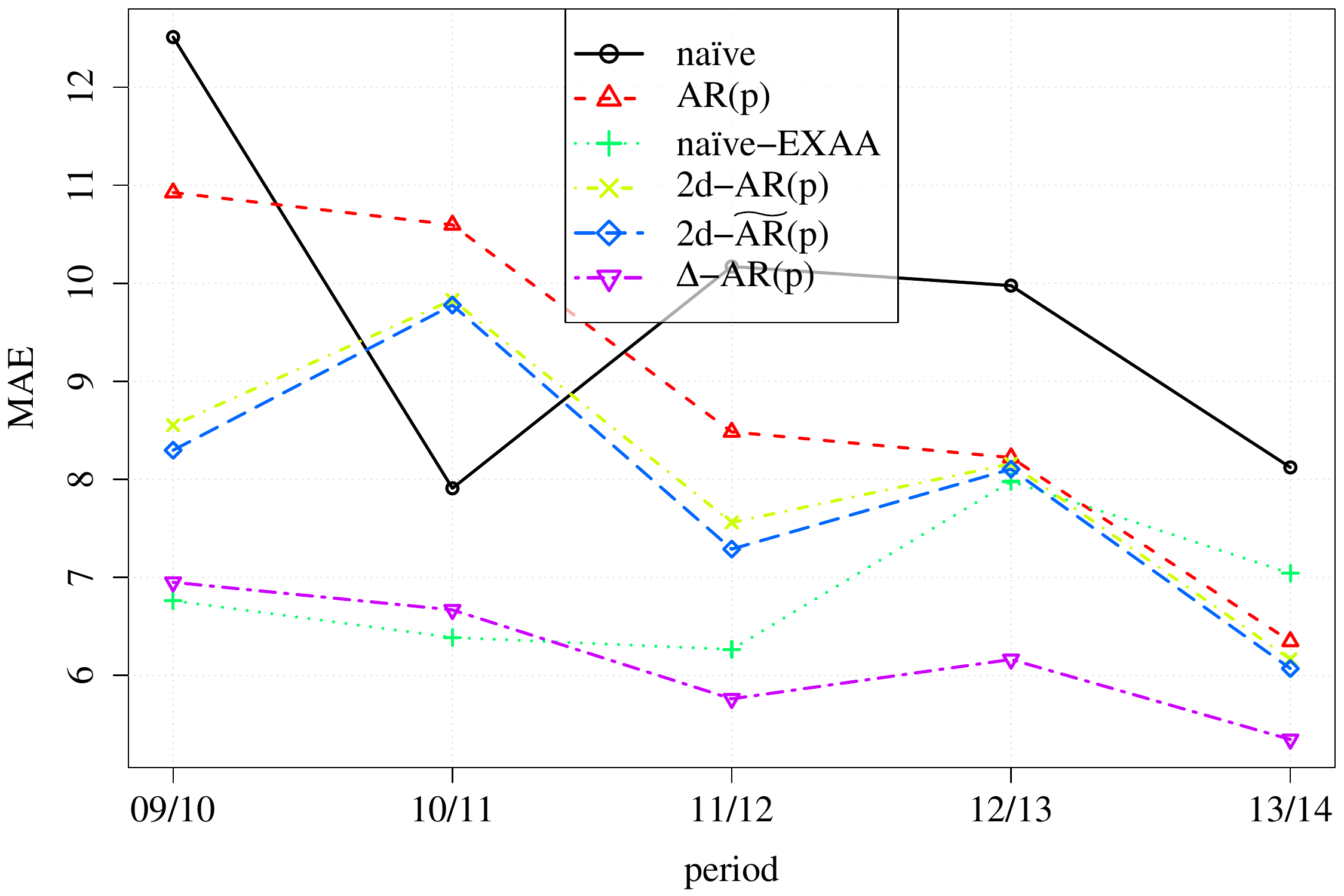} 
  \caption{EPEX FR}
\end{subfigure}
\begin{subfigure}[b]{.49\textwidth}
 \includegraphics[width=1\textwidth, height=.17\textheight]{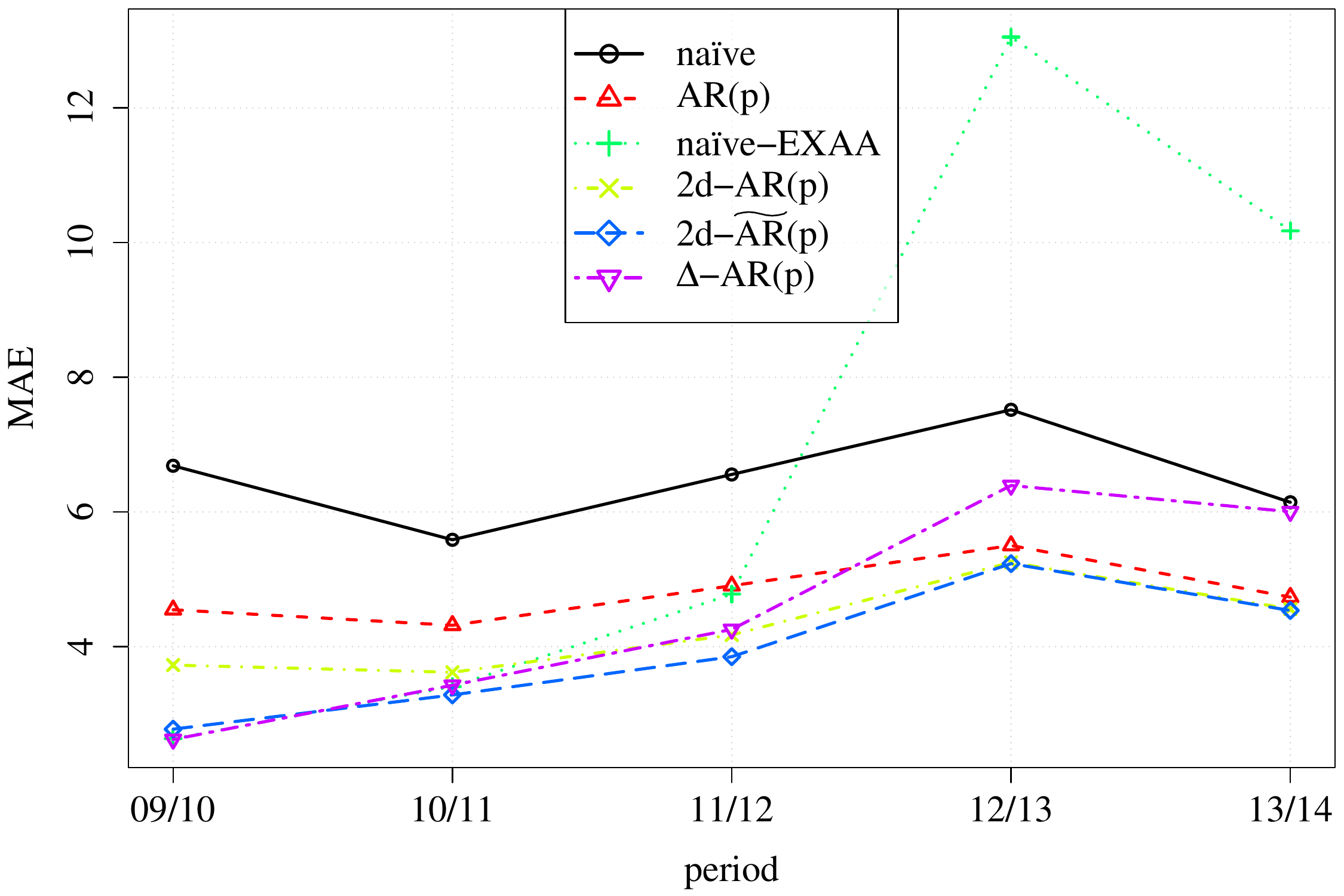} 
  \caption{APX NL}
\end{subfigure}
\begin{subfigure}[b]{.49\textwidth}
 \includegraphics[width=1\textwidth, height=.17\textheight]{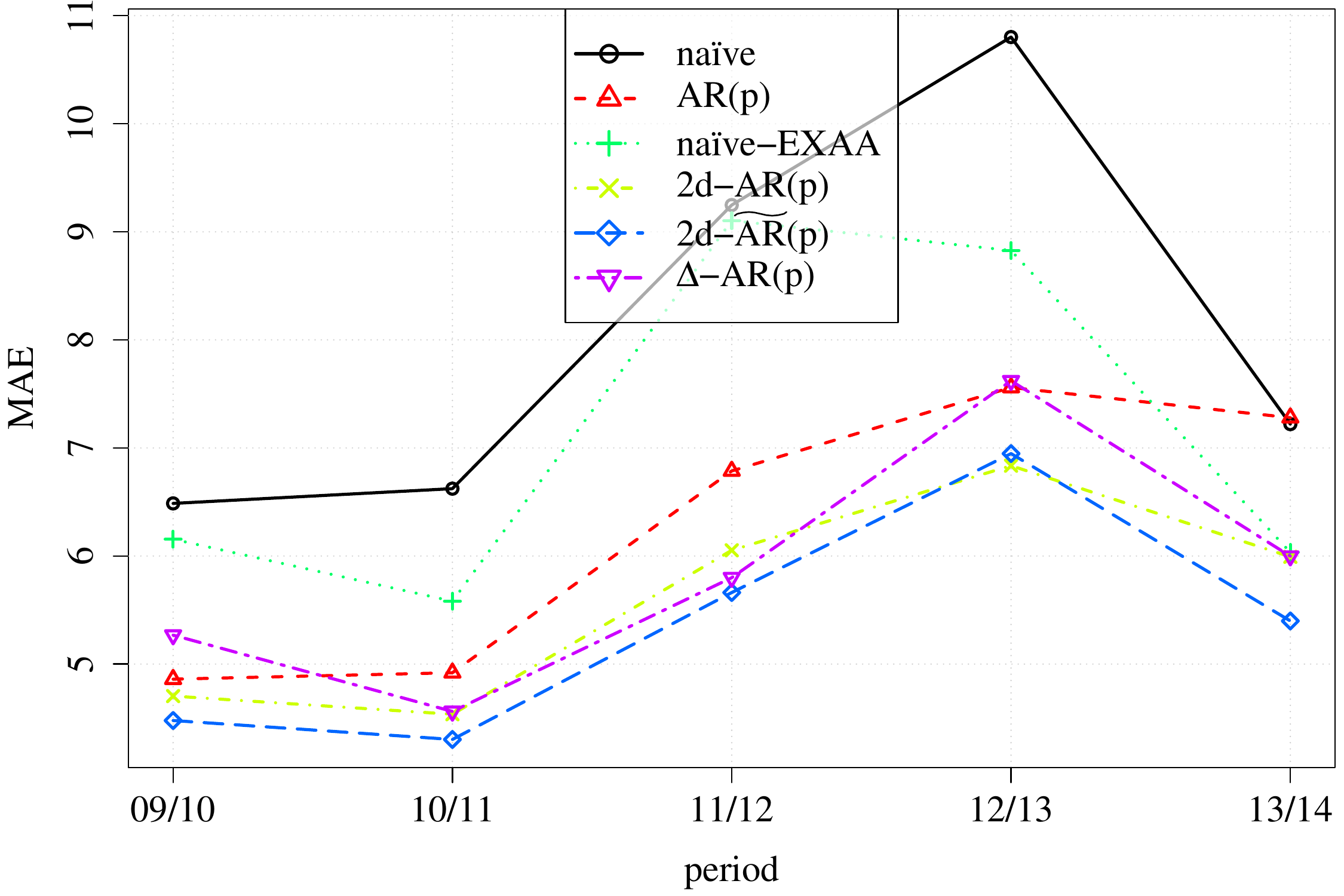} 
  \caption{Nordpool DK West}
\end{subfigure}
\begin{subfigure}[b]{.49\textwidth}
 \includegraphics[width=1\textwidth, height=.17\textheight]{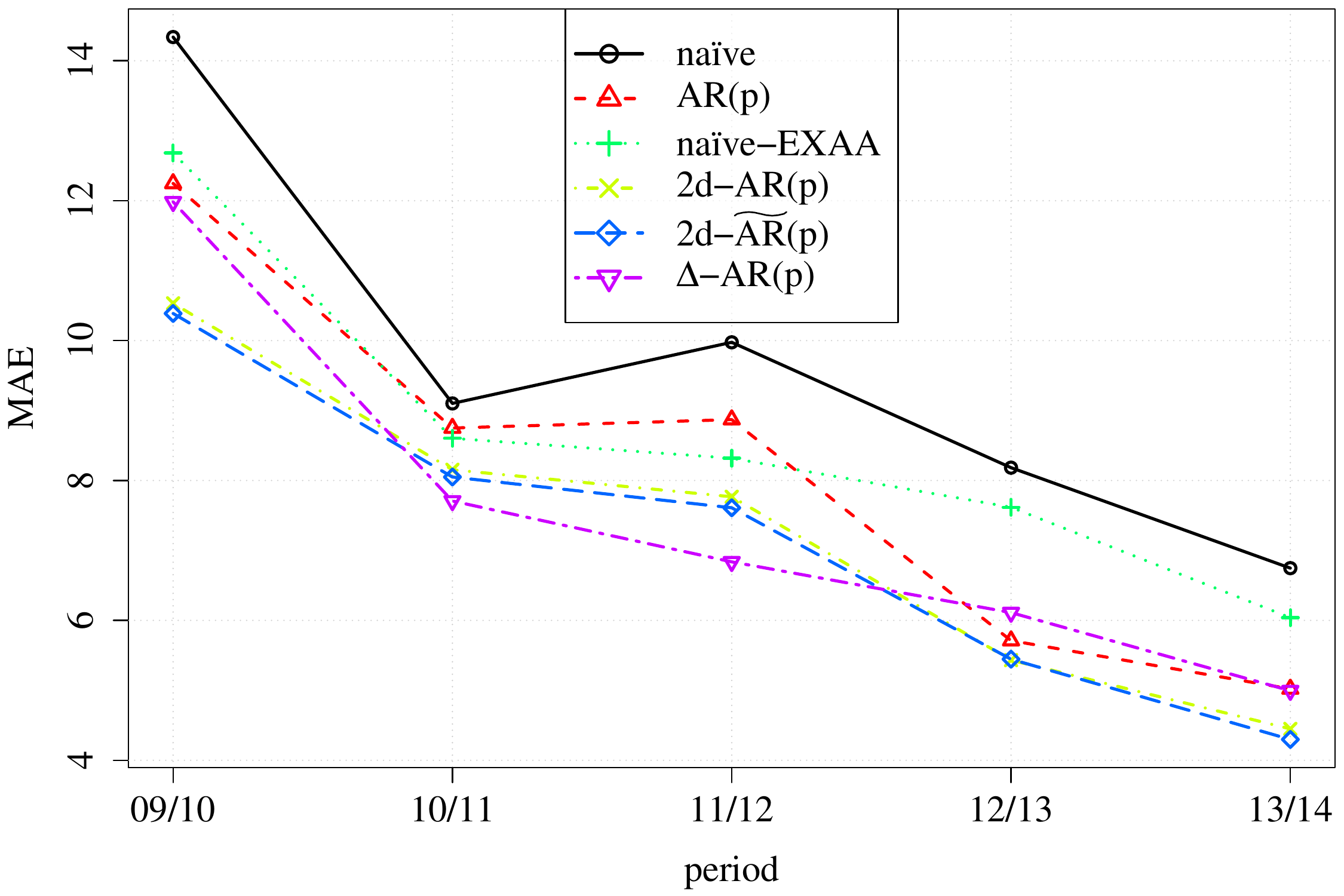} 
  \caption{Nordpool DK East}
\end{subfigure}
\begin{subfigure}[b]{.49\textwidth}
 \includegraphics[width=1\textwidth, height=.17\textheight]{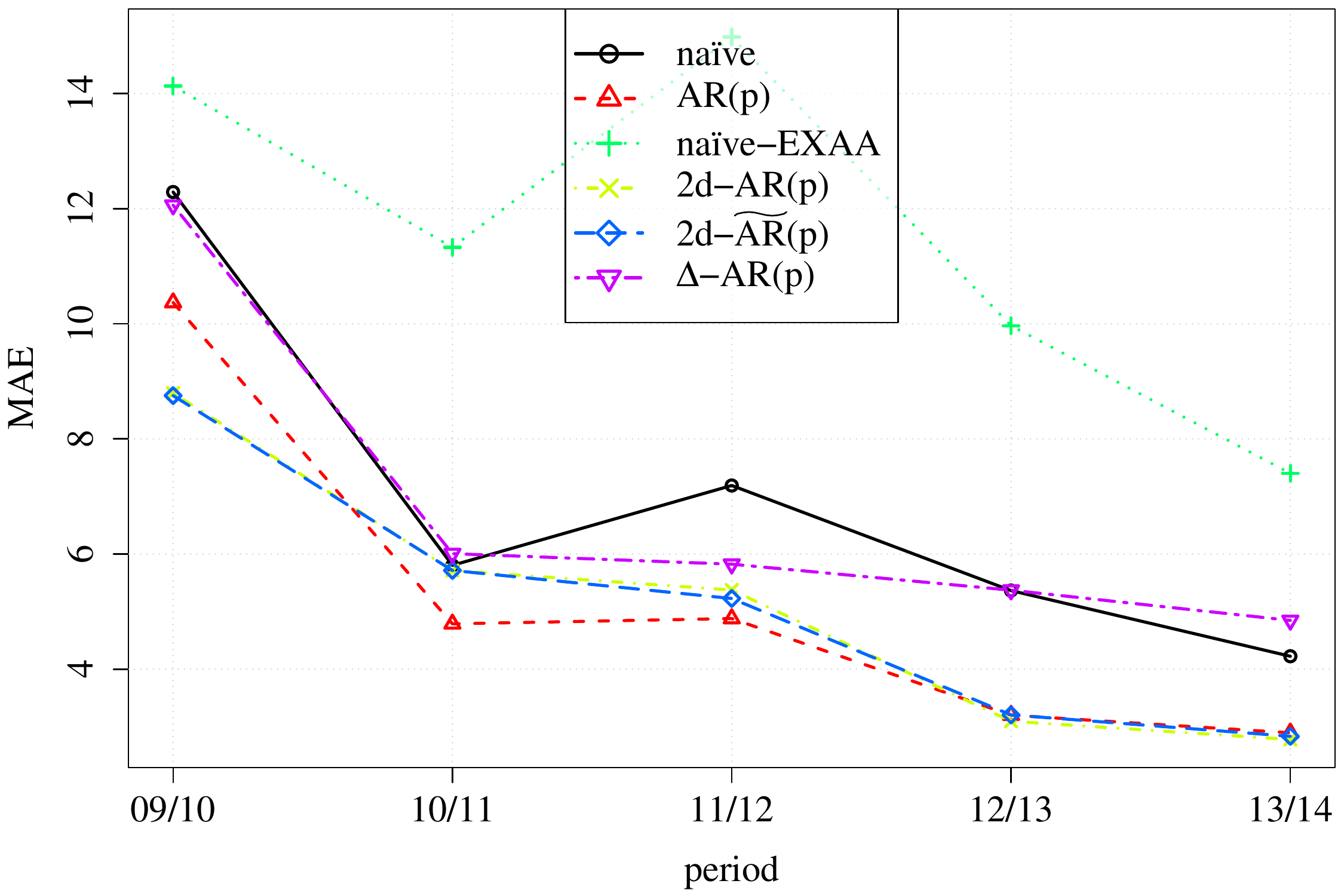} 
  \caption{Nordpool Sweden4}
\end{subfigure}
\begin{subfigure}[b]{.49\textwidth}
 \includegraphics[width=1\textwidth, height=.17\textheight]{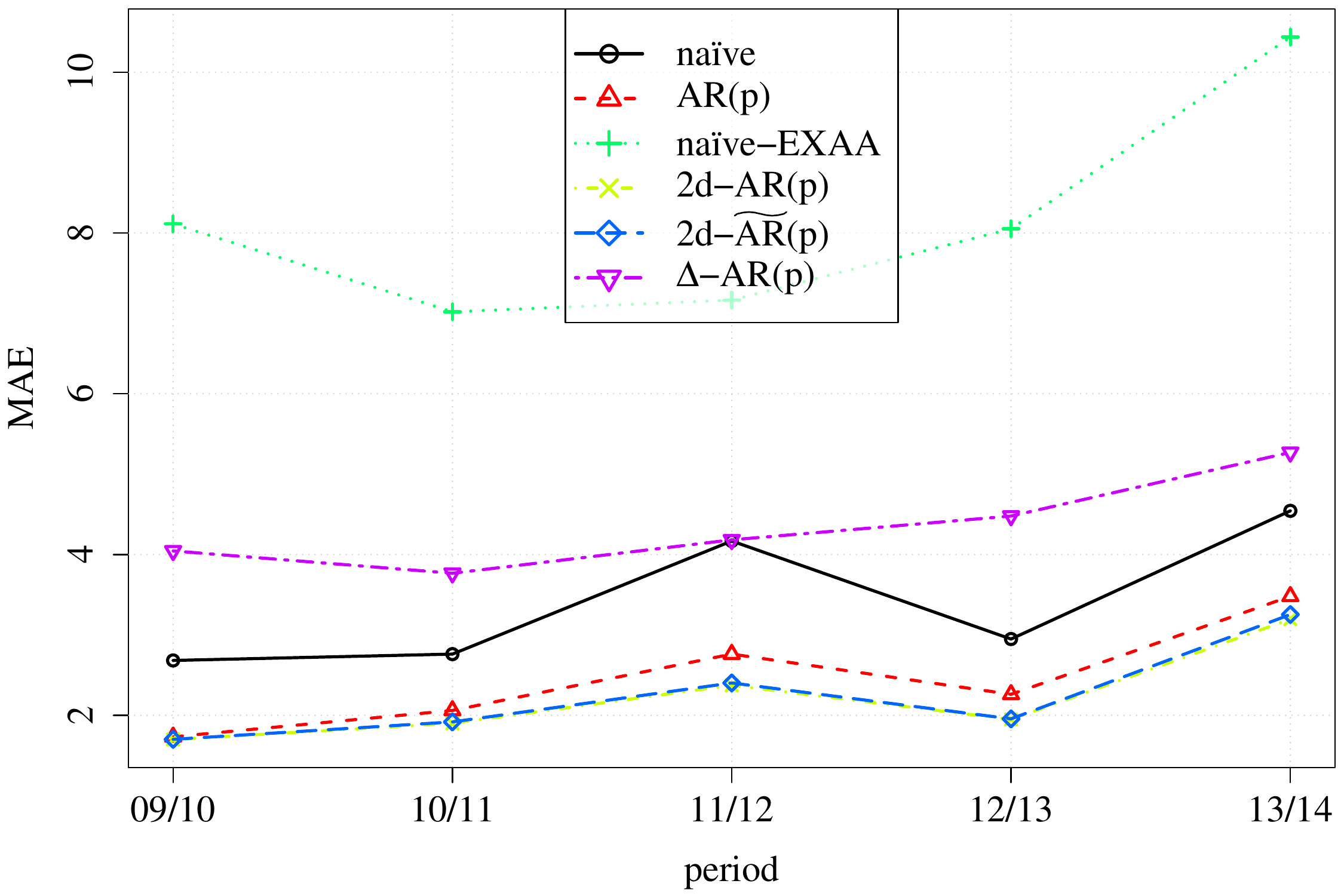} 
  \caption{POLPX PL}
\end{subfigure}
\begin{subfigure}[b]{.49\textwidth}
 \includegraphics[width=1\textwidth, height=.17\textheight]{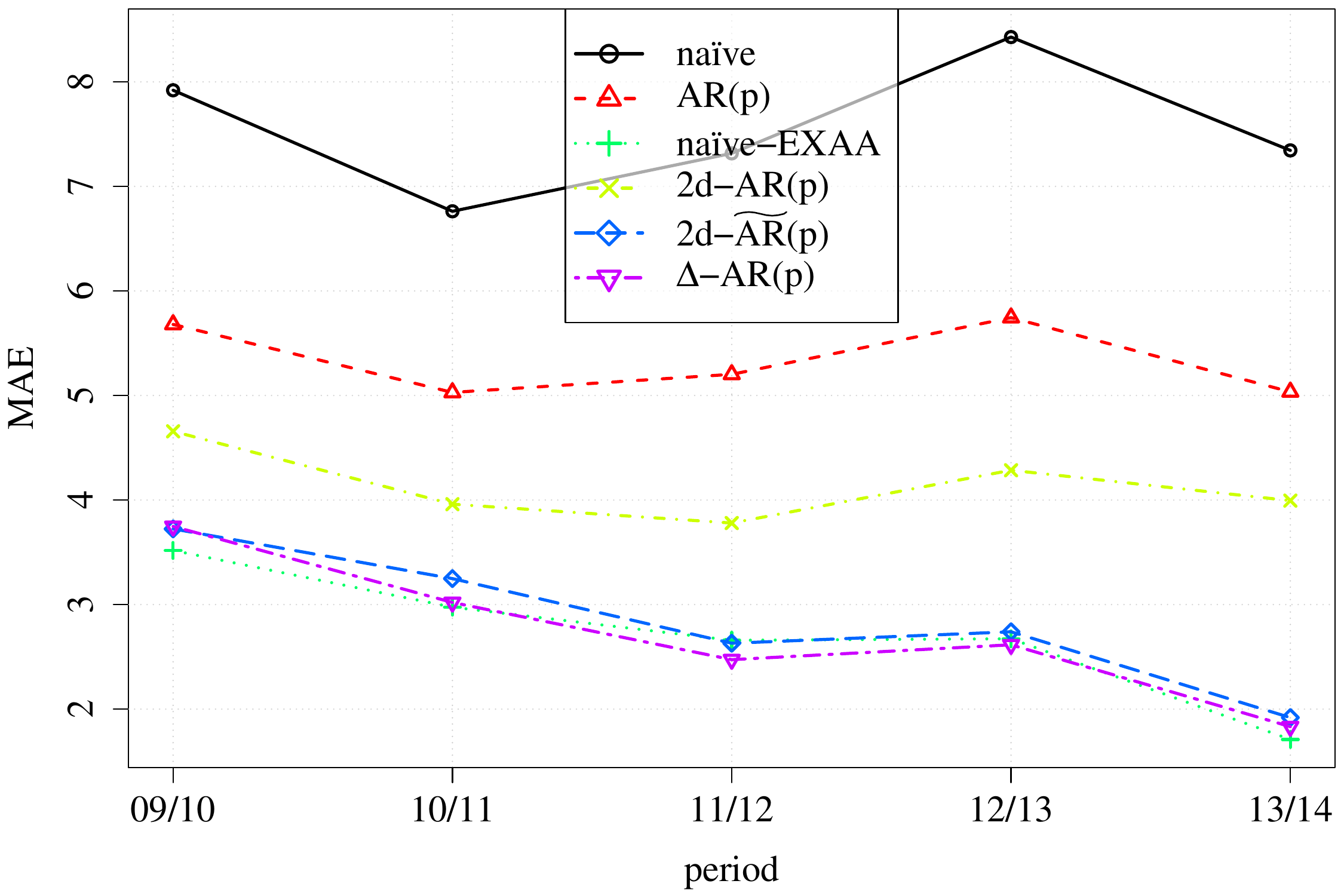} 
  \caption{OTE CZ}
\end{subfigure}
\begin{subfigure}[b]{.49\textwidth}
 \includegraphics[width=1\textwidth, height=.17\textheight]{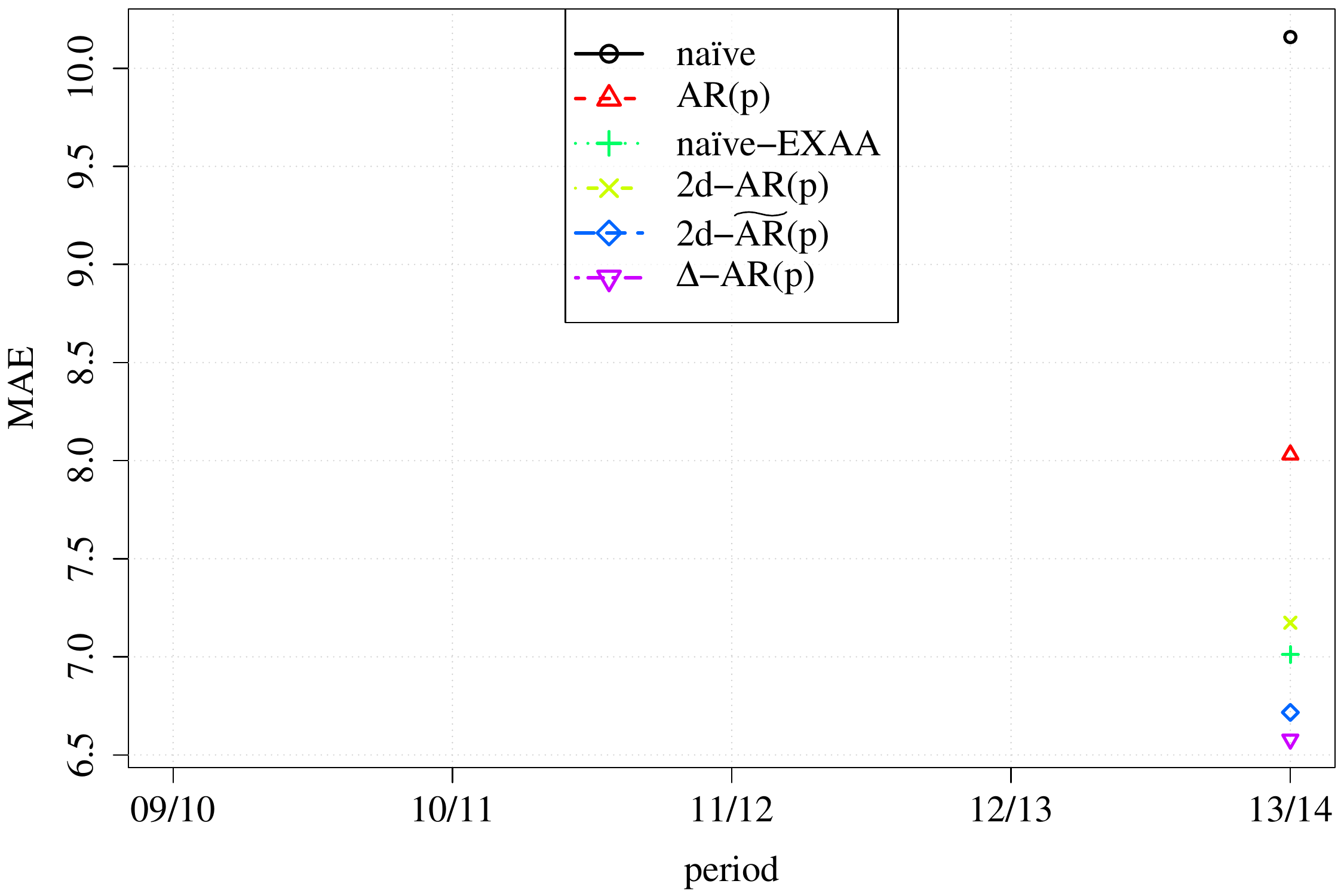} 
  \caption{HUPX HU}
\end{subfigure}
\caption{ $\MAE^\XX_{\text{annual},\T}$ for $\T \in \TT_{\text{annual}}$ (the 5 years of the out-of-sample study). }
 \label{fig_MAE_cp}
\end{figure}

\end{document}